\documentclass{aa}

\usepackage[varg]{txfonts}
\usepackage{sidecap}
\pdfoutput=1
\usepackage{etoolbox}
\makeatletter
\usepackage{adjustbox}
\usepackage{hyperref}
\pdfoutput=1
\usepackage{amssymb}
\usepackage{graphicx}
\usepackage{pdflscape}
\usepackage{tikz}
\def\checkmark{\tikz\fill[scale=0.4](0,.35) -- (.25,0) -- (1,.7) -- (.25,.15) -- cycle;} 

\usepackage{silence}
\usepackage{natbib}
\usepackage{xcolor}
\usepackage[flushleft]{threeparttable}

\DeclareUnicodeCharacter{2212}{-}

\begin{document}

   \title{Duality in spatially resolved star-formation relations in local LIRGs}

   \author{M. S\'anchez-Garc\'ia \inst{1}
         \and
           M. Pereira-Santaella \inst{1}
          \and
           S. Garc\'ia-Burillo \inst{2}
           \and
           L. Colina \inst{1}
           \and
           A. Alonso-Herrero \inst{3}
           \and
           M. Villar-Mart\'in \inst{1}
           \and
           T. Saito \inst{4,5}
           \and
           T. D\'iaz-Santos \inst{6,7}
           \and
           J. Piqueras L\'opez \inst{1}
           \and
           S. Arribas \inst{1}
             \and
           E. Bellocchi \inst{3}
           \and
           S. Cazzoli \inst{8}
           \and
           A. Labiano \inst{3,9}
          }

   \institute{\inst{} Centro de Astrobiolog\'ia (CSIC/INTA), Ctra de Torrej\'on a Ajalvir, km 4, 28850 Torrej\'on de Ardoz, Madrid, Spain \\
        \email{mariasg@cab.inta-csic.es} \and
        \inst{} Observatorio Astron\'omico Nacional (OAN-IGN)-Observatorio de Madrid, Alfonso XII, 3, 28014 Madrid, Spain \and
        \inst{} Centro de Astrobiolog\'ia (CAB, CSIC-INTA), ESAC Campus, E-28692 Villanueva de la Ca\~nada, Madrid, Spain \and
        \inst{} Department of Physics, General Studies, College of Engineering, Nihon University, 1 Nakagawara, Tokusada, Tamuramachi, Koriyama, Fukushima 963-8642, Japan \and
        \inst{} National Astronomical Observatory of Japan, 2-21-1 Osawa, Mitaka, Tokyo 181-8588, Japan \and  
        \inst{} Department of Physics, University of Crete, GR-71003, Heraklion, Greece \and
        \inst{} Institute of Astrophysics, Foundation for Research and Technology−Hellas, Heraklion, GR-70013, Greece \and
     \inst{} Instituto de Astrof\'isica de Andaluc\'ia (IAA-CSIC), Apdo. 3004, E-18008 Granada, Spain \and
        \inst{} Telespazio UK for the European Space Agency (ESA), ESAC, Spain 
 }

   \date{}

 
  \abstract
  {We analyse the star formation (SF) relations in a sample of 16 nearby luminous infrared galaxies (LIRGs) with more than 2800 regions defined on scales of 90 to 500 pc. We used ALMA to map the distribution of the cold molecular gas traced by the J = 2--1 line of CO and archival Pa$\alpha$ HST/NICMOS imaging to trace the recent SF. In four objects, we find two different branches in the Kennicutt-Schmidt relation at 90 pc scales, suggesting the existence of a duality in this relation. The two branches correspond to two different dynamical environments within each galaxy. One branch, which corresponds to 
  the central region of these galaxies (90\% of the regions are located at radii $<$0.85 kpc), shows higher gas and star formation rate surface densities with
  higher velocity dispersion. The other branch, which shows lower molecular gas and SF rate surface densities, corresponds to 
  the more external disk regions (
  r$\sim$1\,kpc). Despite the scatter, the SF efficiency 
  of the galaxies with a dual behaviour increases with increasing boundedness as measured by the $b$ parameter ($b\equiv$ $\Sigma_{H2}$/$\sigma^{2}$ $\propto$ $\alpha_{vir}^{-1}$). At larger spatial scales (250 and 500 pc), the duality disappears. The rest of the sample does not show evidence of this dual behaviour at any scale. }



   \maketitle
\section{Introduction}
The relationship between the rate at which stars form and the amount of gas contained in galaxies is commonly referred to as the star formation (SF) law or as the Kennicutt-Schmidt (KS) relation \citep{Schmidt59, Kennicutt98}. This relation is expressed as
\begin{equation}
    \Sigma_{SFR}\propto\Sigma^{N}_{gas}
\end{equation}
where $\Sigma_{SFR}$ and $\Sigma_{gas}$ are the star formation rate (SFR) and molecular gas surface densities, respectively, and N the power-law index. 
This relation was initially studied in spatially unresolved observations of galaxies, finding a power-law index of 1.4-1.5
\citep{Kennicutt98, Yao2003}. The physical processes that explain the observed power-law index are not clear yet. 
More recently, a duality has been found in the SF laws when normal and starburst galaxies are considered
\citep{daddi2010, genzel2010, burillo2012}. In these studies, normal galaxies show depletion times (t$_{dep}$=M$_{H2}$/SFR) between 4 and 10 times longer than starbursts. 
This duality introduces a discontinuity in the KS relation. In this case, when each galaxy population (normal and starbursts) is treated independently, there is a linear relation (N$\sim$1).

Spatially resolved KS relation studies ($\lesssim$~1~kpc) \citep[e.g.,][]{Leroy2008, Casasola2015, pereira2016, Williams2018, Viaene2018} found a wide range of N values (N$\approx$0.6-3) with a considerable scatter in the relation (0.1-0.4 dex). These results suggest that there is a breakdown in the star-formation law at sub-kpc scales ($\lesssim$ 300 pc), although the correlation is restored at larger spatial scales \citep{Onodera2010, Schruba2010}. This breakdown may be due to the different evolutionary states of individual giant molecular clouds within the galaxies when resolved at sub-kpc scales. In addition to the relation between $\Sigma_{SFR}$ and $\Sigma_{gas}$, 
other parameters, such as the velocity dispersion ($\sigma$)
or boundedness of the gas ($b\equiv\Sigma_{gas}$/$\sigma^{2}\propto\alpha_{vir}^{-1}$, where $\alpha_{vir}$ is the virial parameter), 
have been studied to characterize the local dynamical state of the gas \citep[e.g.,][]{Leroy2017, Sun2018}. These studies suggest that the dynamical environment plays an important role in the ability to form stars within a galaxy.

These previous sub-kpc studies focused on nearby normal and active galactic nuclei (AGN) galaxies. However, more intense local starburst galaxies (i.e., luminous and ultraluminous infrared galaxies; LIRGs and ULIRGs) have been barely studied at sub-kpc scales \citep[e.g.,][]{XU2015, pereira2016, Paraficz2018, Saito2016}. 
In this work, we present a detailed analysis of the SF relations at cloud scales ($\sim$ 100~pc) in a sample of 16 local LIRGs.
 

\section{The sample}\label{sample}
We present new sub-kpc CO(2--1) observations obtained by the Atacama Large Millimeter Array (ALMA) of a representative sample of 16 local LIRGs. 
Our sample is drawn from the volume-limited sample of 34 local LIRGs (40 Mpc $<$ D $<$ 75 Mpc) defined by \cite{AH2006} and contains 85$\%$ of their southern targets which can be observed with ALMA. 
Our sample contains six isolated galaxies, six pre-coalescence systems (interacting galaxies and pairs of galaxies), and four merger objects \citep{Yuan2010, Rich2012, Bellocchi2013}. 
Eight objects are classified as AGN in the optical and/or show evidence of AGN activity from mid-infrared diagnostics \citep{Alonso-Herrero2012}. 
In Table \ref{tab:sample} we present the main properties of the individual galaxies of the sample. 

\begin{table*}[ht!]

    \centering
	
	\caption{The Volume Limited Sample of Local LIRGs}

    \resizebox{\linewidth}{!}{
	\begin{tabular}{ l l c c c c c c c c c c }
		\hline
		\hline
		\multicolumn{2}{c}{Object}& & $\alpha$ & $\delta$ & $z$ & D$_{L}$ & i & log~$L_{IR}$ & Morf. & Spectral & Ref.\\
		
		\cline{1-2}\cline{4-5}
		Galaxy Name & IRAS Name & & J2000.0 & J2000.0 & &  & & & & Class & \\ 

		 &  & & [h m s] & [$^{\circ}$  $\arcmin$ $\arcsec$] & & [Mpc] & [$^{\circ}$] & [L$_{\odot}$] & & & \\ 

			(1) & (2) & & (3) & (4) & (5) & (6) & (7) & (8) & (9) & (10) & (11)\\
		\hline
       ESO~297-G011 & F01341-3735 N & & 01 36 23.40 & -37 19 17.6 & 0.0168  &  73.4  & 38 $\pm$ 11 & 11.13 & 1 & HII & 1,2  \\		
       NGC~1614 & F04315-0840 & & 04 33 59.85 & -08 34 44.0 & 0.0159  & 69.7 & 48 $\pm$ 2 &11.61 & 2 & composite & 3  \\
       NGC~2369 & F07160-6215 & & 07 16 37.73 & -62 20 37.4 & 0.0111 & 49.7 & 66 $\pm$ 6 &11.18 & 0 & composite & 4 \\
       NGC~3110 & F10015-0614 & & 10 04 02.11 & -06 28 29.2 &  0.0163 & 79.8 & 57 $\pm$ 3 &11.37 & 0 & HII & 2,3 \\
       NGC~3256 & F10257-4339 & & 10 27 51.27 & -43 54 13.5 & 0.0093 & 45.7 & - &11.72 & 2 & HII & 5  \\
       ESO~320-G030 & F11506-3851 & & 11 53 11.72 & -39 07 48.9 & 0.0102 & 52.2 & 56 $\pm$ 4 &11.36 & 0 & HII & 4,6 \\
       MCG-02-33-098~W & F12596-1529 & & 13 02 20.00 & -15 46 03.7 & 0.0156 & 75.2 & 54 $\pm$ 6 &11.19 & 1 & composite & 2 \\
       MCG-02-33-098~E & F12596-1529 & & 13 02 20.38 & -15 45 59.7 & 0.0159 & 75.2 & 39 $\pm$ 1 &11.11 & 1 & HII & 2,3,7 \\
       NGC~5135 & F13229-2934 & & 13 25 44.06 & -29 50 01.2 & 0.0136 & 64.8 & 53 $\pm$ 9 &11.33 & 0 & Sy2 & 1,2,3,9,10 \\
       IC~4518~W & F14544-4255 & & 14 57 41.18 & -43 07 55.6 & 0.0160 &  74.6 & 50 $\pm$ 4 &11.16 & 1 & Sy2 & 2,8 \\
       IC~4518~E & F14544-4255 & & 14 57 44.46 & -43 07 52.9 & 0.0154  & 71.2 & 75 $\pm$ 2 &11.12 & 1 & - & 4  \\
       ... & F17138-1017 & & 17 16 35.79 & -10 20 39.4 & 0.0172 & 76.7 & 50 $\pm$ 1 &11.39 & 2 & composite/HII & 2  \\
       IC~4734 & F18341-5732 & & 18 38 25.70 & -57 29 25.6 & 0.0154 & 68.5 & 58 $\pm$ 10 &11.28 & 0 & HII & 2 \\
       NGC~7130 & F21453-3511 & & 21 48 19.52 & -34 57 04.5 & 0.0160 & 67.6 &50 $\pm$ 9 &11.33 & 2 & Sy2 & 3,8,9,10  \\
       IC~5179 & F22132-3705 & & 22 16 09.10 & -36 50 37.4 & 0.0112 & 46.7 & 62 $\pm$ 5 &11.13 & 0  & HII & 3,7 \\
       NGC~7469 & F23007+0836 & & 23 03 15.62 & +08 52 26.4 & 0.0160 & 66.7 & 39 $\pm$ 5 &11.54 & 1 & Sy1 & 8,9,10,11\\ [1ex]
       	\hline
	\end{tabular}}
	\tablefoot{
Col. (1): Galaxy name. Col. (2): IRAS denomination from \cite{Sanders2003}. Col. (3) and (4): right ascension (hours, minutes, seconds) and declination (degrees, arcminutes, arcseconds) from the NASA Extragalactic Database (NED), respectively. Col (5): redshift. Col. (6): Luminosity distance from NED. Col. (7): Inclination. Col. (6): Infrared luminosity (L$_{IR}$(8--1000) $\mu$m) calculated from the IRAS flux densities f$_{12}$, f$_{25}$, f$_{60}$, and f$_{100}$ \citep{Sanders2003}. Col. (8): Morphological class (0 identifies $isolated$ objects, 1 $pre-coalescence$ systems, and 2 stands for $merger$ objects. Col. (9): Nuclear activity class (HII= HII region-like, Sy2=Seyfert 2 y Sy1=Seyfert 1, composite). Col. (10): References of the nuclear classification: 1: \cite{Kewley2001}; 2: \cite{Corbett2003}; 3:\cite{Yuan2010}; 4:\cite{Pereira2011}; 5: \cite{Lipari2000}; 6: \cite{Broek1991}; 7: \cite{Veilleux1995}; 8:\cite{Pereira2010b}; 9: \cite{Alonso-Herrero2012}; 10: \cite{Alonso-Herrero2009}; 11: \cite{Petric2011}. }	
	
\label{tab:sample}
\end{table*}

\section{Observations and data reduction} \label{observationsandreduction}
\subsection{CO(2--1) ALMA data}
We used ALMA Band 6 CO(2--1) observations carried out between August 2014 and August 2018 from several projects (see Table \ref{tab:co}). The observations were obtained using a combination of extended 
and compact 
antenna array configurations, except in the case of the two galaxies part of project 2017.1.00395.S which only used an extended antenna array configuration. The integration time of the sources ranges between $\sim$7 to $\sim$34 min.  
We calibrated the data using the standard ALMA reduction software {\tt CASA}\footnote{\href{http://casa.nrao.edu/}{http://casa.nrao.edu/}} \citep{McMullin2007}. 
We subtracted the continuum emission in the uv plane using an order 0 baseline. For the cleaning, 
we used the Briggs weighting with a robustness parameter of 0.5 \citep{Briggs1995}, providing a spatial resolution of 48--106 pc (0.19$\arcsec$--0.37$\arcsec$). The maximum recoverable scales (MRS) for the compact+extended configuration data range between $\sim$8$\arcsec$ and $\sim$11$\arcsec$ (1.7--2.3\,kpc). In the case of the only extended configuration observations, the MRS is $\sim$3$\arcsec$ (1.1 kpc). In this paper, we study spatial scales between 90 and 500 pc, which are 2 to 25 times smaller than the MRS, so we expect that the missing flux due to the absence of short spacing is low at these scales. In addition, for two of these systems with single-dish CO(2--1) observations, the integrated ALMA and single-dish fluxes agree within 15\% \citep{pereira2016, pereira2016b}.

The final data cubes have channels of 
7.8 MHz ($\sim$ 10 km~s$^{-1}$) for the sample, except ESO320-G030 and NGC 5135, which have channels of $\sim$4 MHz ($\sim$5 km~s$^{-1}$) and $\sim$23 MHz ($\sim$30 km~s$^{-1}$) respectively. The field of view (FoV) of the ALMA single pointing data has a diameter of $\sim$24$\arcsec$ ($\sim$ 5-8~kpc). The three mosaics (MCG-02-33-098, NGC 3256 and NGC 7469) have a diameter between $\sim$38 and $\sim$48$\arcsec$ ($\sim$11 and 17~kpc). We applied the primary beam correction to the data cubes. Further details on the observations for each galaxy are listed in Table \ref{tab:co}.

\begin{table*}[ht!]
    \centering
	\caption{CO(2--1) observations of the sample.}
	\begin{threeparttable}
	\begin{tabular}{ l l c c c c c c c c}
		\hline
		\hline
		\multicolumn{2}{c}{Object} & $\theta_{maj}\times\theta_{min}$ & $\theta_{m}$ &  P.A. & Sensitivity & Project & Mosaics & MRS & SINFONI  \\
		\cline{1-2}
		Galaxy Name & IRAS Name & [$\arcsec$]~~~~~~[$\arcsec$] & [$\arcsec$, pc] & [$^{\circ}$] & [mJy beam$^{-1}$] & PI & & [$\arcsec$] & \\ [0.5ex]

		(1) & (2) & (3) & (4) & (5) & (6) & (7) & (8) & (9) \\
		\hline
       ESO~297-G011 & F01341-3735 & 0.21$\times$0.16 & 0.19, 67 & -72  & 0.53 & MPS & & 9.4 & \\		
       NGC~1614 & F04315-0840  & 0.22$\times$0.15 & 0.19, 62 & -74 & 0.43 & MPS & & 10.8 & \\
       NGC~2369 & F07160-6215 & 0.24$\times$0.21 & 0.22, 53 & 88 & 0.51 & MPS & & 8.8 & \checkmark \\
       NGC~3110 & F10015-0614 & 0.26$\times$0.21 & 0.23, 89 & -83 & 0.35 & MPS & & 8.8 & \checkmark\\
       NGC~3256 & F10257-4339 & 0.23$\times$0.21 & 0.22, 48 & 63 & 0.43 & KS & \checkmark & 5.4 & \checkmark \\
       ESO~320-G030 & F11506-3851 & 0.30$\times$0.24 & 0.27, 68 & 63 & 0.89 & LC1 & & 8.7 & \checkmark\\
       MCG-02-33-098~W & F12596-1529 & 0.23$\times$0.17 & 0.20, 73 &  89 & 0.48 & MPS & \checkmark & 9.4 & \\
       MCG-02-33-098~E &  F12596-1529 & 0.23$\times$0.17 & 0.20, 72 &  89 & 0.48 & MPS & \checkmark & 9.4 & \\
       NGC~5135 &  F13229-2934 & 0.31$\times$0.22 & 0.26, 82 & 63&  0.21 & LC2 & & 9.4  & \checkmark \\
       IC~4518~W & F14544-4255  & 0.23$\times$0.20 & 0.21, 75 & -86 & 0.46 & MPS & & 10.3 & \\
       IC~4518~E & F14544-4255  & 0.23$\times$0.20 & 0.21, 73 & -87 & 0.47 & MPS & & 10.3 & \\
       ... &  F17138-1017 & 0.26$\times$0.22 & 0.24, 87  & -62 & 0.75  & MPS & & 7.6 & \checkmark \\
       IC~4734 & F18341-5732 & 0.25$\times$0.21 & 0.23, 75 & -73 &  0.77 & TDS & & 2.7 & \\
       NGC~7130 & F21453-3511 & 0.36$\times$0.29 & 0.32, 105  & 69 & 0.29 & MPS &  & 9.9 & \checkmark \\
       IC~5179 & F22132-3705 & 0.40$\times$0.34 & 0.37, 82 & 44 &  0.45 & MPS &  & 9.8 & \checkmark \\
       NGC~7469 & F23007+0836 & 0.23$\times$0.18 & 0.21, 65 & -39 & 0.29 & TDS &  \checkmark & 2.8 &\\ [1ex]
       	\hline
	\end{tabular}

	\begin{tablenotes}
	\item \textbf{Notes}: Col. (1): Galaxy name. Col. (2): IRAS denomination from \cite{Sanders2003}. Col. (3): major ($\theta_{maj}$) and minor ($\theta_{min}$) FWHM beam sizes. Col. (4): mean FWHM beam size ($\theta_{m}$) in arcseconds and parsecs, respectively. Col. (5): position angle (P.A.) in degrees. Col. (6): 1$\sigma$ line sensitivity of the CO(2--1) observations.  Col. (7): Principal investigator of the ALMA project: MPS: Miguel Pereira-Santaella (2017.1.00255.S),  KS: Kazimierz Sliwa (2015.1.00714.S), LC1: Luis Colina (2013.1.00271.S), LC2: Luis Colina (2013.1.00243.S) and TDS: Tanio D\'iaz-Santos (2017.1.00395.S). Col. (8): \checkmark galaxies with mosaics data.  Col. (9): Maximum recoverable scales. Col. (10): \checkmark galaxies with SINFONI data.
	\end{tablenotes}
	\end{threeparttable}
    \label{tab:co}
\end{table*}

A common spatial scale of about 70-90 pc was defined to have a homogeneous data set. 
We convolved to 80 pc the data cubes of the galaxies with spatial resolutions better between 48 and 68 pc (ESO297-G011, NGC1614, NGC2369, NGC3256, ESO320-G030 and NGC7469). For the remaining objects, we directly used cleaned data cubes with spatial resolutions between 72 and 89 pc. 
For NGC~7130 the original spatial 
resolution was $\sim$110 pc. 
We used this slightly larger spatial scale for the SF law in this galaxy.
We obtained the CO(2--1) moment 0 and 2 maps  
by doing the following: 
to identify the CO(2--1) emission in each channel of data cube, we selected pixels with fluxes $>$ 5$\sigma_{CO}$. 
We estimated the sensitivity $\sigma_{CO}$ in a spectral channel without evident CO(2--1) emission and with no primary beam correction. In addition to the 5$\sigma_{CO}$ criterion, and to ensure that the emission of data cubes does not include noise spikes, we did not consider spatial pixels that have emission from less than three spectral channels. Finally, for each pixel 
meeting the above criteria, we expanded the spectral range to include a channel before and after the emission to ensure that line profile wings below 5$\sigma_{CO}$ are also considered.
In addition to the nominal 90 pc resolution, we smoothed the data to 240 and 500 pc resolutions to study the effect of the spatial scale on the SF laws.

\subsection{Ancillary HST/NICMOS data}
We used the continuum subtracted near-infrared narrow-band Pa$\alpha$ 1.87 $\mu m$ images taken with the NICMOS instrument on board the Hubble Space Telescope (HST) to map the distribution of recent star formation in the galaxies of the sample (see \citealt{AH2006}).

We downloaded the raw data from the $Hubble$ Legacy Archive (HLA) \footnote{\href{http://hla.stsci.edu/hlaview.html}{http://hla.stsci.edu/hlaview.html}}. The individual frames were combined using the PyDrizzle package with a final pixel size (0.03$\arcsec$) half of the original to improve the spatial sampling. 
The FoV of the images is approximately 19$\arcsec$.5$\times$19$\arcsec$.5 ($\sim$ 4.2-7.4 kpc). 
To obtain the final images, we subtracted the background emission and corrected the astrometry using stars within the NICMOS FoV in the F110W ($\lambda_{\rm eff} = 1.13 \mu m$) or F160W ($\lambda_{\rm eff} = 1.60 \mu m$) filters and the Gaia DR2 catalogue \footnote{\href{http://www.cosmos.esa.int/web/gaia/dr2}{http://www.cosmos.esa.int/web/gaia/dr2}}. Three objects (ESO297-G011, MCG-02-33-098~E/W, and IC4518~E) do not have Gaia stars in their NICMOS images FoV. In these cases, we adjusted the astrometry using likely NICMOS counterparts of the regions detected in the ALMA continuum and CO(2--1) maps. 
After that, the images were rotated to have the standard north-up, east-left orientation.
The Pa$\alpha$ maps (spatial resolutions of 25-50 pc) 
were convolved with a Gaussian kernel 
to match the angular resolution of the ALMA maps. 

\subsection{Region selection}
We defined circular apertures centred on local maxima in the CO(2--1) moment 0 maps with a diameter of 90 pc, 240 pc and 500 pc, depending on the spatial resolution of the maps. To do so, we first sorted the CO moment 0 pixel intensities. Then, we defined circular regions using as centre the pixels in descending intensity order preventing any overlap between the regions.
With this method, we end up having independent non-overlapping regions centred on local emission maxima that cover all the CO emission in each galaxy. 
In total, we defined 4802 regions for the whole sample.

We estimated the cold molecular gas mass using the Galactic CO-to-H$_{2}$ conversion factor, $\alpha_{CO}^{1-0}$=4.35 M$_{\odot}$/K/(km/s)/pc$^{2}$ \citep{Bolatto2013} and the CO(2--1)/CO(1--0) ratio (R$_{21}$) of 0.7 obtained from the single-dish CO data of LIRG IC4687
\citep{Albrecht2007}. The R$_{21}$ value used is within the range found by \cite{Garay:1993} in infrared galaxies and is similar to the one found by \cite{Leroy:2013} in nearby spiral galaxies.
We explore the variation of the CO-to-H$_{2}$ conversion factor in Sect. \ref{factor_co}.   
We calculated the molecular gas mass surface density ($\Sigma_{H_{2}}$) taking into account the area of the selected regions.

Once we have the regions in CO(2--1) emission maps, we selected the regions in the Pa$\alpha$ maps. These regions are at the same spatial coordinates as the CO(2--1) regions. 
In this case, we considered Pa$\alpha$ detections when the line emission is above 3$\sigma_{Pa\alpha}$. The $\sigma_{Pa\alpha}$ in these images corresponds to the background noise. 
. 
The regions that are below 3$\sigma$ correspond to the upper limits. Pa$\alpha$ emission is detected in 2783 regions (58\% of the total). 
Then, we estimated the SFR surface density ($\Sigma_{SFR}$) of the regions. We used the H$\alpha$ \cite{KE2012} calibration, which assumes a \cite{Kroupa2001} initial mass function, and an H$\alpha$/Pa$\alpha$ ratio of 8.6 \cite[case B at $T_{e}$=10.000~K and $n_{e}$ = 10$^{4}$ cm$^{-3}$,][]{Osterbrock:2006}. The variation of this ratio is $\sim$15$\%$ due to changes in the physical properties of the ionized gas (i.e., $T_{e}$ = 5–20 $\times$ 1000~K and  $n_{e}$ = 10$^{2}$-10$^{6}$ cm$^{-3}$.
We took into account the area of the selected regions, obtaining the SFR surface density.
All these $\Sigma_{H2}$ and $\Sigma_{SFR}$ values are corrected for the inclination of each galaxy (see Table~\ref{tab:sample}). 

Both the SFR and the cold molecular gas surface density estimates are affected by the flux calibration errors. We assume an uncertainty of about 10\% in the ALMA data (see ALMA Technical Handbook \footnote{\href{http://almascience.eso.org/documents-and-tools/latest/documents-and-tools/cycle8/alma-technical-handbook}{http://almascience.eso.org/documents-and-tools/latest/documents-and-tools/cycle8/alma-technical-handbook}}), and $\sim$15-20\% in the case of NICMOS data \citep{AH2006,Boker1999}. 

\subsection{Extinction correction}
To correct the Pa$\alpha$ emission for extinction, we used the Br$\delta$ and Br$\gamma$ line maps observed at  240 pc scales with the SINFONI instrument on the Very Large Telescope (VLT) in eight objects from our sample 
\citep[effective FoV between 8$\arcsec\times$8$\arcsec$ and 12$\arcsec\times$12$\arcsec$; ][]{Piqueras2013} to derive A$_{K}$ (see Table \ref{tab:co}).
We calculated the Br$\delta$/Br$\gamma$ ratio in circular regions with a diameter of 240 pc. We assumed an intrinsic Br$\delta$/Br$\gamma$ ratio of 1.52 \citep{HS1987}
and the \cite{Fitzpatrick1999} extinction law. In each 240\,pc region, we determined A$_{K}$ (A$_{K}$=0.11$\times$A$_{V}$) and the column density from the CO(2--1) 240 pc maps (N$_{H_{2}}$). 

The N$_{H_2}$ values were divided in five equally spaced ranges between $\log N_{H_2}\slash {\rm cm^{-2}}$= 22.55 and 23.88.
For each range, we estimated the mean and standard deviation of A$_{K}$ obtaining slightly increasing values with $N_{H_2}$ between 0.95$\pm$0.6 and  1.98$\pm$1.29 mag. 
To obtain an estimation of the extinction, 
we measured N$_{H_{2}}$ in the circular apertures with a diameter of 90 (110), 240 and 500 pc in our entire sample and assigned them the mean A$_{K}$ corresponding to their N$_{H_{2}}$ range. We assume that the galaxies without SINFONI data (half of the sample) follow the same trend found between A$_{K}$ and N$_{H_{2}}$ in the other eight galaxies. 

\subsection{The effects of CO--to--H$_{2}$ conversion factor}
\label{factor_co}

The obtained cold molecular gas masses depend on the conversion factor ($\alpha_{CO}$) used. In this paper we assume a Galactic $\alpha_{CO}$ conversion factor to derive molecular gas masses. As argued in the following, we do not expect that a lower conversion factor, typical of ULIRGs \citep{Papadopoulos2012} is appropriate for our targets. 

Our sample does not contain strongly interacting objects or compact mergers like most local ULIRGs. The galaxies of our sample have a mean infrared luminosity of log(L$_{IR}$/L$_{\odot}$)=11.30. 
In addition, galaxies of our sample show a mean effective radius of the molecular component (R$^{eff}_{CO}$) of 740 pc (Bellocchi et al. in prep), while local ULIRGs show a mean value of R$^{eff}_{CO}$=340 pc \citep{Pereira:2021}. Therefore, it is likely that $\alpha_{CO}$ of our sample differs from that of local ULIRGs. 

The CO-to-H$_{2}$ conversion factor can be affected by the metallicity of the galaxies, showing higher values with decreasing metallicity \citep[$\alpha_{CO}$ = 4.35~(Z/Z$_{\odot}$)$^{-1.6}$ M$_{\odot}$pc$^{-2}$(K~km~s$^{-1}$)$^{-1}$,][]{Accurso2017}. 
\cite{Rich2012} studied the metallicity in some local (U)LIRGs, showing a decrease in the abundance with increasing radius. In the case of the metallicity in local disks, \cite{Ssanchez2014} observed a similar behaviour. Based on these works, the expected variation of the conversion factor due to metallicity gradients at $r<4$~kpc is small, 20-30$\%$.

\section{Results and discussion} \label{resultsanddiscussion}
\subsection{The star formation relation for individual galaxies}

\begin{figure*}[ht!]
   \centering
    \includegraphics[width=.7\linewidth]{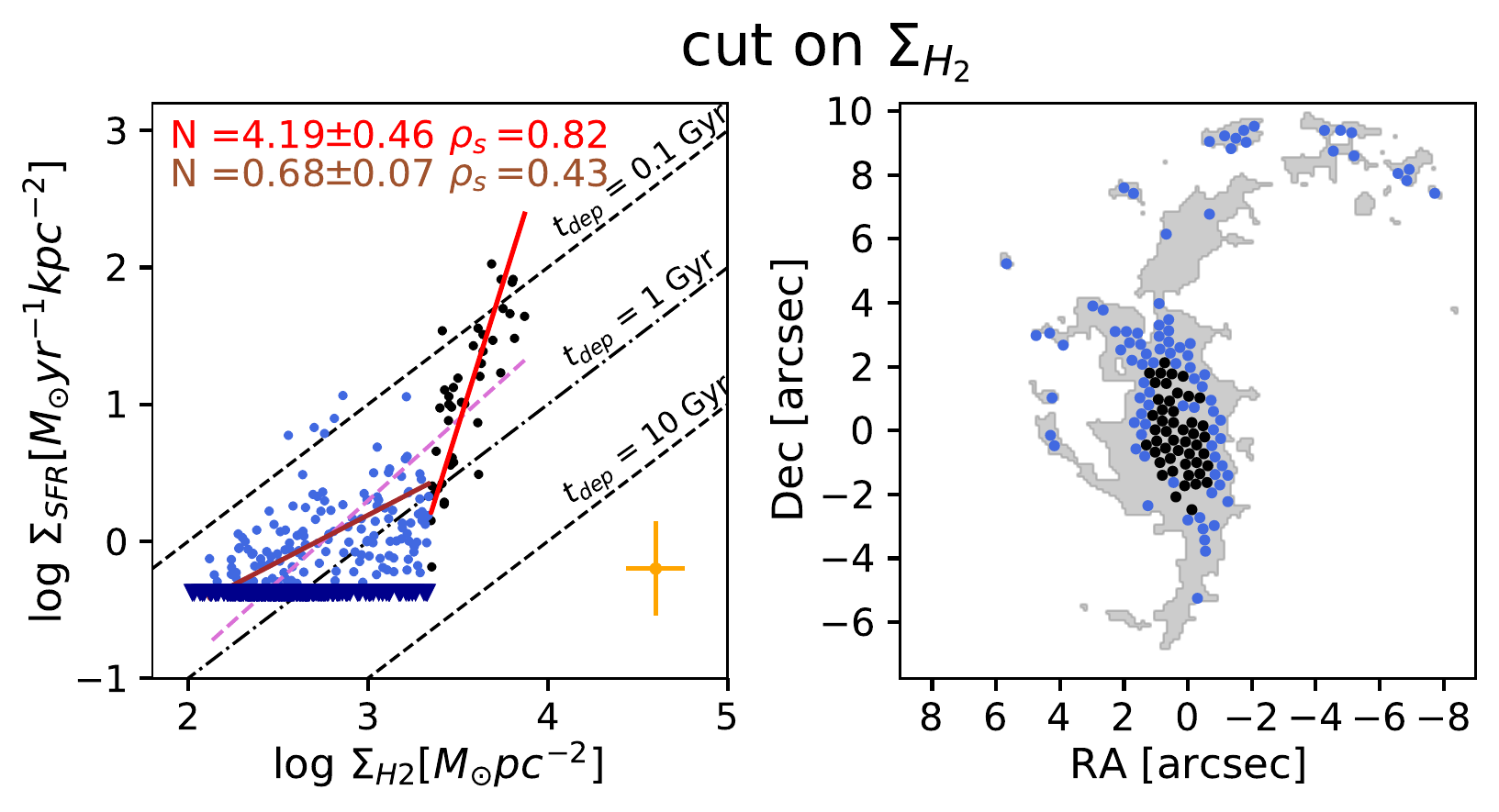}
 \caption{$Left$ $panel$: SFR surface density ($\Sigma_{SFR}$) as a function of the molecular gas surface density ($\Sigma_{H_{2}}$) derived from CO(2--1) in NGC~7130 at 110 pc scale. The blue and black points show the two branches derived applying the MARS method with breaking points in the $\log \Sigma_{\rm H_{2}}$ axis. The black dashed lines mark constant depletion times (t$_{dep}$ = $\Sigma_{H_{2}}$/$\Sigma_{SFR}$). The red and brown solid lines are the best fit for each branch. We indicate the Spearman's rank correlation coefficients ($\rho_{s}$) and the power-law indices (N) of the derived best-fit KS relations. The pink dashed line is the best fit for whole points. The inverted triangles indicate upper limits. The errorbars indicate the mean systematic uncertainties in $\Sigma_{H_{2}}$ of $\pm$0.14 dex (horizontal) and the extinction correction in $\Sigma_{SFR}$ of $\pm$0.21 dex (vertical). 
$Right$ $panel$: Location of the regions on the 
CO(2--1) map (grey). The black and blue circles correspond to regions in each of the two branches.
 }
    \label{ks:ngc7130x}%
\end{figure*}
    

We studied the molecular KS relation for each LIRG at scales of 90 (110) pc. As an example, 
Fig. \ref{ks:ngc7130x} shows the SFR surface density as a function of molecular gas surface density for %
NGC~7130 (similar figures for the rest of the sample are presented in Appendix \ref{app:figures_regions}). The KS diagram suggests that the regions follow two different power-laws. These two branches were identified using the Multivariate Adaptive Regression Splines (MARS) fit \citep{Friedman} in $\log \Sigma_{SFR}$ and $\log \Sigma_{H2}$, which gives the position of the breaking points (cut-points) for a linear regression with multiple slopes. We obtained the adjusted coefficient of determination, the cut-points, and their errors using MARS fit in 100 realizations of the data based on the uncertainties in both axes. 

We consider that the MARS breaking point is significant when the adjusted coefficient of determination found by MARS ($\bar{R}_{MARS}^{2}$) is larger than that of the linear fit ($\bar{R}_{linear}^{2}$). 
The adjusted coefficient of determination is used to compare the linear and MARS fits since it takes into account both the number of terms in the model and the number of data points. 
In this galaxy, the break of a linear regression occurs at $\log \Sigma_{H_{2}}\slash ({\rm M_{\odot}\, yr^{-1} kpc^{-2})}$=3.35 (for cut on $\Sigma_{SFR}$, see Figure \ref{anexo:ngc7130y}).

\begin{table*}[ht!]
    \centering
	\caption{Statistical parameters for dual galaxies.}
	\begin{tabular}{lc cc c cc}
	\hline
	\hline
	galaxies & 	& \multicolumn{2}{c}{cut on log$_{10}\Sigma_{H_{2}}$} & & \multicolumn{2}{c}{cut on log$_{10}\Sigma_{SFR}$}\\ 
	\cline{3-4} \cline{6-7}
	
	 & $\bar{R}_{linear}^{2}$ & cut-point & $\bar{R}_{MARS}^{2}$ &  & cut-point & $\bar{R}_{MARS}^{2}$  \\ 
	[0.5ex]	
	\hline
	NGC~1614 & 0.69 & 3.31$\pm$0.12 & 0.75$\pm$0.04 & & 1.68$\pm$0.32 & 0.71$\pm$0.03  \\
	NGC~3110 & 0.36 & 3.21$\pm$0.15 & 0.50$\pm$0.08 && 0.81$\pm$0.11 & 0.49$\pm$0.07 \\
	NGC~7130 &0.49  & 3.35$\pm$0.11 & 0.71$\pm$0.05 && 0.53$\pm$0.15 & 0.52$\pm$0.03 \\ 
	IC~5179 & 0.48 & 3.12$\pm$0.18 & 0.62$\pm$0.05 && 0.61$\pm$0.37 & 0.52$\pm$0.03 \\

	\hline
	\end{tabular}
	
	\tablefoot{The adjusted coefficient of determination ($\bar{R}_{linear}^{2}$) obtained by linear fit and the breaking point, the adjusted coefficient of determination ($\bar{R}_{MARS}^{2}$) obtained by MARS fit.}
	\label{tab:r2_duals}
\end{table*}

\begin{table*}[ht!]
    \centering
	\caption{Statistical parameters for the non dual galaxies}
	\begin{tabular}{lc cc c cc}
	\hline
	\hline
	galaxies & &	\multicolumn{2}{c}{cut on log$_{10}\Sigma_{H_{2}}$} & & \multicolumn{2}{c}{cut on log$_{10}\Sigma_{SFR}$} \\ 
	\cline{3-4} \cline{6-7}
	
	 &  $\bar{R}_{linear}^{2}$ & cut-point & $\bar{R}_{MARS}^{2}$  & & cut-point & $\bar{R}_{MARS}^{2}$  \\ 
	[0.5ex]	
	\hline
	ESO~297-G011 & 0.22 & 3.25$\pm$0.12 & 0.13$\pm$0.05 & & 0.54$\pm$0.15 & 0.11$\pm$0.05 \\
	NGC~2369 & 0.63 & 3.45$\pm$0.20 & 0.60$\pm$0.03 & & 0.62$\pm$0.22 & 0.60$\pm$0.04  \\
	NGC~3256 & 0.36 & 3.92$\pm$0.16 & 0.24$\pm$0.06 & & 2.17$\pm$0.40 & 0.26$\pm$0.06 \\
	ESO~320-G030 & 0.18 & 2.89$\pm$0.10 & 0.17$\pm$0.03 & & 0.95$\pm$0.12 & 0.16$\pm$0.05 \\
	MCG-02-33-098~W & 0.65 & 3.50$\pm$0.13 & 0.63$\pm$0.05 && 1.72$\pm$0.29  & 0.64$\pm$0.04 \\
	MCG-02-33-098~E & 0.27 & 3.02$\pm$0.14 & 0.11$\pm$0.05 & & 0.95$\pm$0.15 & 0.08$\pm$0.03 \\
	NGC~5135 & 0.35 &3.44$\pm$0.18 & 0.33$\pm$0.09 & & 0.33$\pm$0.19 & 0.33$\pm$0.08 \\
	IC~4518~W & 0.18 & 3.57$\pm$0.11 & 0.15$\pm$0.06 &  & 0.44$\pm$0.21 &0.15$\pm$0.07  \\
	IC~4518~E & 0.65 & 2.85$\pm$0.09 & 0.63$\pm$0.02 &  & 0.30$\pm$0.20 & 0.52$\pm$0.05 \\
	IRAS~F17138-1017 & 0.50 & 2.75$\pm$0.21 & 0.46$\pm$0.07 & & 0.21$\pm$0.18 & 0.47$\pm$0.05  \\
    IC~4734 & 0.79 & 3.55$\pm$0.14 & 0.83$\pm$0.05 && 1.24$\pm$0.29 & 0.81$\pm$0.04 \\
	NGC~7469 & 0.62 & 3.47$\pm$0.11 & 0.63$\pm$0.04 && 0.93$\pm$0.12 & 0.62$\pm$0.05  \\
	\hline
	\end{tabular}
	\tablefoot{The adjusted coefficient of determination ($\bar{R}_{linear}^{2}$) obtained by linear fit, and the breaking point, the adjusted coefficient of determination ($\bar{R}_{MARS}^{2}$) obtained by MARS fit to the log$_{10}\Sigma_{H_{2}}$ and $log_{10}\Sigma_{SFR}$ axes in the KS relation for the non-dual galaxies.}
	\label{tab:r2_noduals}
\end{table*}

We fit the two branches using the orthogonal distance regression (ODR) method. This fit gives power-law indices of N=4.19$\pm$0.46 and N=0.68$\pm$0.07.  
The $right$ panel of Fig.~\ref{ks:ngc7130x} shows that the branch with higher gas and SFR densities ($left$ panel) is located in the central region of the galaxy (at radii up to 0.85~kpc), 
while the other branch with lower gas and SFR densities is located in the more external disk regions. The duality is reinforced if we consider a factor $\alpha_{CO}$=0.8 \citep{Downes:1998} typical of ULIRGs in the central regions of our galaxies.

We do not include in our analysis the upper limits. \cite{Pessa2021} studied the influence of the non-detections in several resolved scaling relations. In general, the non-detections could artificially flatten the relations at small spatial scales, resulting in a steepening when the analysis is carried out at larger spatial scales compared to the small scales. This is because the pixels with signal are averaged with the non-detection pixels at larger scales. However, they obtained that the no consideration of the non-detections in the star formation relation has a small impact on the measured slope.

\subsection{The star-formation relation across the sample} \label{section:ksplotsample}


  \begin{figure*}[htp!]
   \centering
   \includegraphics[width=.999\linewidth]{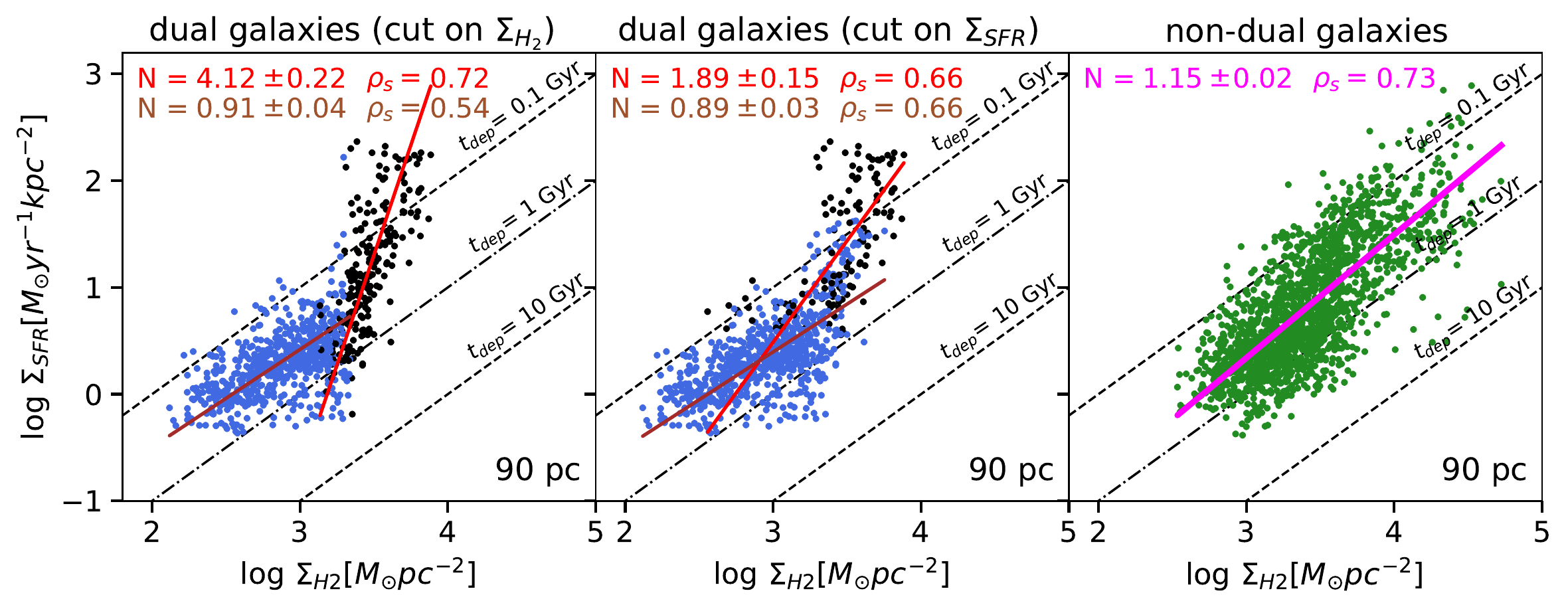}
      \includegraphics[width=.72\linewidth]{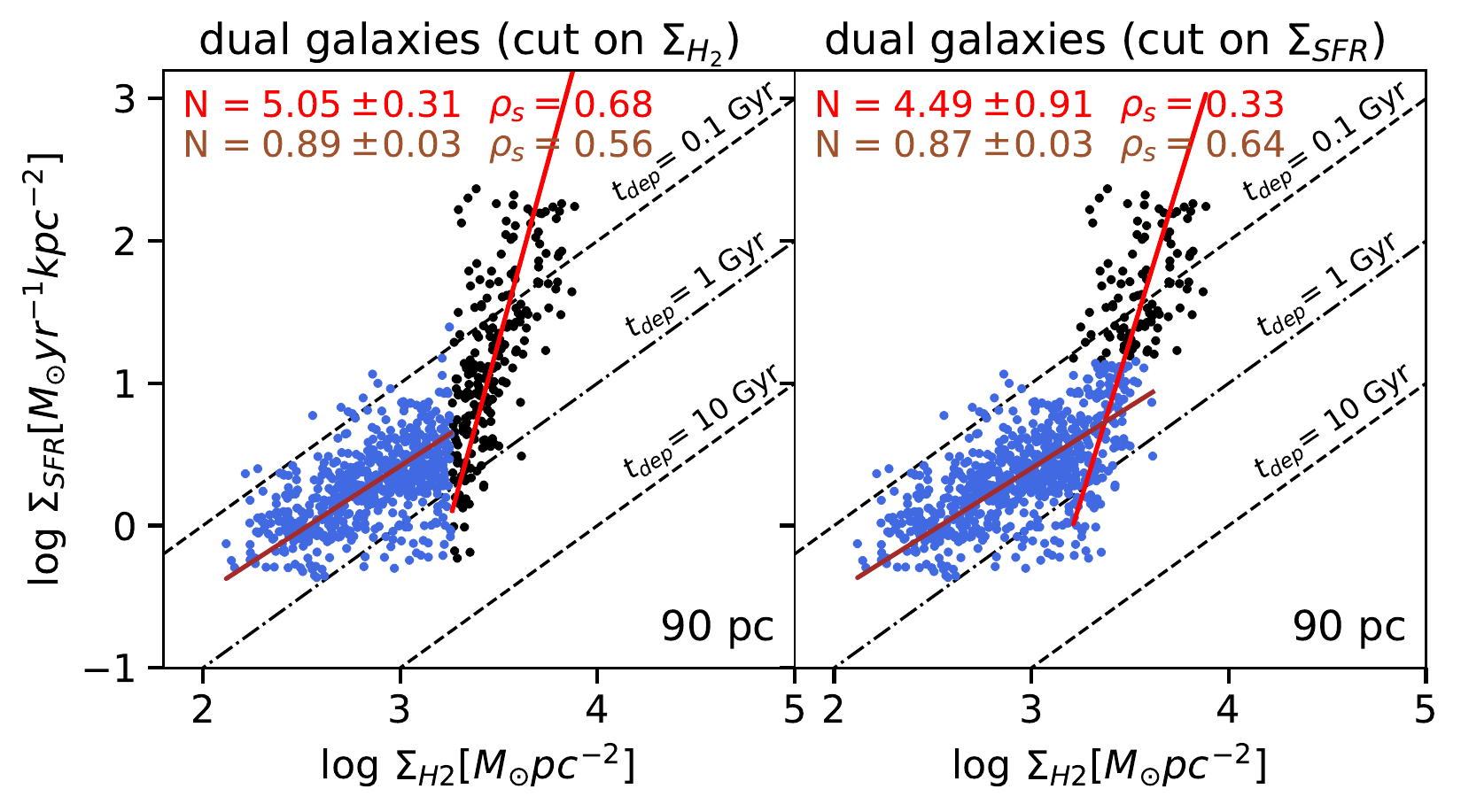}
     \caption{SFR surface density as a function of the molecular gas surface density using 90 (110) pc regions. 
     $Top$ $left$ and $top$ $middle$ $panels$: The black and blue points correspond to the two different regimes (branches) identified in the four dual galaxies using the MARS method in each individual galaxy. 
     $Top$ $right$: The green circles correspond to the regions in the remaining twelve galaxies. 
     $Bottom$ $row$: The black and blue points correspond to the two branches identified using the MARS method in all the regions from the dual galaxies. 
     The red, brown, and magenta solid lines indicate the best fit for each regime. The dashed lines mark constant t$_{dep}$.
     }
        \label{ks:total}%
    \end{figure*}

\begin{figure*}[htp]
   \centering
    \includegraphics[width=.82\linewidth]{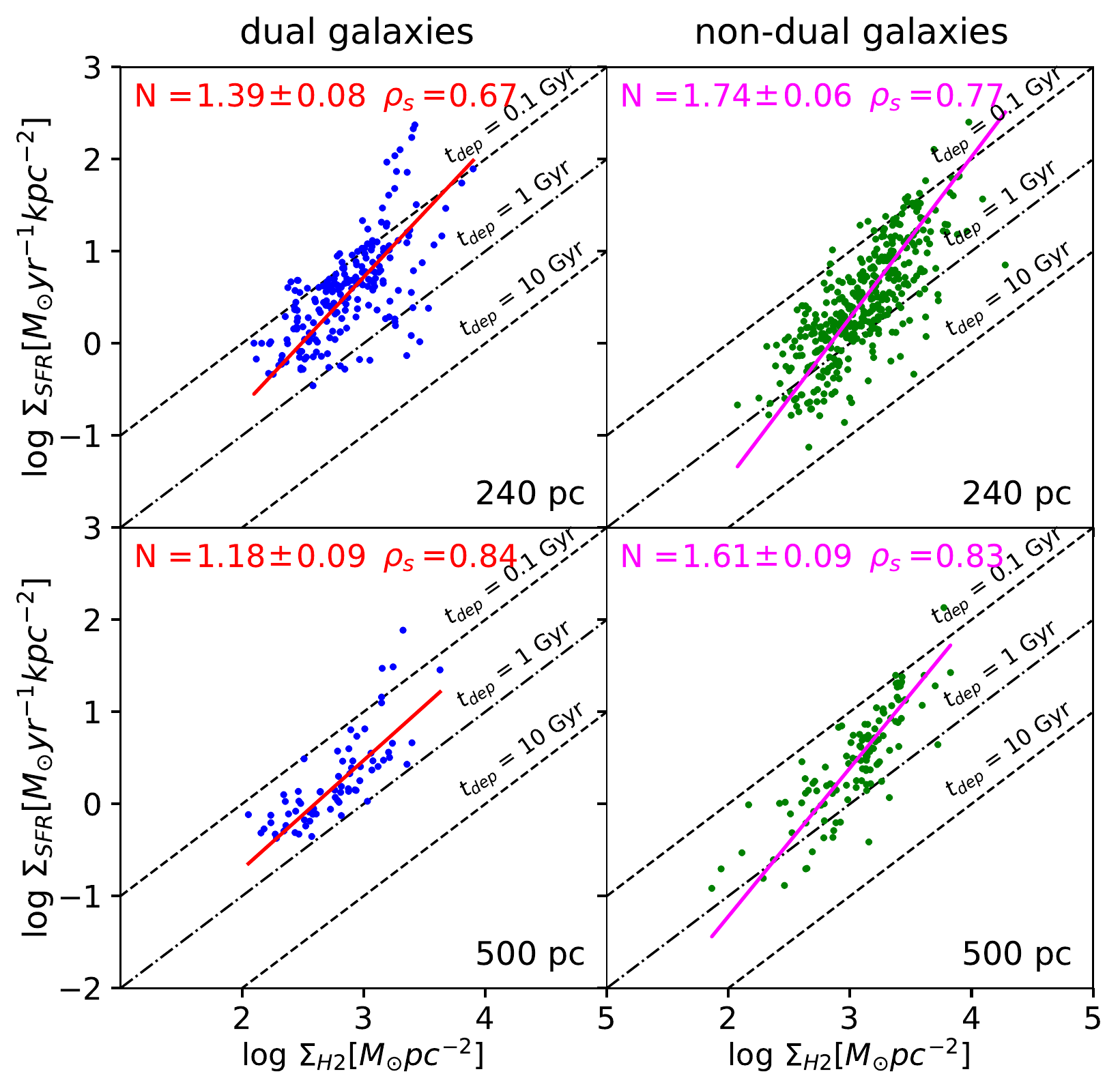}

   \caption{SFR surface density as a function of the molecular gas surface density using 240 ($top$ $row$) and 500 ($bottom$ $row$) pc regions. The blue and green circles correspond to the regions of the galaxies with two regimes at 90 pc and the non-dual galaxies, respectively. The dark orange and magenta solid lines indicate the best fit. The dashed lines mark constant t$_{dep}$.}
    \label{anexo:fig_resolutions} %
\end{figure*}

We repeated the same analysis for the rest of the sample finding two different regimes (branches) in the KS relations in four galaxies (25$\%$ of the sample; dual galaxies hereafter; see Table \ref{tab:r2_duals} and Fig. \ref{anexo:figuresdual}). At larger spatial scales (240 and 500~pc), the duality disappears and a standard single power-law KS relation is recovered (see Fig. \ref{anexo:fig_resolutions}).

The remaining twelve galaxies (75$\%$ of the sample) can be modelled with a single power-law (non-dual galaxies hereafter; see Table~\ref{tab:r2_noduals} and Fig.~\ref{anexo:figuresnondual}) at 90 pc scales.

For the four dual galaxies, the cut-points on both the $\Sigma_{H2}$ and $\Sigma_{SFR}$ axes are similar ($\log \Sigma_{H2}\slash {\rm (M_{\odot}\,pc^{-2})} \approx$ 3.25 and $\log \Sigma_{SFR}\slash {\rm ( M_{\odot}\,yr^{-1}\,kpc^{-2})}\approx$ 0.91). 
 
Therefore, in the $top$ $left$ and $top$ $middle$ panels of Fig.~\ref{ks:total}, we combine all the regions of the dual galaxies with a cut on both axes obtained in each individual dual galaxy.
We find that the power-law for the regions above the cut-points (hereafter referred to as high-N regions) is steeper than for the regions below them (hereafter referred to as low-N regions).  The indices of the best power-law fits are N=4.12$\pm$0.22 
(high-N regions) and N=0.91$\pm$0.04 
(low-N regions) when using the $\Sigma_{H2}$ cut-point (Fig.~\ref{ks:total} $top$ $left$) 
and N=1.89$\pm$0.15
(high-N regions) and N=0.89$\pm$0.03
(low-N regions) when we consider the $\Sigma_{SFR}$ cut-point (Fig.~\ref{ks:total} $top$ $middle$). 

When we fit all the regions from the dual galaxies using the MARS method (Fig.~\ref{ks:total} $bottom$ $row$), we obtain a value of $\bar{R}^{2}_{MARS}$ (0.67$\pm$0.06 and 0.59$\pm$0.03 on log$_{10}\Sigma_{H_{2}}$ and log$_{10}\Sigma_{SFR}$, respectively) higher than $\bar{R}^{2}_{linear}$ (0.55) and similar cut-point values to the ones in the individual dual galaxies ($\log \Sigma_{H2}\slash {\rm (M_{\odot}\,pc^{-2})}$ = 3.27 $\pm$ 0.17 and $\log \Sigma_{SFR}\slash {\rm ( M_{\odot}\,yr^{-1}\,kpc^{-2})}$ = 1.16 $\pm$ 0.19).

\begin{figure*}[htp!]
   \centering

    \includegraphics[width=.82\linewidth]{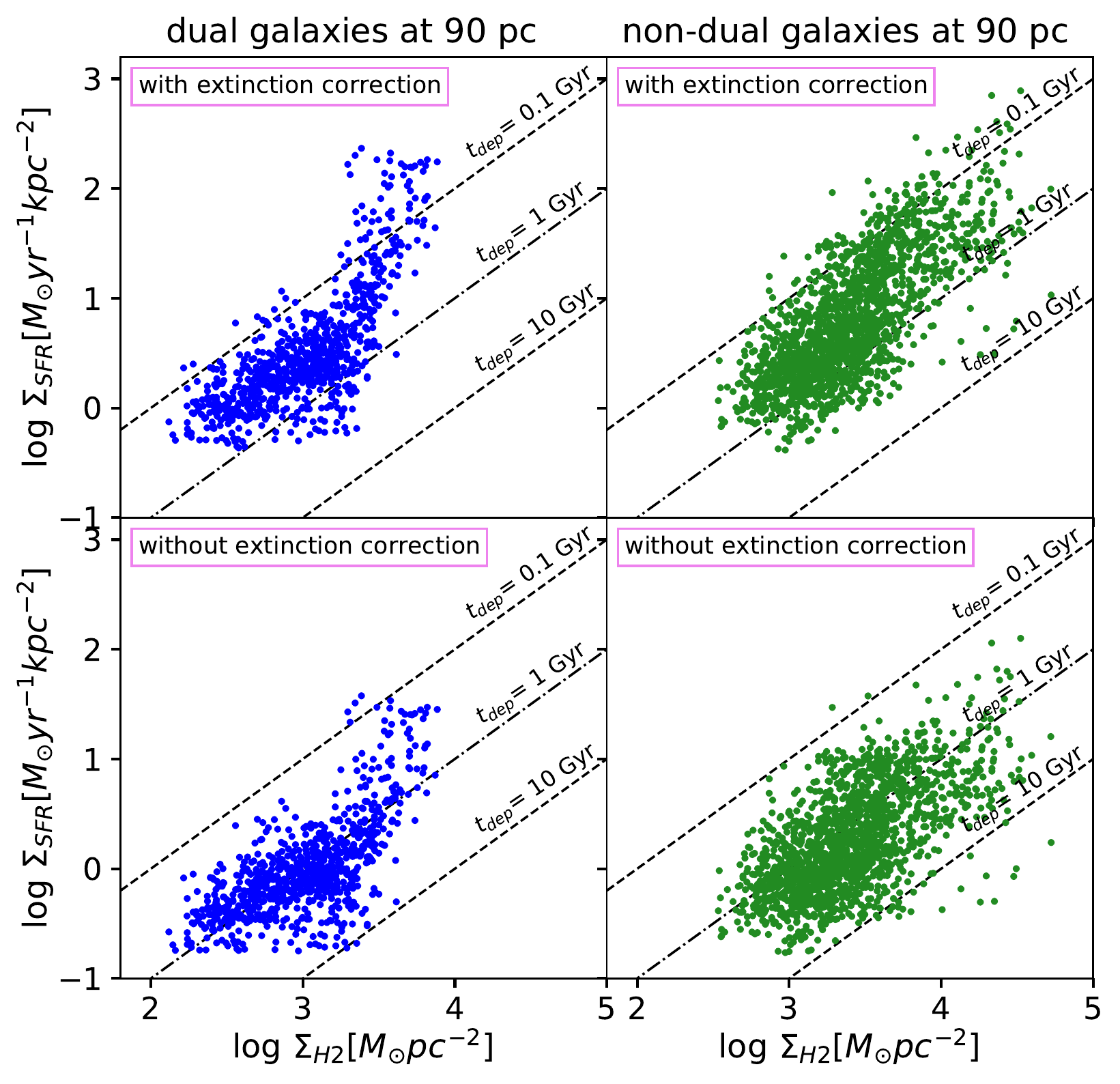}

   \caption{$Left$ $panels$: KS relation of galaxies with two branches, taking into account the extinction correction ($top$) and without this correction ($bottom$). $Right$ $panels$: the same for the non-dual galaxies.}
   \label{anexo:fig_extinction}
\end{figure*}

For the twelve non-dual galaxies, we find a single linear power-law with an index 
N=1.15$\pm$0.02 
(Fig.~\ref{ks:total} $top$ $right$). 
The dual and non-dual behaviours are also present before applying the extinction correction in Figure  \ref{anexo:fig_extinction}.

\subsection{Radial distribution of the two regimes}
\label{sect33}

  \begin{figure}[ht!]
   \centering

  \includegraphics[width=.96\linewidth]{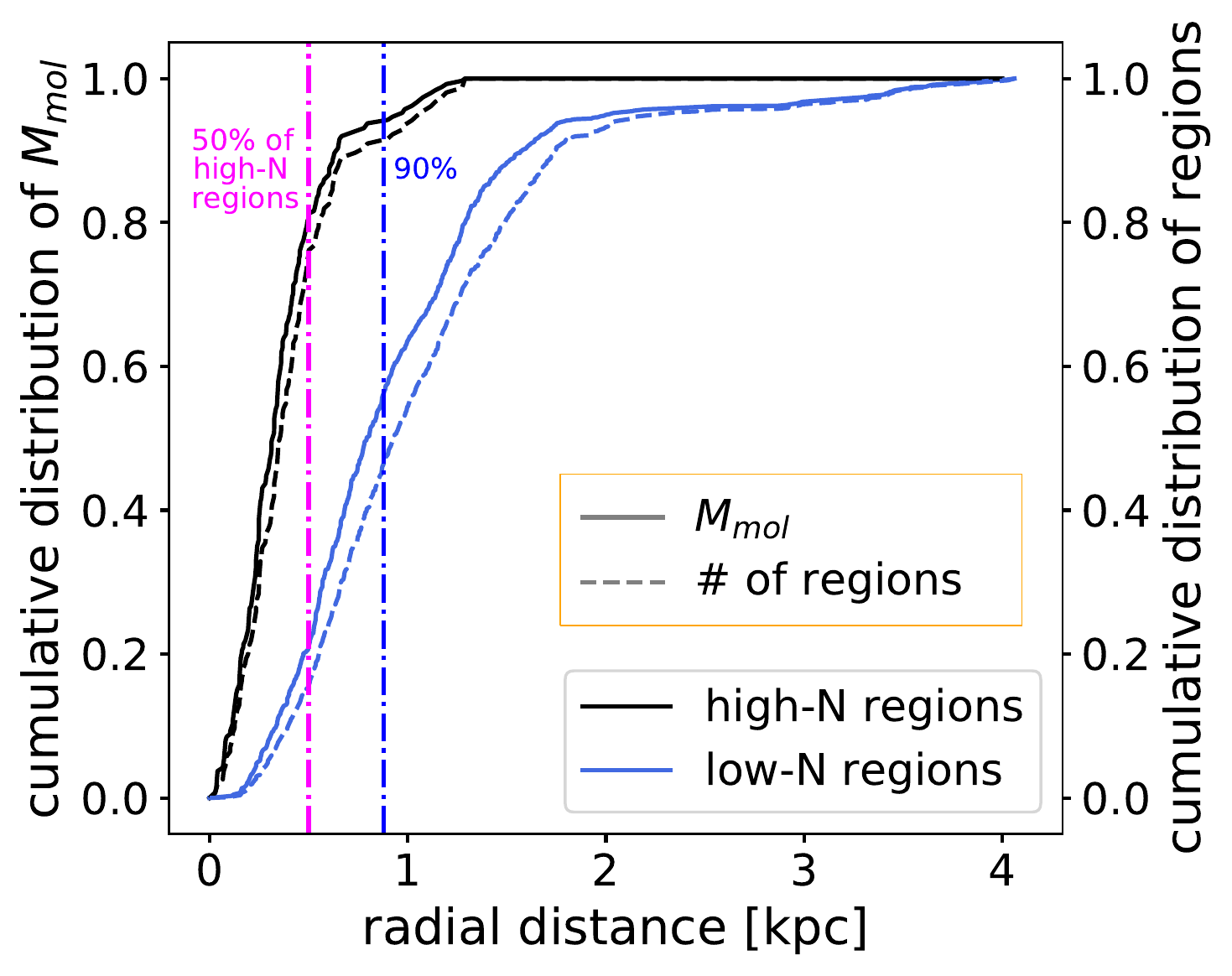}
   \caption{Cumulative distribution of the molecular mass ($left$ $vertical$ axis and solid lines) of the 90 pc regions in the dual galaxies, selected applying a cut on $\Sigma_{H_{2}}$, as a function of the radial distance. The dashed lines ($right$ $vertical$ axis) show the cumulative distribution of the number of regions in each branch. Colours are as in Fig. \ref{ks:ngc7130x}. The dash-dot lines correspond to the radii that include a percentage of regions.} 
        \label{cumulative_duals}%
    \end{figure}
To identify what causes the two branches in the SF laws for these four galaxies, we first investigate their spatial distribution. 
Fig. \ref{cumulative_duals} shows the cumulative distribution of the molecular gas mass of the regions, based on the $\Sigma_{H_{2}}$ cut selection, in the dual galaxies as a function of the radial distance. 
We find that the high-N regions are located in the central region of the galaxies, 50$\%$ (90$\%$) at radii smaller than 0.50\,kpc (0.85\,kpc) from the centre.
The molecular mass in the high-N regions follows the same radial distribution. 
The low-N regions are located at larger radii with a median radius of $\sim$ 1\,kpc and only $\sim$ 45\% of the regions are at radii lower than 0.88\,kpc.
We find the same trends using the cut on $\Sigma_{SFR}$.

\subsection{Self-gravity of the gas}\label{s:self-gravity}


  \begin{figure*}[htp]
   \centering
 \includegraphics[width=.6\linewidth]{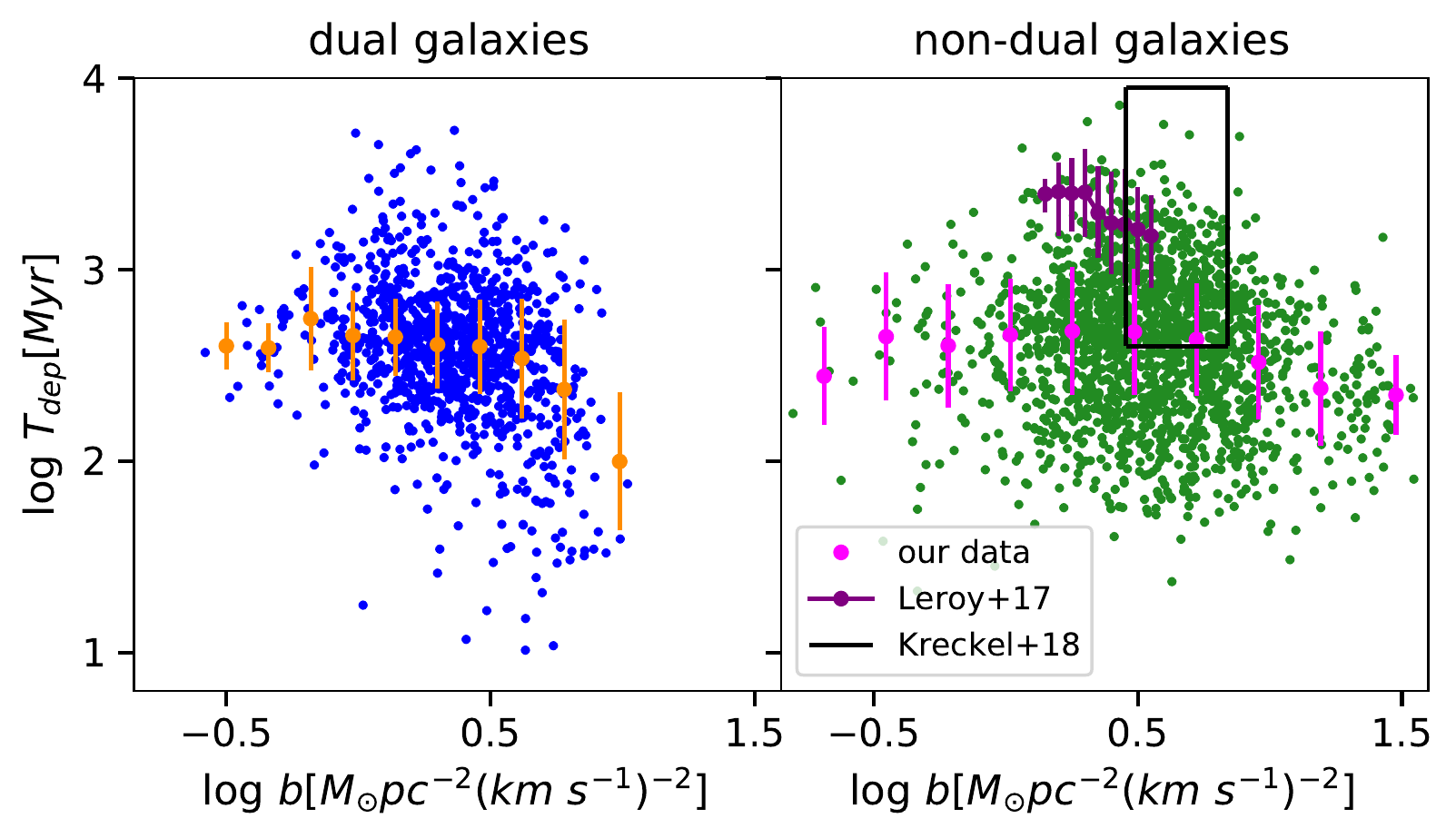}
 \includegraphics[width=.6\linewidth]{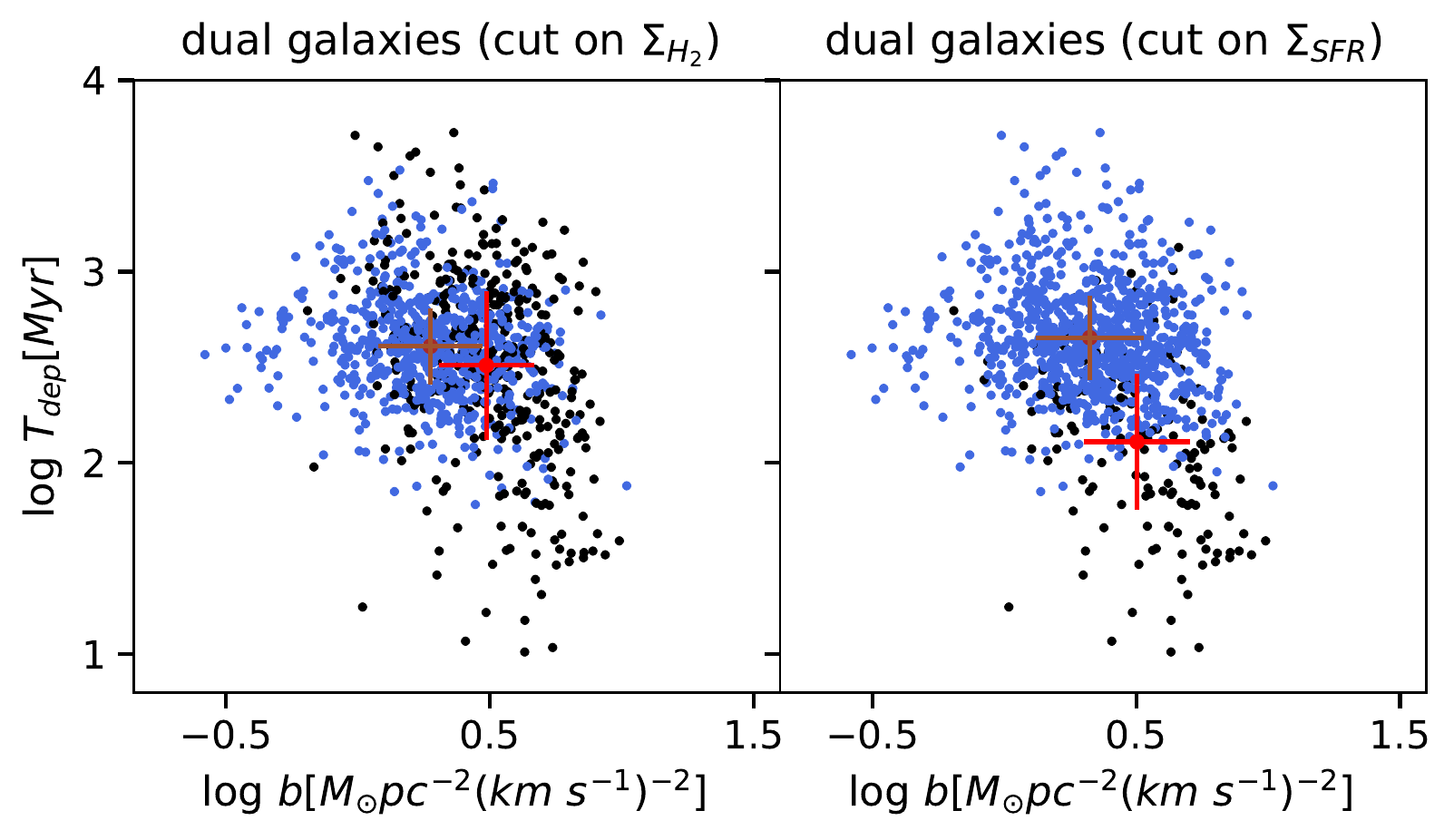}
   \caption{Cold molecular gas depletion time, t$_{dep}$, as a function of the self-gravity of the regions (parameter $b$) at 90 pc scales for the dual ($top$ $left$) and the non-dual galaxies ($top$ $right$). The orange and magenta circles show the median t$_{dep}$ in bins of $b$. The errorbars indicate the mean absolute deviation of the points in the bins. We show in purple the values of these parameters estimated in M51 (Leroy+17) and the open black rectangle represents the range of values in NGC~628 (Kreckel+18). The black and blue circles correspond to regions in the high- and low-N regimes, respectively, for the four dual objects ($bottom$ $row$). The red and brown lines indicate mean the mean absolute deviation (MAD) for the regions in high- and low-N regimes, respectively.}
              \label{Fig4}%
    \end{figure*}

We explored the dynamical state of molecular gas in the regions using the boundedness parameter ($b\equiv\Sigma_{mol}$/$\sigma^{2}\propto\alpha_{vir}^{-1}$, where $\sigma$ is the velocity dispersion and $\alpha_{vir}$ the virial parameter). We obtained the velocity dispersion from the CO(2--1) moment 2.
Fig. \ref{Fig4} shows the cold molecular gas depletion time (t$_{dep}$ = $\Sigma_{H_{2}}$/$\Sigma_{SFR}$) as a function of the boundedness parameter ($b$) 
at 90 pc scales. 
Despite the scatter ($\sim$2.5\,dex in t$_{dep}$), at these scales there is a weak trend 
 with decreasing t$_{dep}$ for increasing $b$ in the dual galaxies ($top$ $left$).

 When we consider the low- and high-N regions separately ($bottom$ $row$), 
  we find that the high-N regions in both cuts show a slightly better correlation between t$_{dep}$ and $b$, 
 while for the low-N regions the trend disappears.  
 The high-N regions show gas with larger $b$ (with a mean $b$ parameter of log$b$/M$_{\odot}$pc$^{-2}$(km~s$^{-1}$)$^{-2}\approx$0.52) and have shorter $t_{\rm dep}$ in both cuts than the low-N regions (with a mean $b$ 
 of log$b$/M$_{\odot}$pc$^{-2}$(km~s$^{-1}$)$^{-2}\approx$0.30). 
The non-dual galaxies do not show a clear relation. Table \ref{tab:statistical} summarizes the correlations. %
\begin{table*}[ht!]
\centering
	\caption{Spearman rank correlation coefficients}
	\begin{tabular}{l cc c cc}
	\hline
	\hline
	 & \multicolumn{2}{c}{$t_{dep}$ vs. $b$} & & \multicolumn{2}{c}{SFE vs. $\sigma$} \\ 
	\cline{2-3} \cline{5-6}
	
	 &  $\rho_{s}$ & p-value &   & $\rho_{s}$ & p-value  \\ 
	[0.5ex]	
	\hline
	dual & -0.26 & 6.13e-17 &  & -0.08 & 0.01 \\
	high-N (cut on $\Sigma_{H{2}}$) & -0.38 & 2.79e-14 &  & 0.32 & 5.79e-7 \\
	low-N (cut on $\Sigma_{H{2}}$) &-0.18 & 4.12e-6  &  & 	- & - \\

	high-N (cut on $\Sigma_{SFR}$) & -0.48 & 2.51e-10 &  & 0.13 & 0.16 \\
	low-N (cut on $\Sigma_{SFR}$)&  -0.15 & 1.22e-5 &  & - &  - \\ 
	non-dual & -0.13 & 9.78e-8 &  & 0.02 & 0.34 \\

	\hline
	\end{tabular}
	\label{tab:statistical}
	\tablefoot{Spearman $\rho_{sp}$ rank correlation coefficients (two-sided p-values). We exclude upper limits from the analysis. We consider the correlations are statistically significant when $\rho_{sp}\gtrsim$0.3 and p-value<3\%.}
\end{table*}

\cite{Leroy2017} found, from the intensity weighted average on scales of 40 pc within regions of 370 pc, that gas with larger $b$ (more bound) exhibits shorter $t_{\rm dep}$ in the spiral galaxy M51. This means that when b increases the system is more gravitationally bounded.  
However, \cite{Kreckel2018} did not find any correlation between $b$ and $t_{\rm dep}$ 
in another spiral (NGC~628) at 50 pc scales within 500 pc regions, which is in agreement with our results for the non-dual galaxies and the low-N regions in dual galaxies. For the high-N regions in the dual galaxies, $t_{\rm dep}$ seems to decrease for increasing $b$ although the scatter is large.
The depletion times in our sample are between 4-8 times shorter than in these two spirals. This difference is consistent with what was found in previous works for starbursts \citep{daddi2010, genzel2010, burillo2012}. However, 
for similar $b$, there is a factor of 10 in $t_{\rm dep}$. 
As a consequence, it is not clear if a universal relation between $t_{\rm dep}$ and $b$ exists. 

\subsection{Velocity dispersion of the gas} \label{s:velocity-dispersion}

  \begin{figure*}[htp]
   \centering
    \includegraphics[width=.92\linewidth]{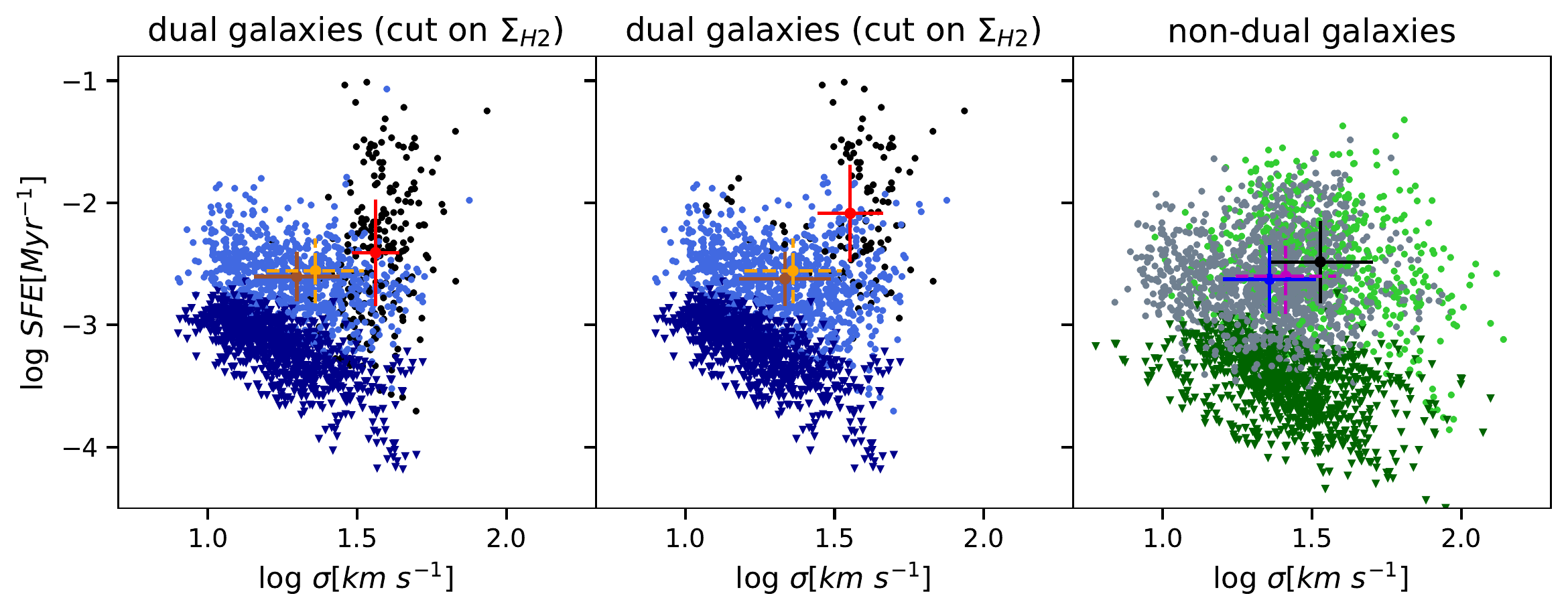}
   \caption{SF efficiency of the molecular gas, as a function of the velocity dispersion of the gas ($\sigma$) at 90 pc scales. 
   The $left$ and $middle$ panels correspond to the cut on $\Sigma_{H2}$ and $\Sigma_{SFR}$ axis for each galaxy, respectively. Colours are as in previous figures. 
   The red and brown points indicate the mean and mean absolute deviation (MAD) for the regions in high- and low-N regimes, respectively. 
   $Right$ panel: Similar to the $left$ and $middle$ panels but for non-dual galaxies.
   The central regions (r<0.50\,kpc) are represented in green and the more external regions in grey. The black and blue points correspond to the mean and MAD value of the central and external regions, respectively. The orange and magenta points are the mean and MAD for all the regions of each panel. The inverted triangles indicate upper limits. }
    \label{Fig5}%
    \end{figure*}

We explore the behaviour of the velocity dispersion in our sample. 
Fig. \ref{Fig5} shows the SF efficiency of the cold molecular gas (SFE=$\Sigma_{SFR}$/$\Sigma_{H_{2}}$) as a function of the velocity dispersion ($\sigma$). The velocity dispersion was
obtained from the CO(2--1) moment 2.  
We find that the global mean values of the $\sigma$ and SFE for the dual galaxies 
($\log \sigma \slash {\rm km\,s^{-1}}$ = 1.36 $\pm$ 0.16 and $\log {\rm SFE\slash Myr^{-1}}$ = $-$2.56 $\pm$ 0.26) and for the non-dual galaxies ($\log \sigma \slash {\rm km\,s^{-1}}$ = 1.41 $\pm$ 0.18 and $\log {\rm SFE\slash Myr^{-1}}$= $-$2.60 $\pm$ 0.31) are similar. However, when we consider the low- and high-N regions independently, the mean values are different. The high-N regions show higher mean values ($\log \sigma \slash {\rm km\,s^{-1}} \sim$ 1.56 for both cuts 
and $\log {\rm SFE\slash Myr^{-1}}$ = $-$2.41 $\pm$ 0.44 for cut on $\Sigma_{H_{2}}$ and $\log {\rm SFE\slash Myr^{-1}}$ = $-$2.10 $\pm$ 0.40 on $\Sigma_{SFR}$) than the low-N regions ($\log \sigma \slash {\rm km\,s^{-1}} \sim$ 1.30 
and $\log {\rm SFE\slash Myr^{-1}} \sim$  $-$2.62 
 for both cuts). Also, for the high-N regions, the SFE increases with increasing $\sigma$, though the scatter is large ($\sim$2\,dex).

The high-N regions are located in the central regions of the four dual objects, so we also investigate if the central regions of the non-dual galaxies have different SFE and\slash or $\sigma$. To do this, 
we consider the regions at radii $<$500\,pc, which is where most of the high-N regions are located in the dual galaxies (see Sect.\,\ref{sect33}). 
As opposed to the dual galaxies, we find that for the non-dual galaxies, the internal (r$<$500\,pc) and external regions have similar mean SFE ($\log {\rm SFE\slash Myr^{-1}}$ = $-$2.48 $\pm$ 0.34 and $-$2.66 $\pm$ 0.28 for the internal and external regions respectively) and just slightly higher $\sigma$ ($\log \sigma \slash {\rm km\,s^{-1}}$ = 1.53 $\pm$ 0.18 and 1.36 $\pm$ 0.16, respectively). 

The large scatter at these scales may be due to the fact that we can resolve individual regions, obtaining information from the clouds in different evolutionary phases \citep{Kruijssen2014}. 
Several SF models suggest that the dynamical state of the cloud, and not only its density, affects its ability to collapse and form stars \citep[e.g.,][]{K2005,H2011,F2013}. These models focus on the properties of turbulent molecular clouds, proposing that the supersonic and compressive turbulence induces the formation of stars.
In this case, we would expect the SFE to increase with increasing gas velocity dispersion \citep{Orkisz2017}. 
This is consistent with our findings for the high-N regions in the dual galaxies.
Cloud-cloud collisions could be enhanced near the location of the bar resonances in the central regions of these galaxies (S\'anchez-Garc\'ia et al. in prep). These collisions could result in an increased turbulence, 
which may induce a greater compression of the gas (increasing its density), and finally lead to an enhanced star formation. Moreover, 
the increase in gas density compensates for the high turbulence, causing, together, $b$ to increase in these central regions.

\section{Conclusions} \label{conclusions}

We have presented a high-resolution study of the star-formation relation in a sample of 16 local LIRGs on spatial scales of $\sim$ 90 pc. We have combined the SFR calculated from the HST/NICMOS Pa$\alpha$ emission with cold molecular gas from ALMA CO(2--1) data to probe the star-formation relations.

We find that four galaxies from our sample show a dual behaviour in their KS relation at 90\,pc scales.
The regime with higher gas and SFR surface densities is characterized by a steeper power-law index in the central region of the galaxies (r$<$0.85\,kpc). The other regime, which shows lower values of gas and SFR surface densities, is located in the more external disk regions. This dual behaviour disappears  
at large spatial scales (240 and 500 pc). 

The gas in the central region of the dual galaxies shows greater turbulence (higher $\sigma$) and slightly stronger self-gravity (higher $b$) than the external region. These dynamical conditions of the gas might lead to more efficient star formation in the central region. The rest of the galaxies do not show a clear trend between these two parameters.
These variations within each galaxy and among the galaxies of the sample suggest that the local dynamical environment plays a role in the star formation process. 
The fraction of AGN and bars is similar for dual and non-dual galaxies, although a larger sample is needed to evaluate their impact on the SF law at 90\,pc scale.

\begin{acknowledgements}
We thank referee for the useful comments and suggestions. 
MSG acknowledges support from the Spanish Ministerio
de Economía y Competitividad through the grants BES-2016-078922, ESP2017-83197-P. LC and MSG acknowledge support from the research project PID2019-106280GB-100. MPS acknowledges support from the Comunidad de Madrid through the Atracción de Talento Investigador Grant 2018-T1/TIC-11035 and PID2019-105423GA-I00 (MCIU/AEI/FEDER,UE). SGB and AAH acknowledge support from PGC2018-094671-B-I00 (MCIU/AEI/FEDER,UE). 
SGB acknowledges support from the research project PID2019-106027GA-C44 of the Spanish Ministerio de Ciencia e Innovación.   
JPL acknowledges support from PID2019-105423GA-I00. 
EB acknowledges the support from Comunidad de Madrid through the Atracción de Talento grant 2017-T1/TIC-5213. 
SC acknowledge financial support from the State Agency for Research of the Spanish MCIU through the ‘Centre of Excellence Severo Ochoa’ award to the Instituto de Astrofísica de Andalucía (SEV-2017-0709).

AL acknowledges the support from Comunidad de Madrid through the Atracción de Talento Investigador Grant 2017-T1/TIC-5213, and PID2019-106280GB-I00 (MCIU/AEI/FEDER,UE)

This paper makes use of the following ALMA data: ADS/JAO.ALMA\#2017.1.00255.S, ADS/JAO.ALMA\#2013.1.00271.S, ADS/JAO.ALMA\#2013.1.00243.S, ADS/JAO.ALMA\#2015.1.00714.S and ADS/JAO.ALMA\#2017.1.00395.S. ALMA is a partnership of ESO (representing its member states), NSF (USA) and NINS (Japan), together with NRC (Canada) and NSC and ASIAA (Taiwan) and KASI (Republic of Korea), in cooperation with the Republic of Chile. The Joint ALMA Observatory is operated by ESO, AUI/NRAO and NAOJ. The National Radio Astronomy Observatory is a facility of the National Science Foundation operated under cooperative agreement by Associated Universities, Inc.

\end{acknowledgements}

\bibliographystyle{aa}
\bibliography{SFrelations_lirgs}

\clearpage
\begin{appendix}

\section{Figures}    

\subsection{Star formation relation for individual galaxies and CO(2--1) maps and HST/NICMOS images}\label{app:figures_regions}
In this appendix we present the KS relations, the regions considered in this work, the ALMA CO(2--1) maps and the HST/NICMOS Pa$\alpha$ images for the whole sample.

\begin{figure*}[ht!]
   \centering

    \includegraphics[width=.7\linewidth]{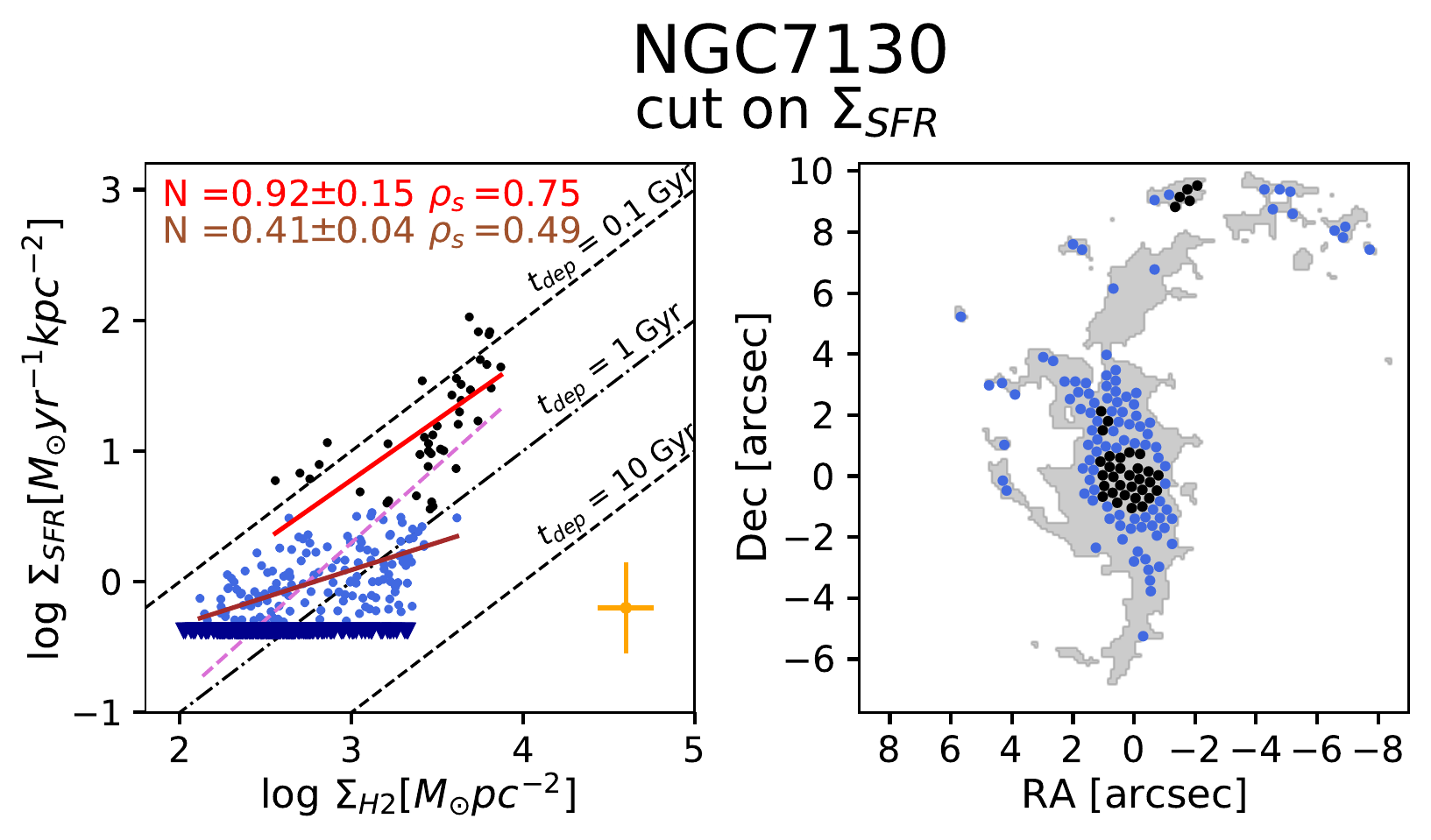}
   \includegraphics[width=.82\linewidth]{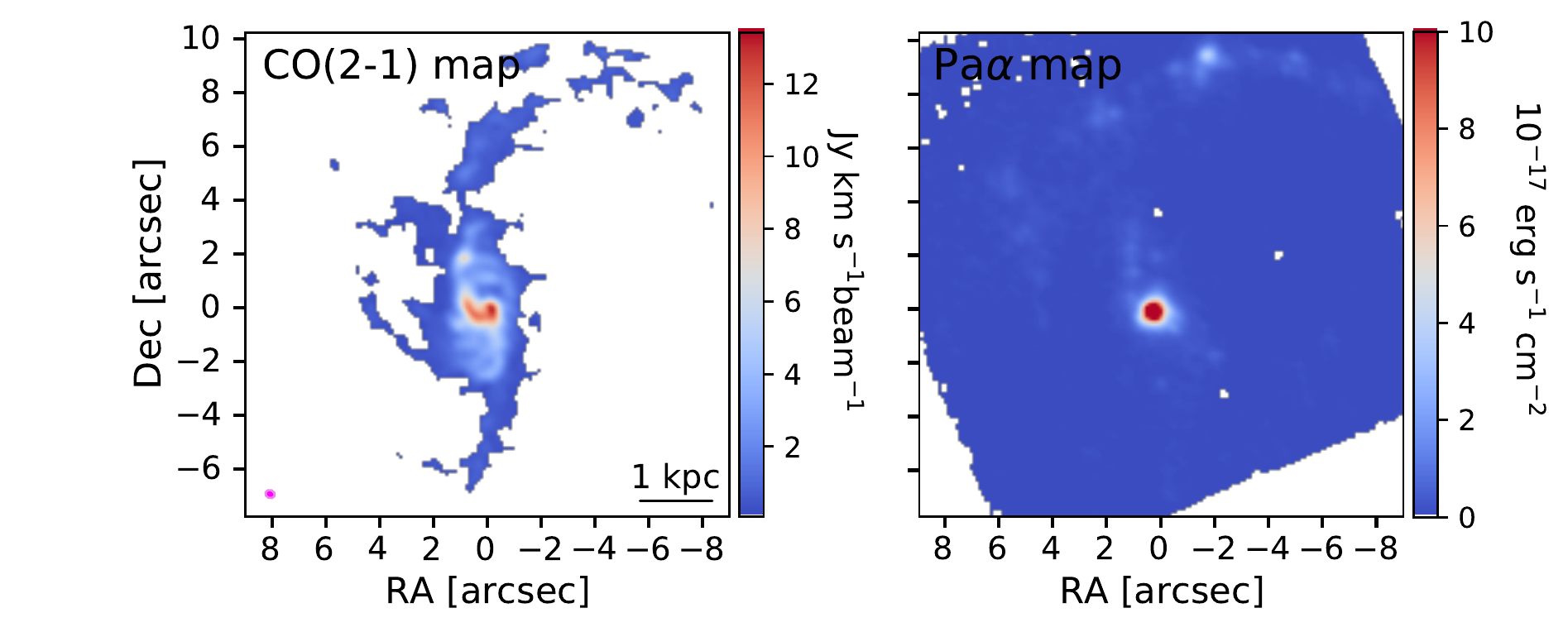}
  \caption{$Top$ panels: Same as Fig. \ref{ks:ngc7130x}, applying the MARS method to the log$_{10}\Sigma_{H_{2}}$. $Bottom$ panels: ALMA CO(2--1) ($left$) and HST/NICMOS Pa$\alpha$ maps ($right$). This last map is smoothed to the ALMA resolution. The magenta filled ellipse ($bottom$ $left$) represents the beam size (0$\arcsec$.36$\times$0$\arcsec$.29 PA=69$^\circ$).} 
              \label{anexo:ngc7130y}%
\end{figure*}    
    
\begin{figure*}[ht!]
   \centering
    \includegraphics[width=.61\linewidth]{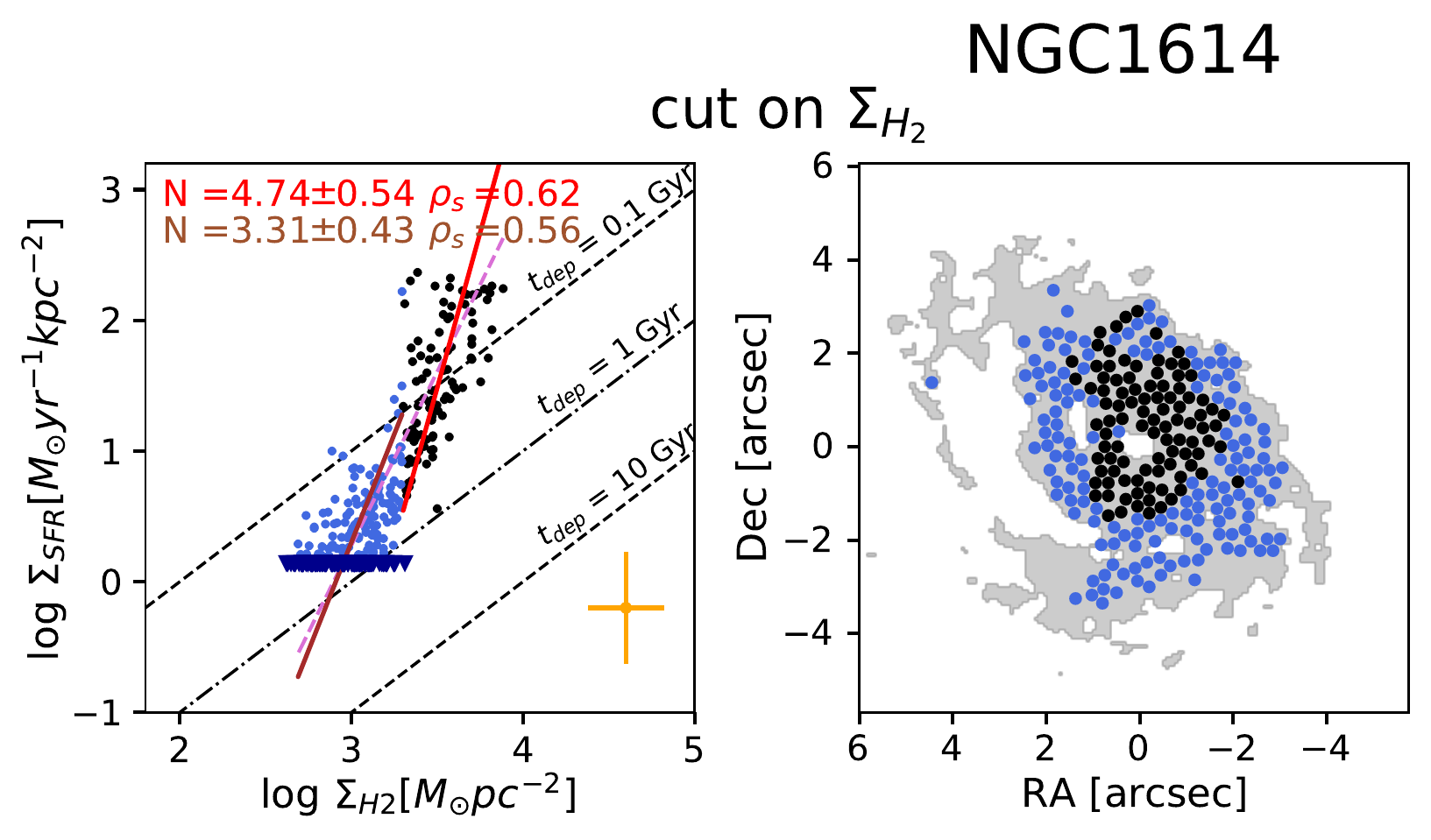}
   \includegraphics[width=.3245\linewidth]{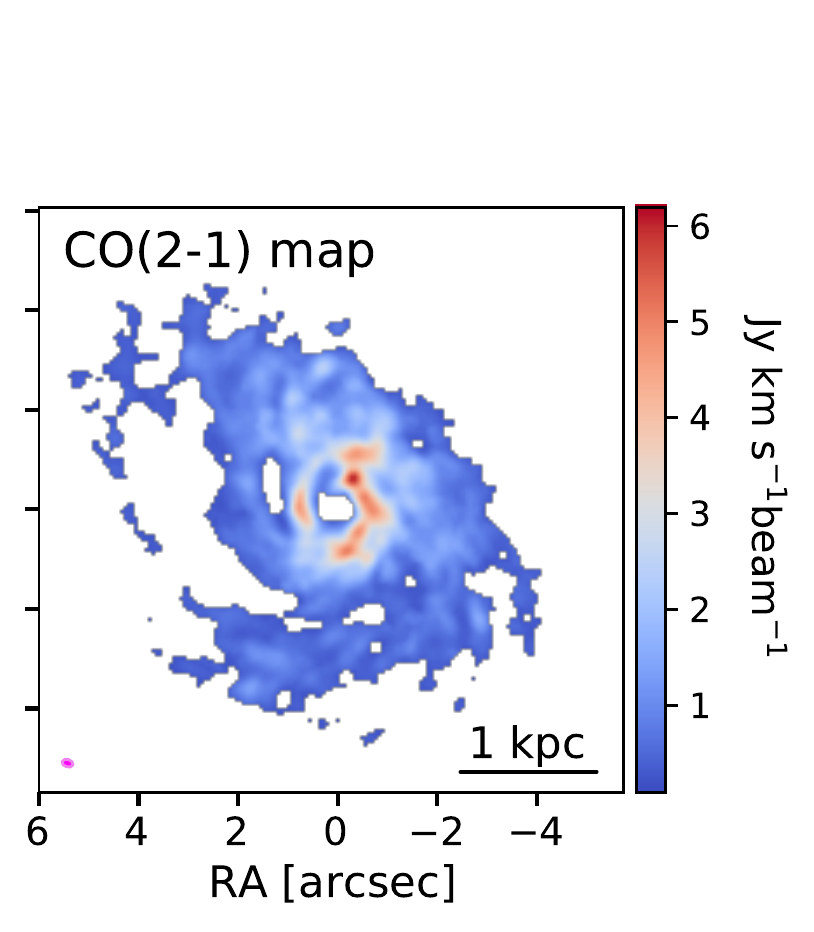}
   \includegraphics[width=.61\linewidth]{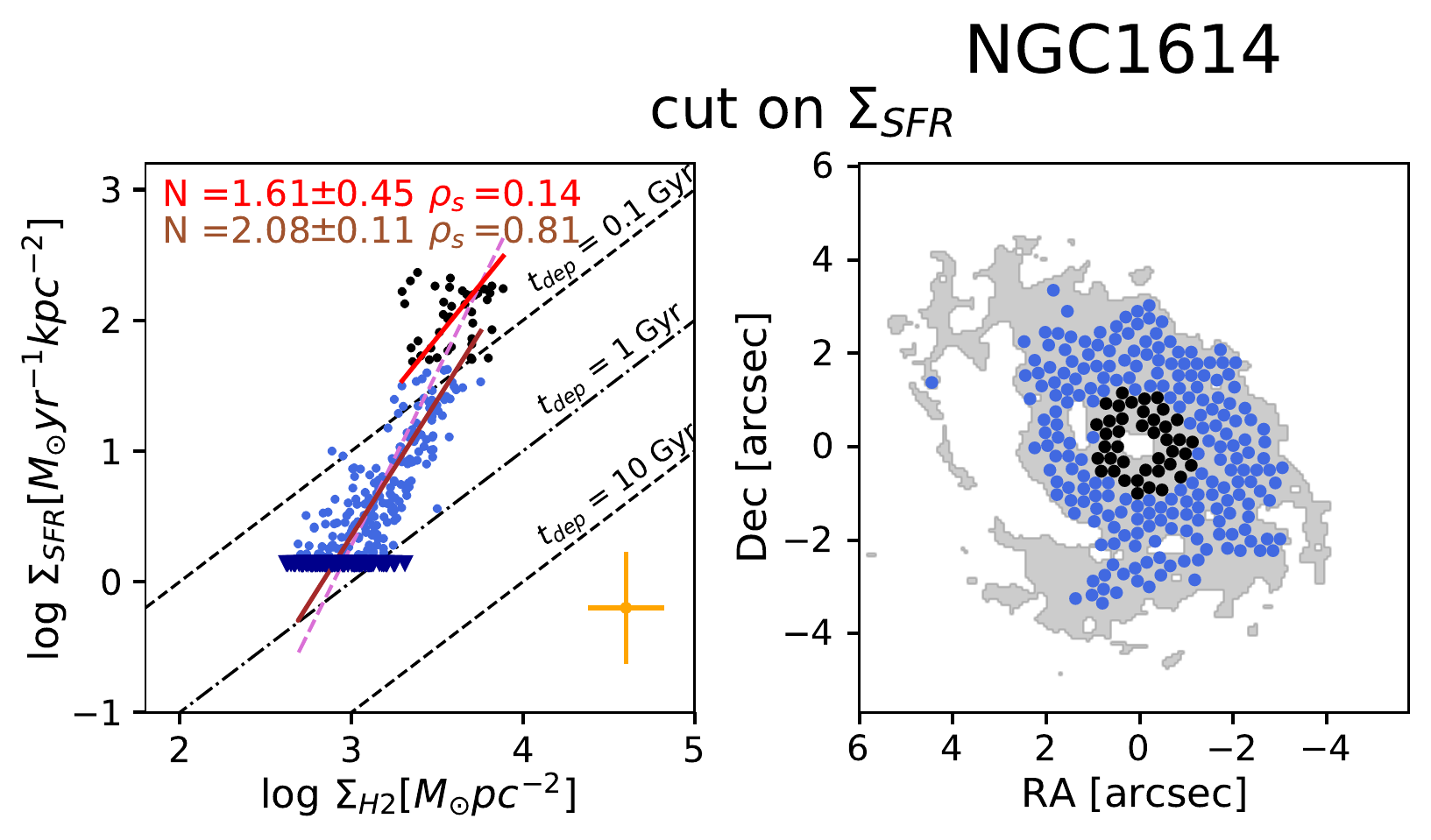}
   \includegraphics[width=.3235\linewidth]{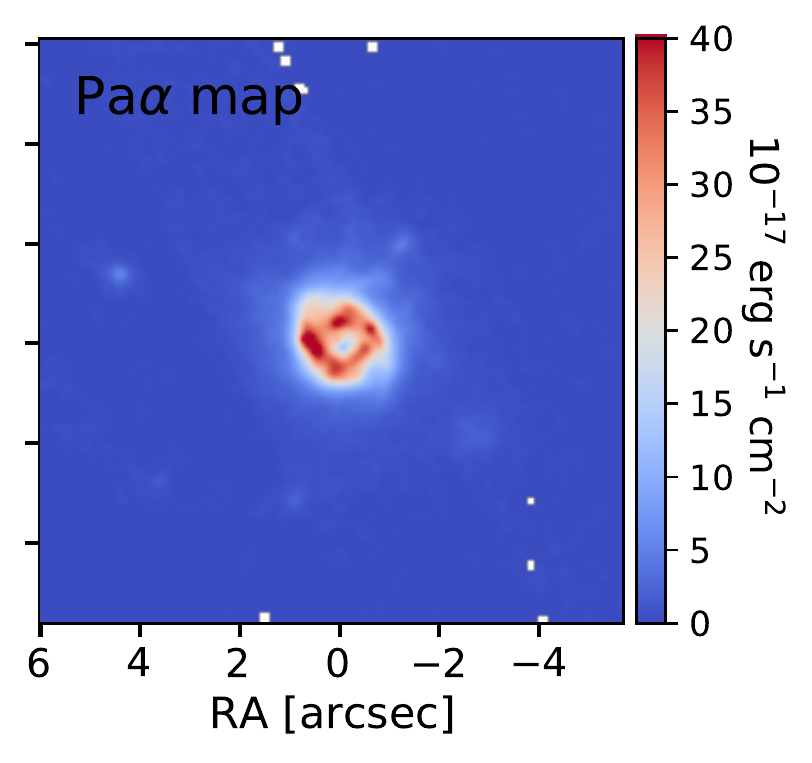}
   \caption{Same as Fig. \ref{ks:ngc7130x}, applying the MARS method to the log$_{10}\Sigma_{H_{2}}$ ($top$ $left$ and $middle$) and log$_{10}\Sigma_{SFR}$ ($bottom$ $left$ and $middle$). $Right$ panels: ALMA CO(2--1) ($top$) and HST/NICMOS Pa$\alpha$ maps ($bottom$). This last map is smoothed to the ALMA resolution. The magenta filled ellipse represents the beam size.} 
    \label{anexo:figuresdual}%
\end{figure*}

\begin{figure*}[ht!]
   \centering
   \addtocounter{figure}{-1}
    \includegraphics[width=.61\linewidth]{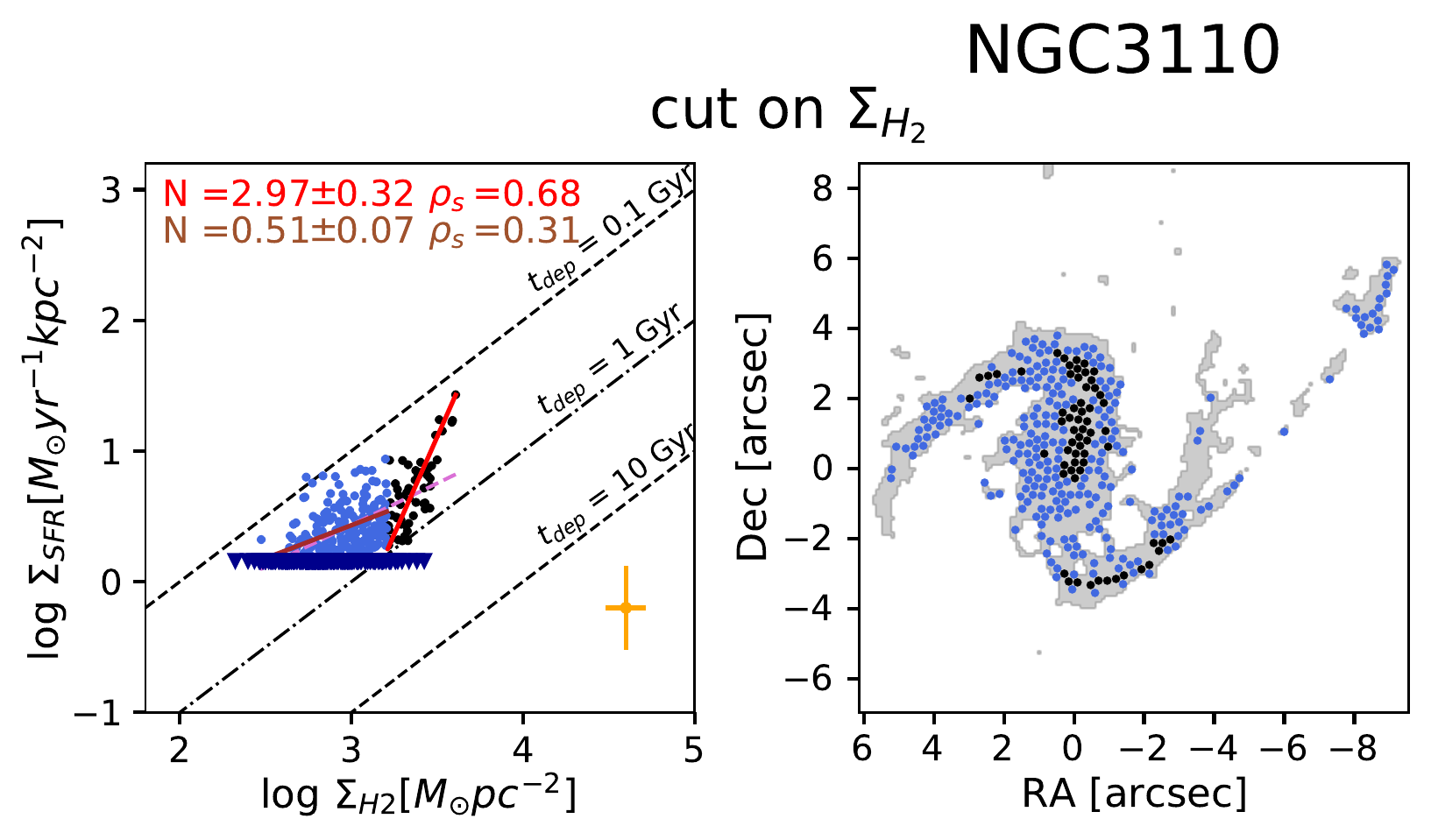}
   \includegraphics[width=.327\linewidth]{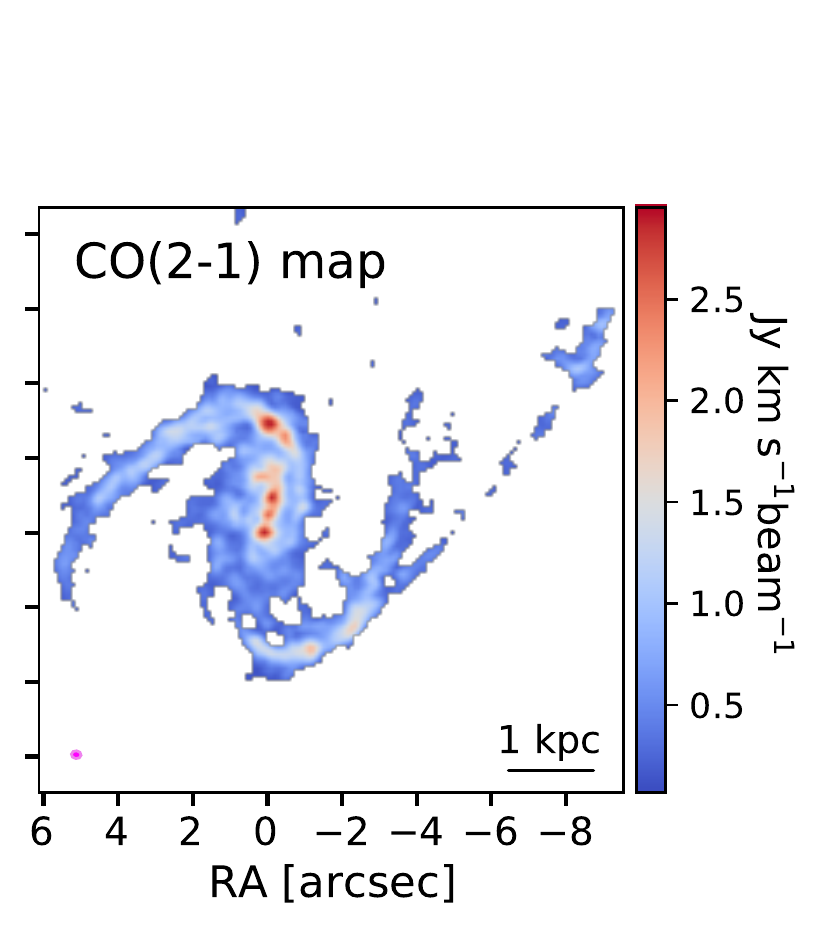}
    \includegraphics[width=.61\linewidth]{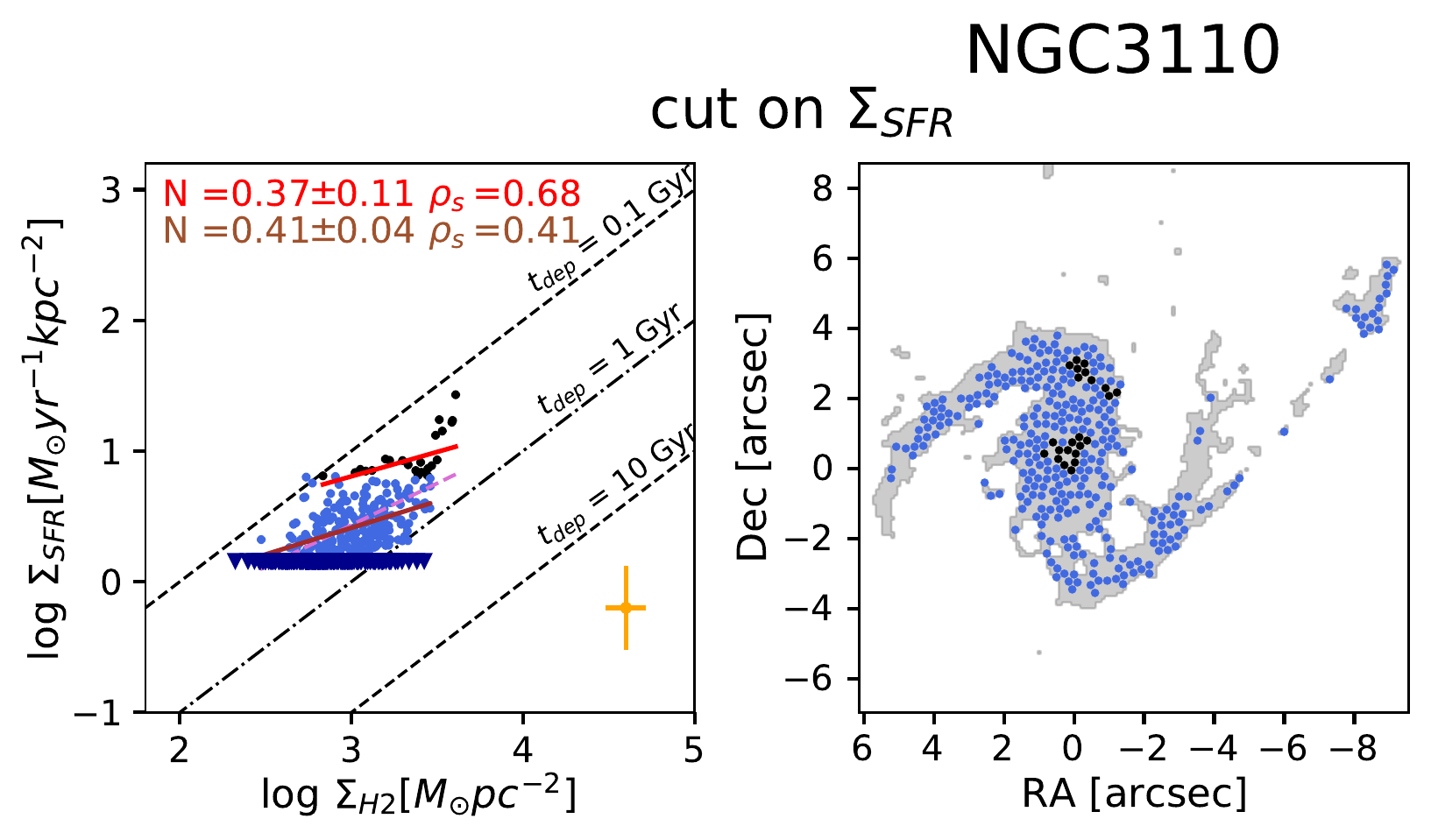}
   \includegraphics[width=.3313\linewidth]{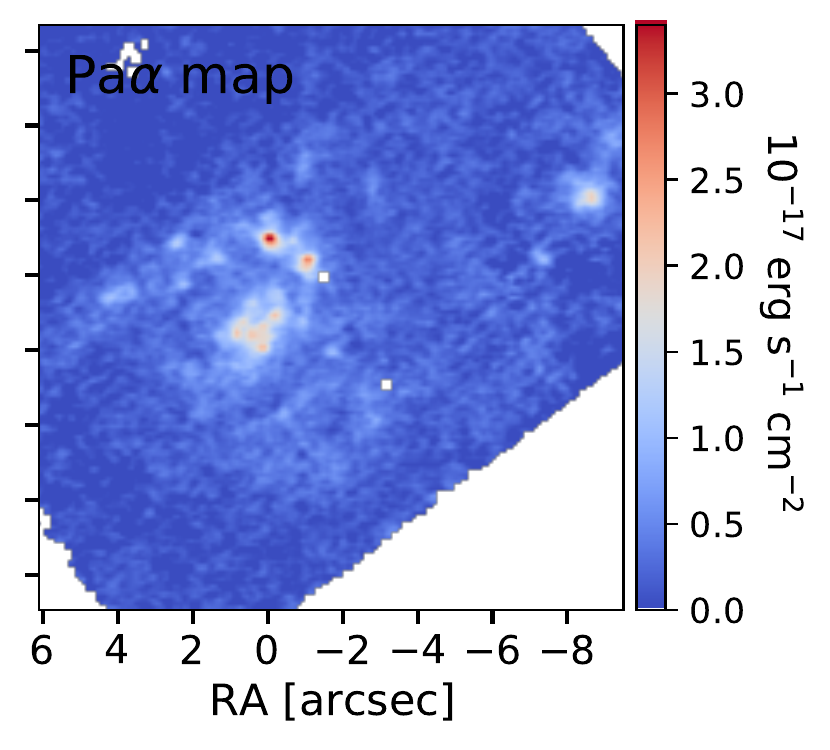}
   \caption{Continued.} 
   
\end{figure*}

\begin{figure*}[ht!]
   \centering
   \addtocounter{figure}{-1}
    \includegraphics[width=.61\linewidth]{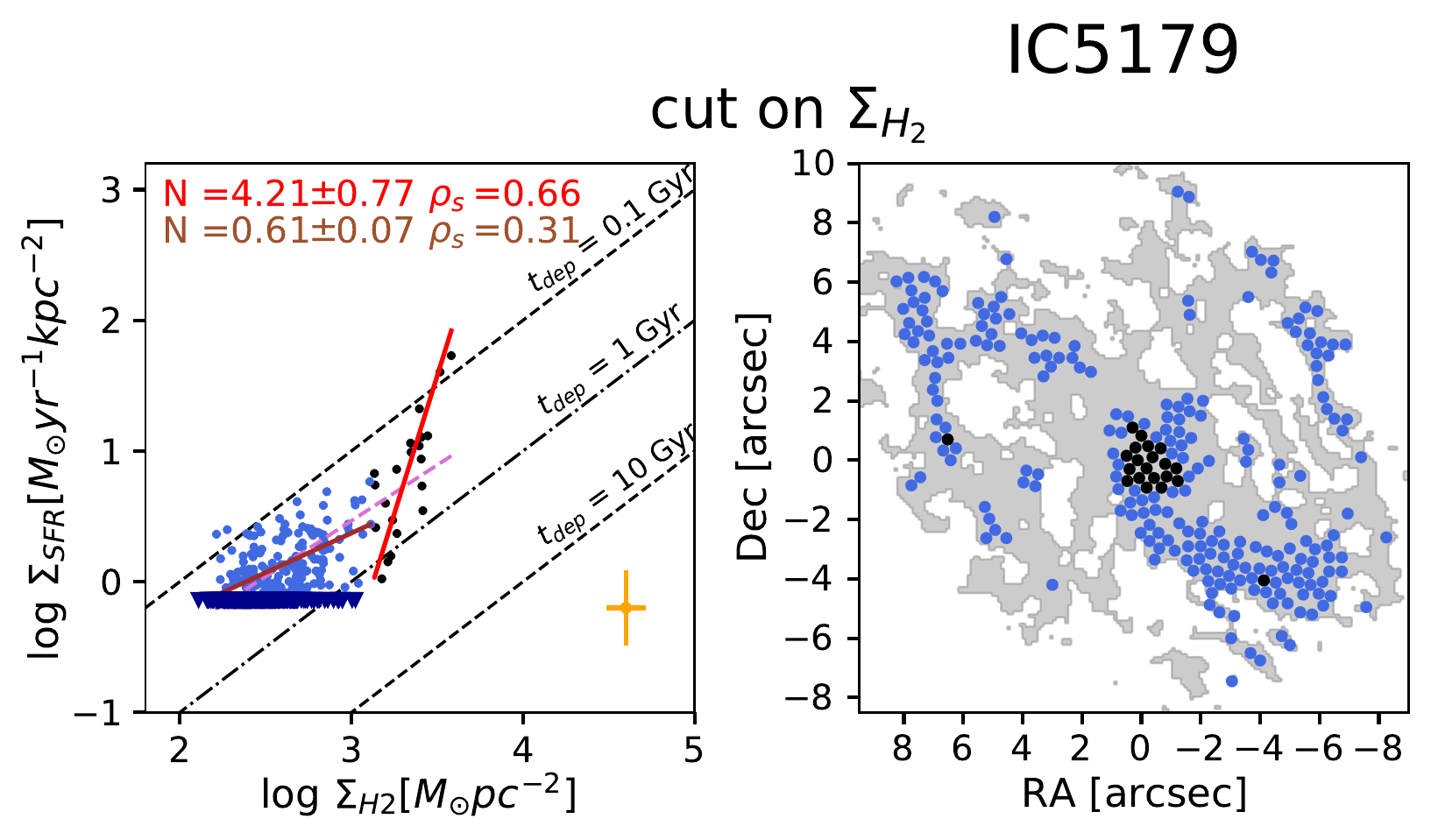}
   \includegraphics[width=.32\linewidth]{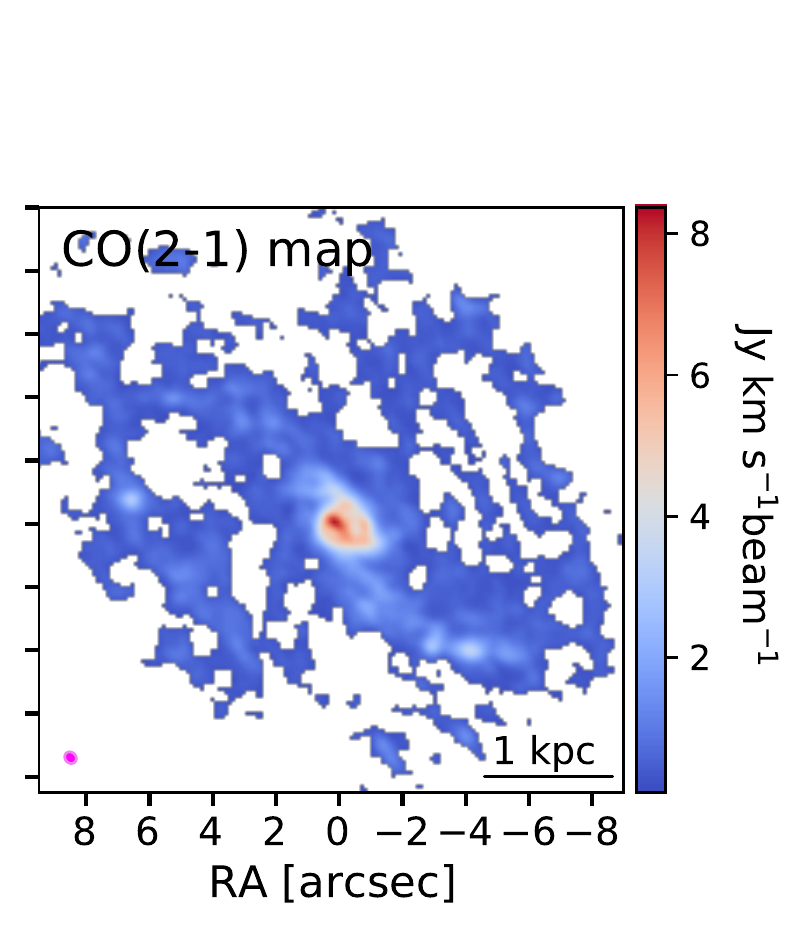}
    \includegraphics[width=.609\linewidth]{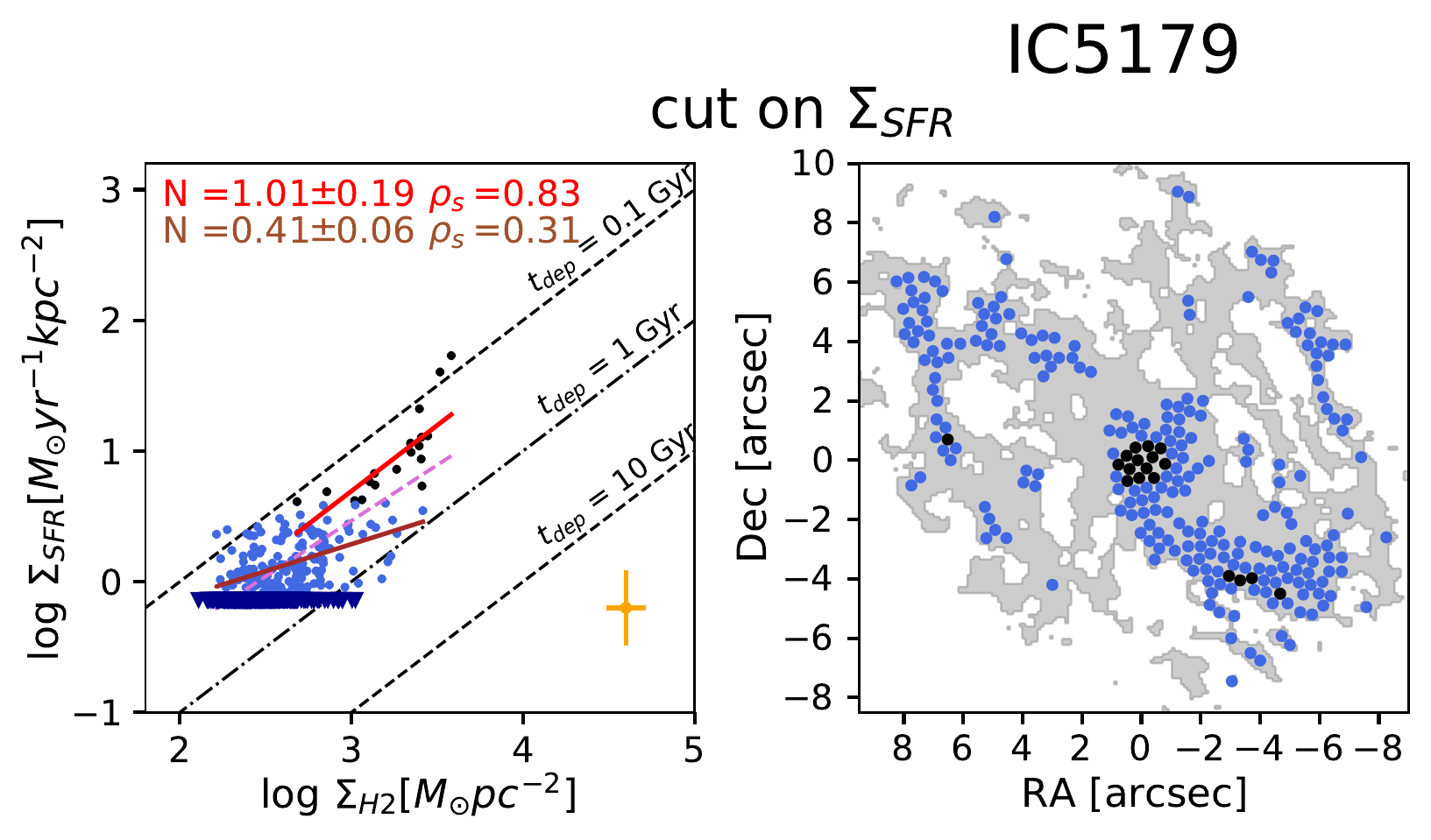}
   \includegraphics[width=.316\linewidth]{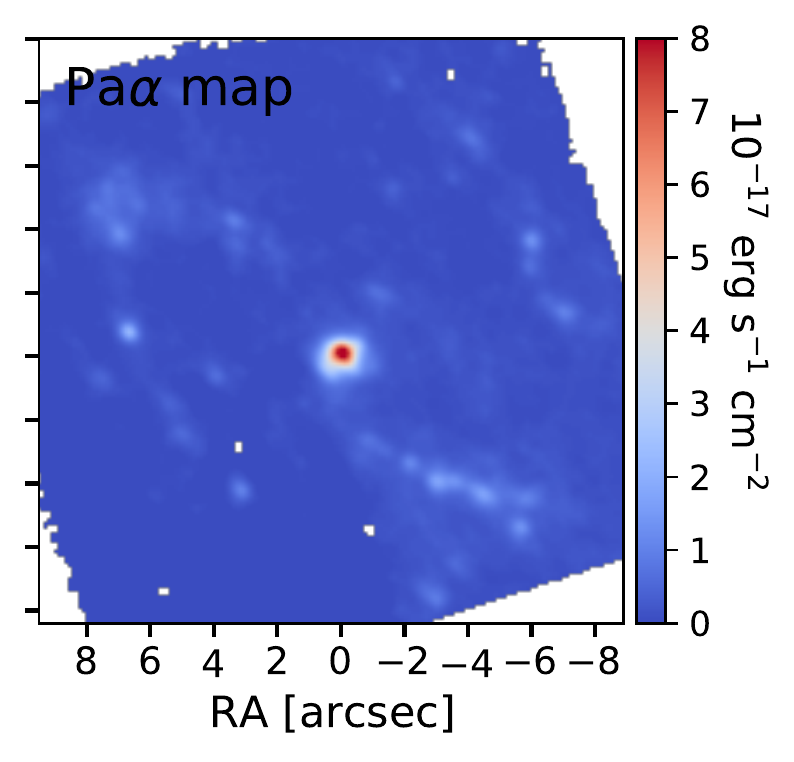}
   \caption{Continued.} 
 
\end{figure*}
    

\begin{figure*}[htp]
   \centering
    \includegraphics[width=.7\linewidth]{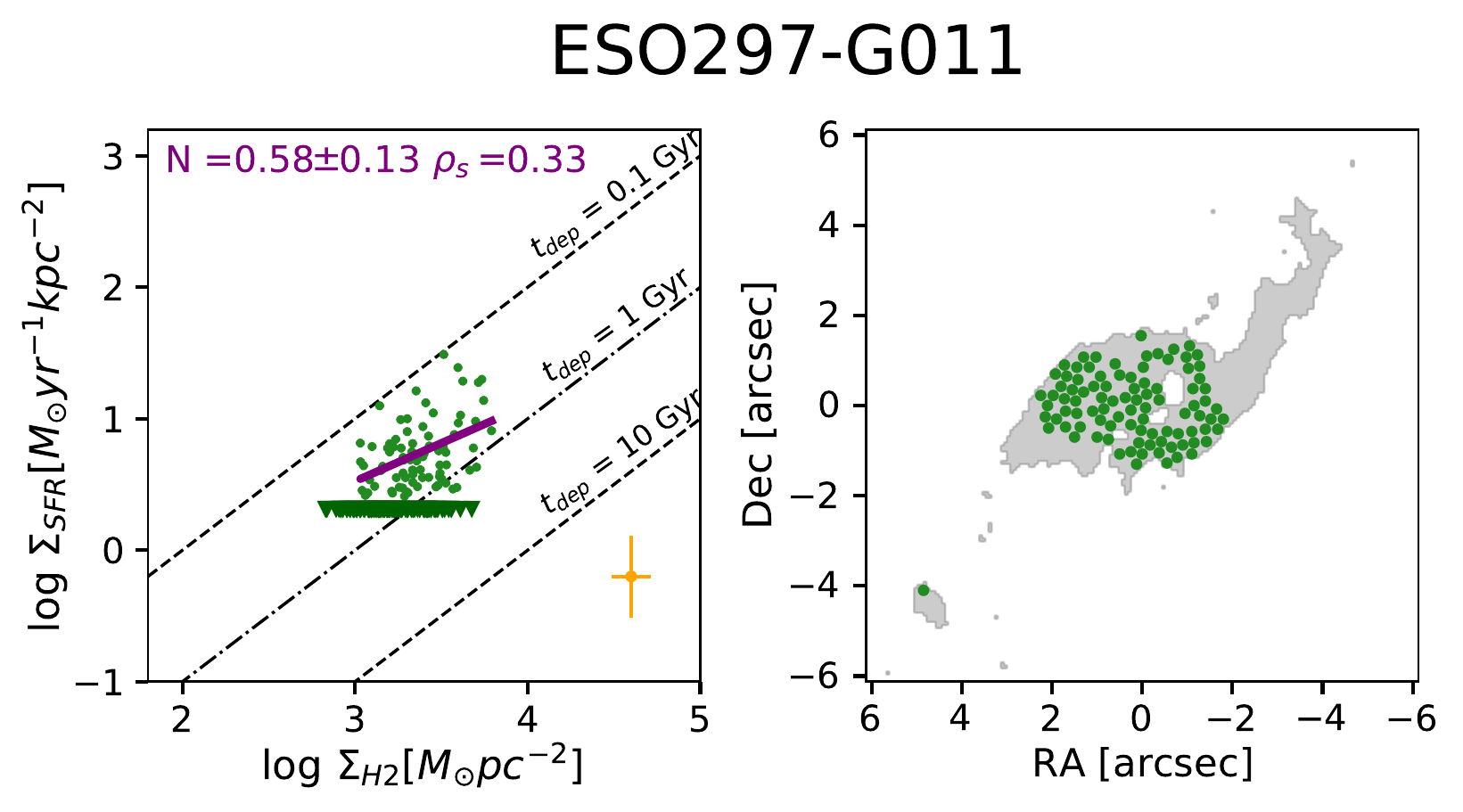}
    \includegraphics[width=.785\linewidth]{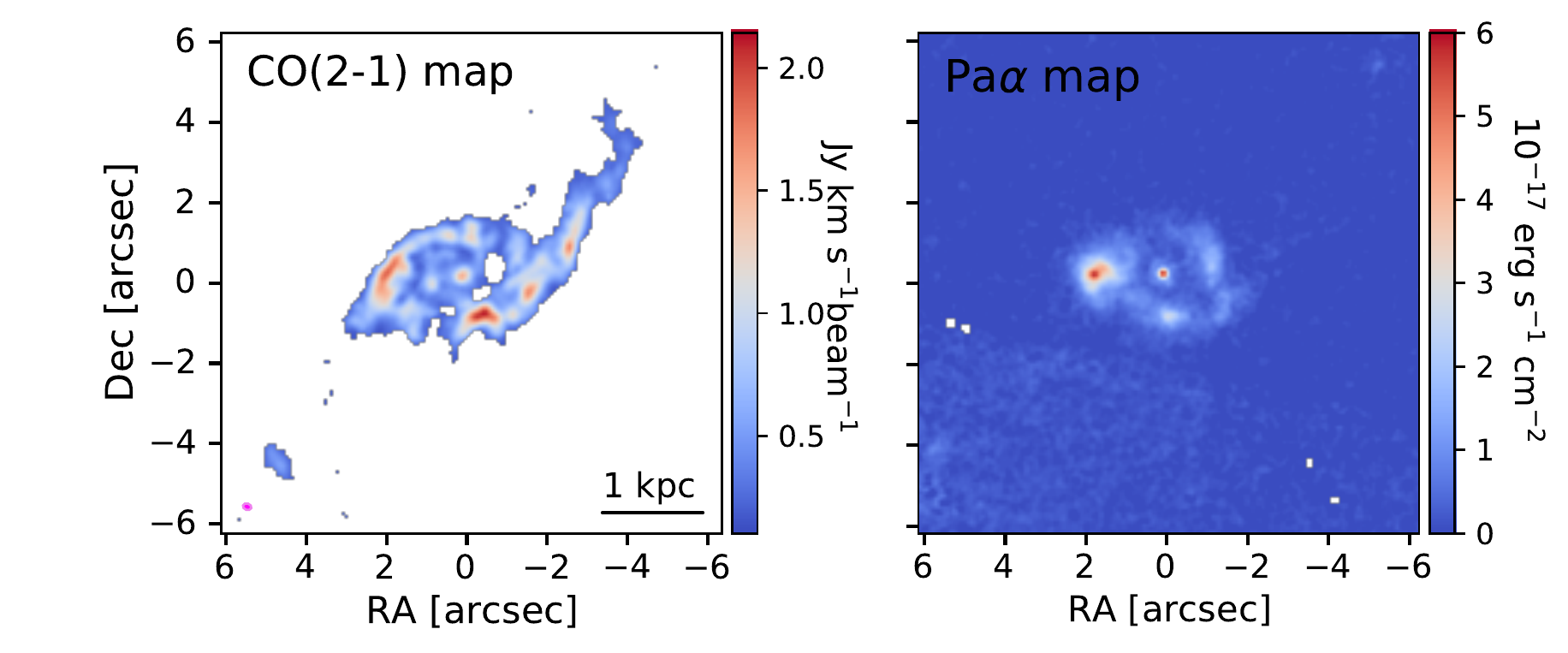}
    \caption{$Top$ $left$ panel: SFR surface density ($\Sigma_{SFR}$) as a function of the molecular gas surface density ($\Sigma_{H_{2}}$) for the non dual galaxies of the sample using 90 pc regions. The green circles indicate data points in each galaxy. The solid purple line indicates the best fit. The Spearman's rank correlation coefficients ($\rho_{s}$) and the power-law index (N) of the derived best-fits KS relation are reported in the figure. The dashed lines mark constant t$_{dep}$ values. $Top$ $right$: Location of the regions on the CO(2--1) map (grey). $Bottom$ panels: ALMA CO(2--1) ($left$) and HST/NICMOS Pa$\alpha$ ($right$) maps.}
    \label{anexo:figuresnondual} %
\end{figure*}

\begin{figure*}[htp]
   \centering
   \addtocounter{figure}{-1}
    \includegraphics[width=.7\linewidth]{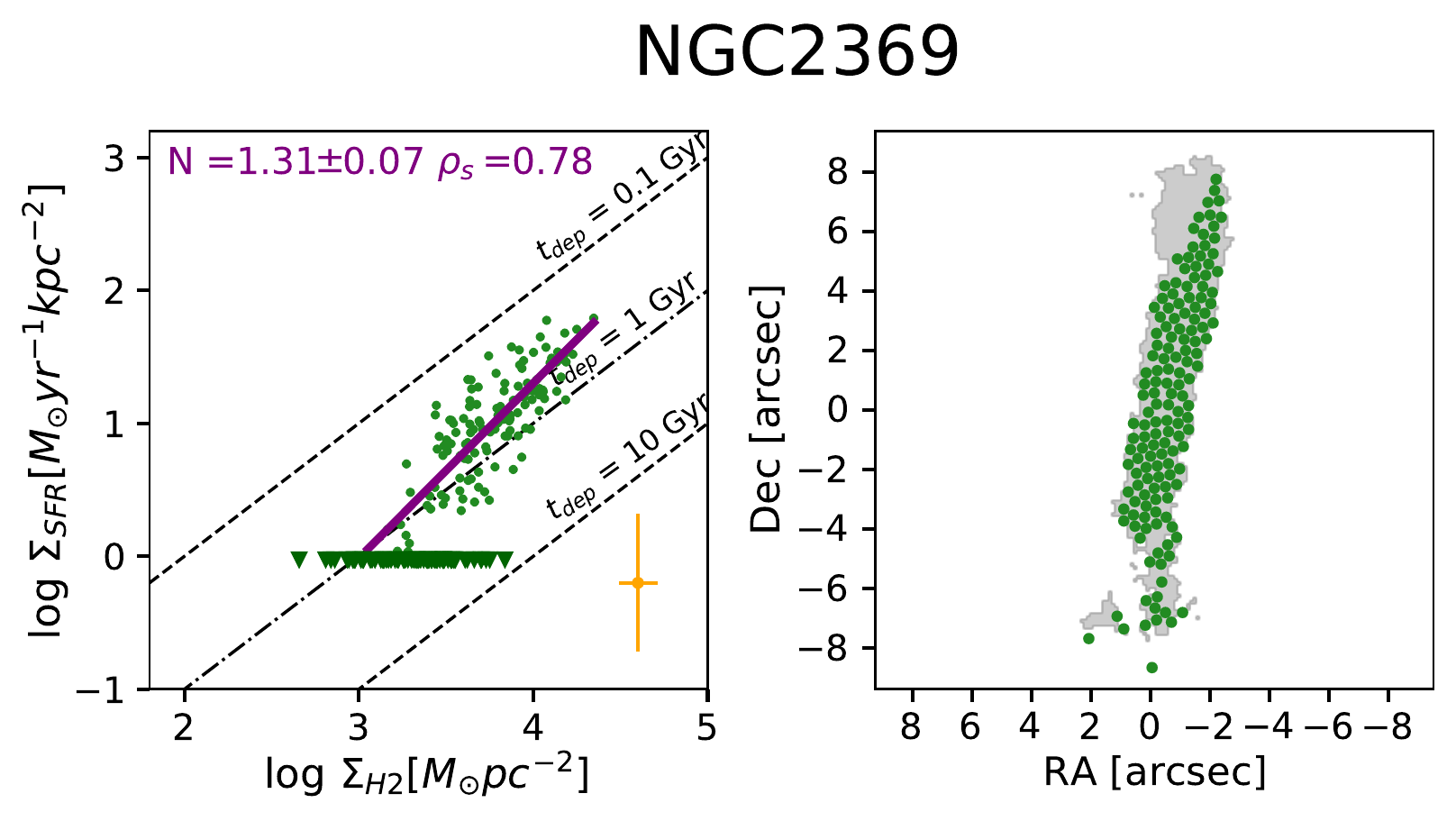}
    \includegraphics[width=.785\linewidth]{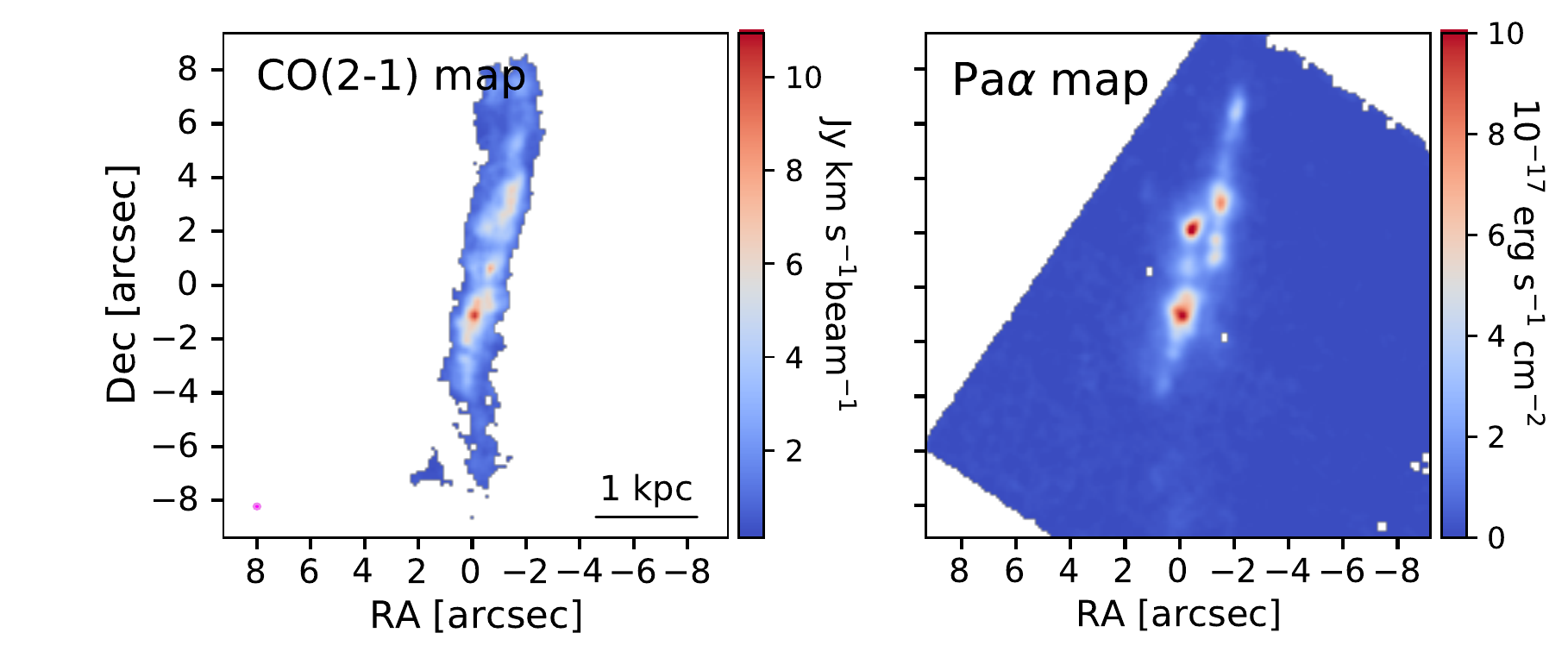}
 \caption{Continued.} 
\end{figure*}

\begin{figure*}[htp]
   \centering
   \addtocounter{figure}{-1}
    \includegraphics[width=.7\linewidth]{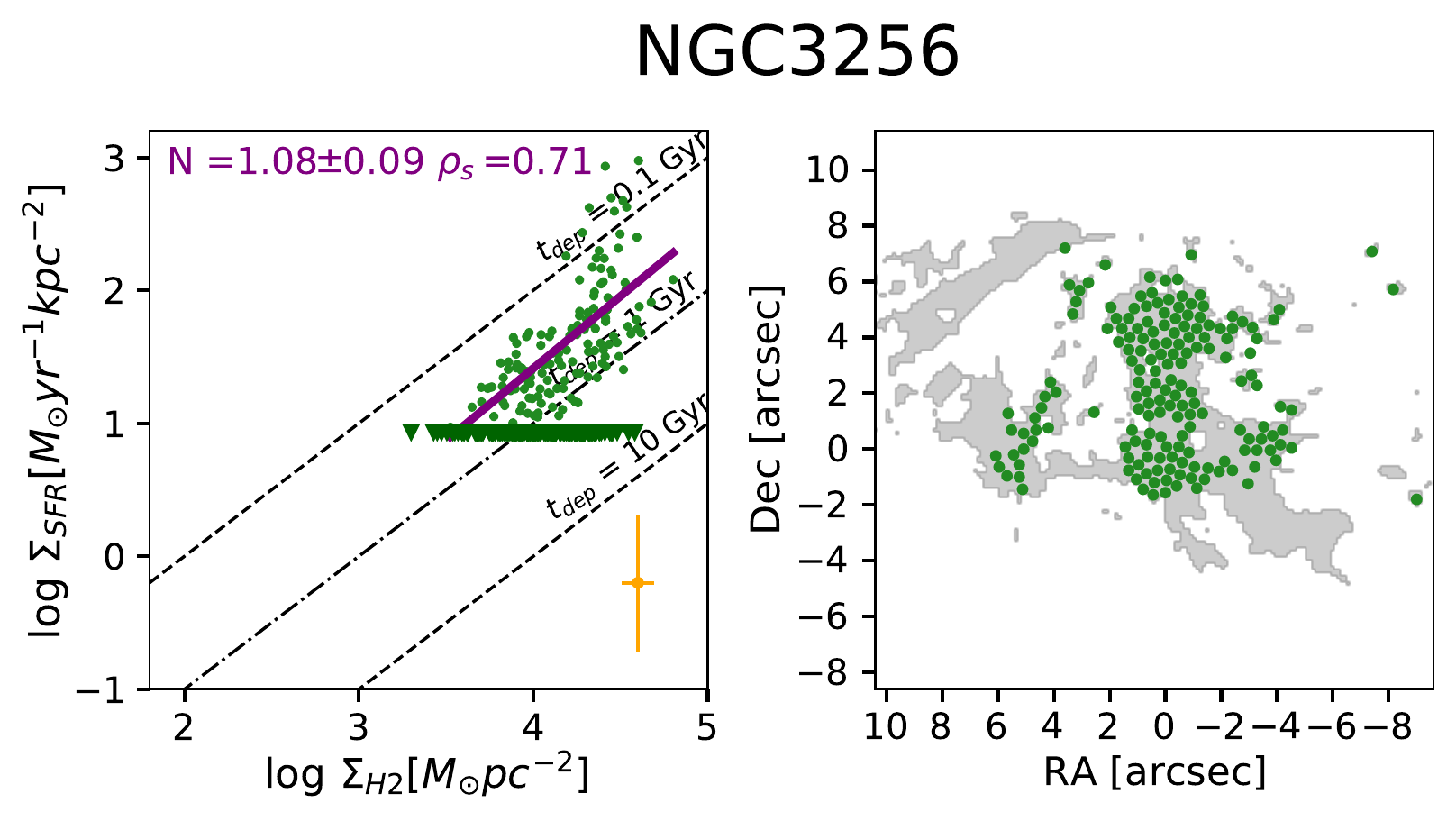}
    \includegraphics[width=.824\linewidth]{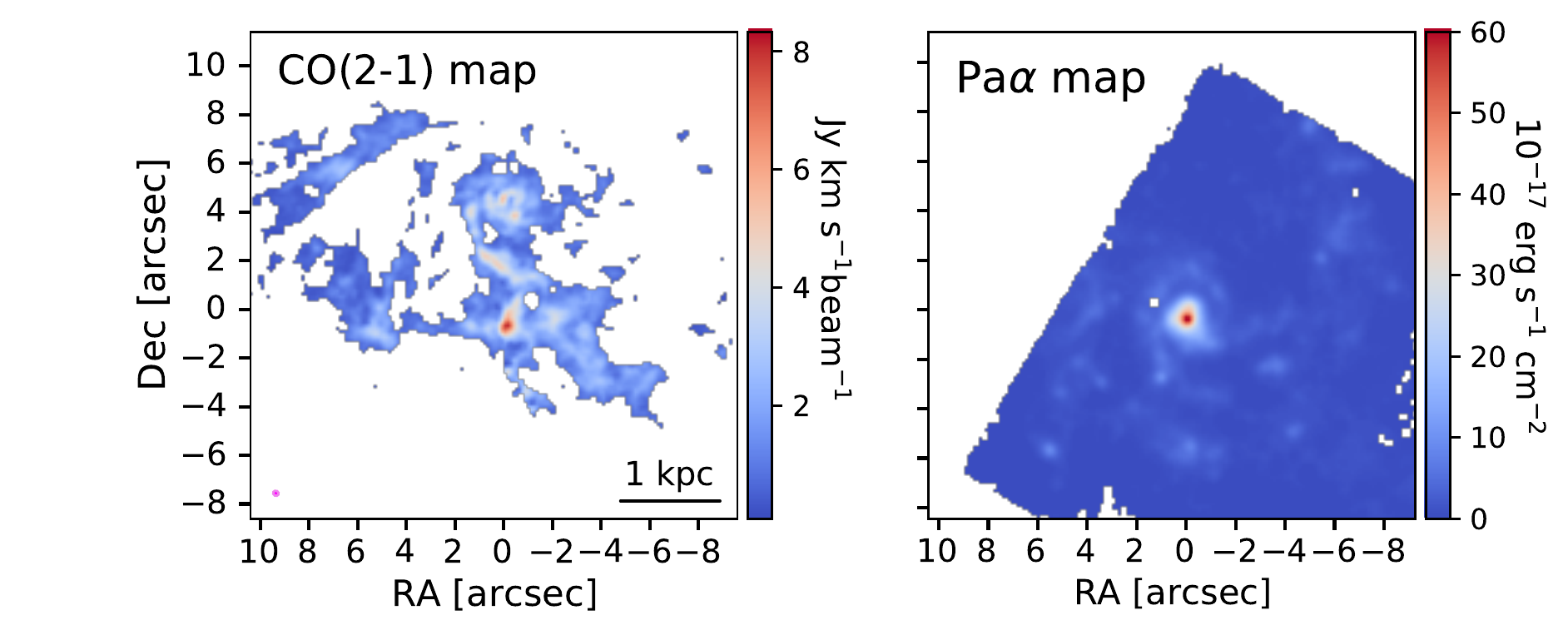}
\caption{Continued.}
   
\end{figure*}

\begin{figure*}[htp]
   \centering
    \addtocounter{figure}{-1}
    \includegraphics[width=.7\linewidth]{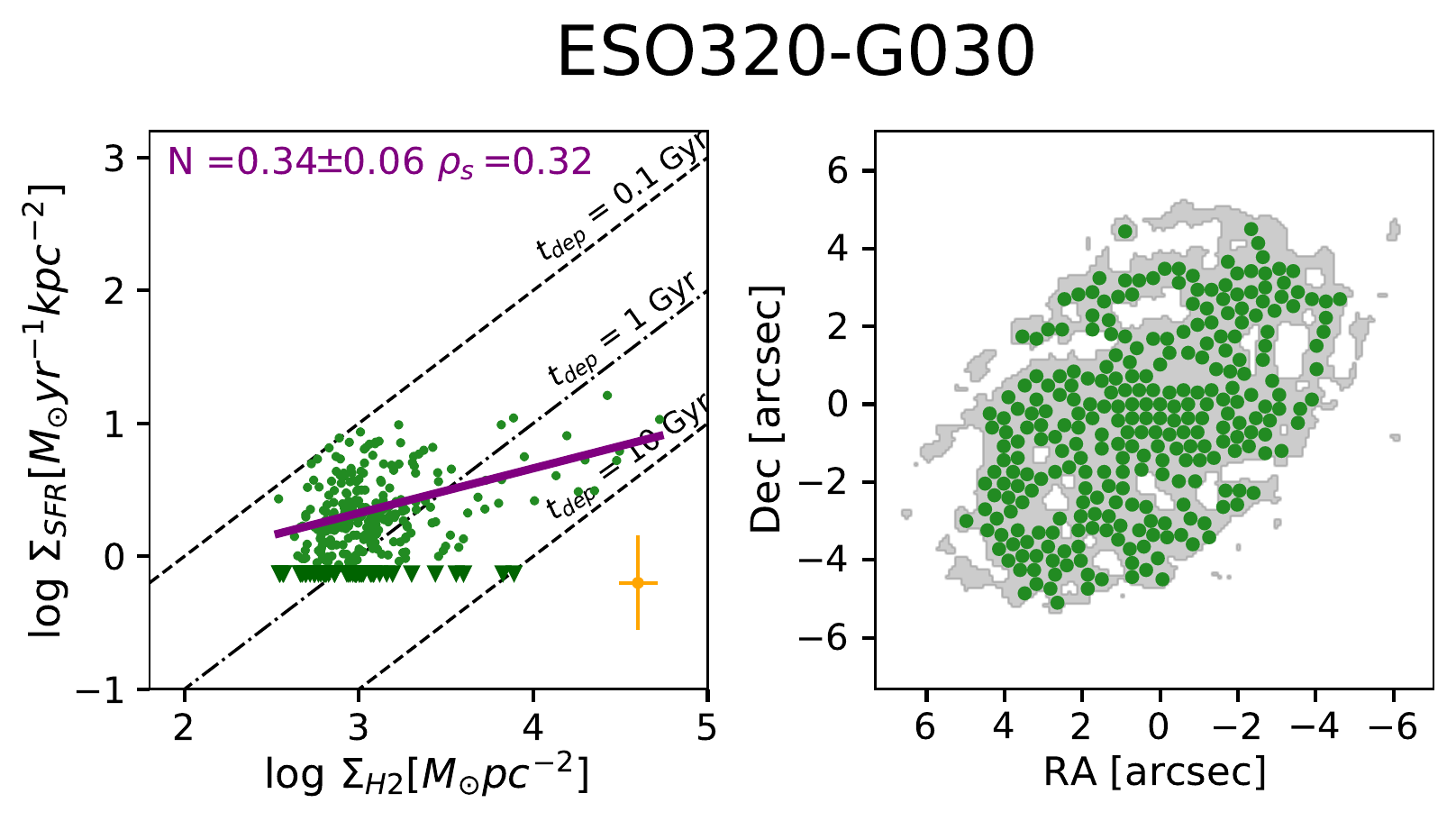}
    \includegraphics[width=.83\linewidth]{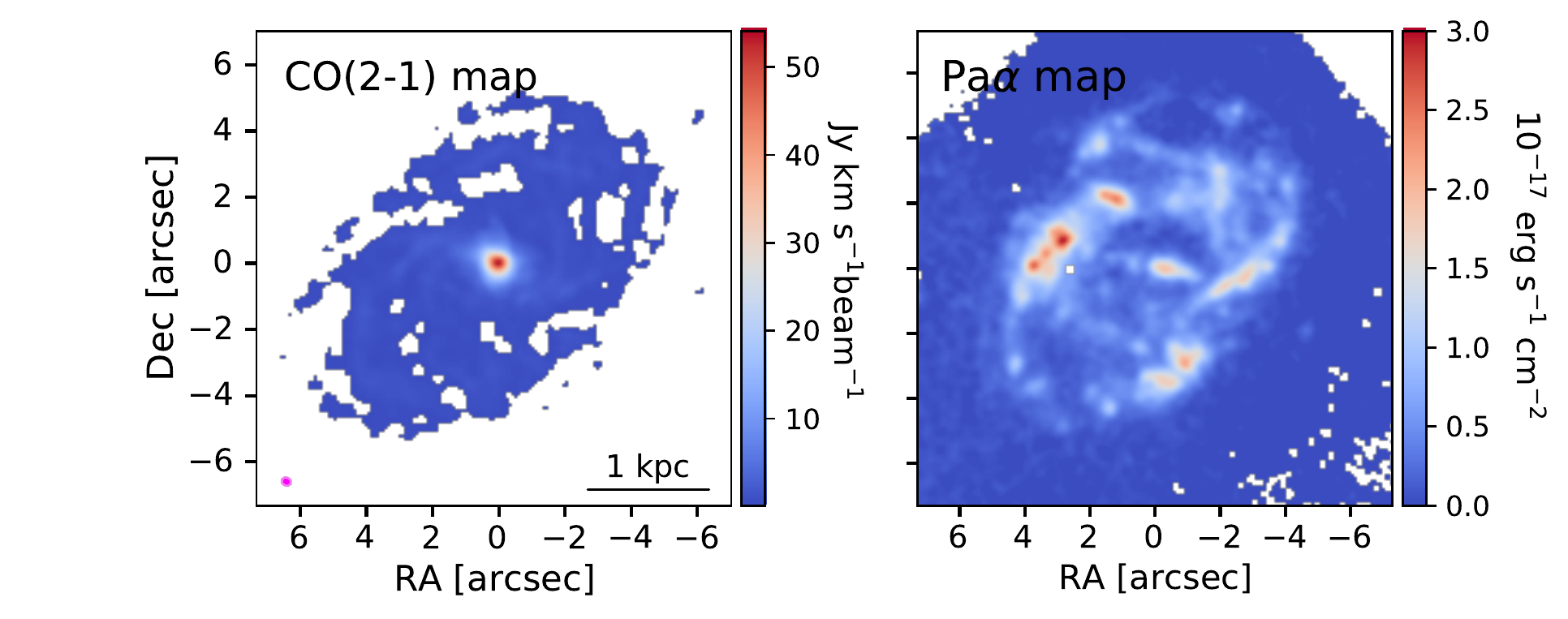}
    
  \caption{Continued.}   
\end{figure*}

\begin{figure*}[htp]
   \centering
   \addtocounter{figure}{-1}
    \includegraphics[width=.7\linewidth]{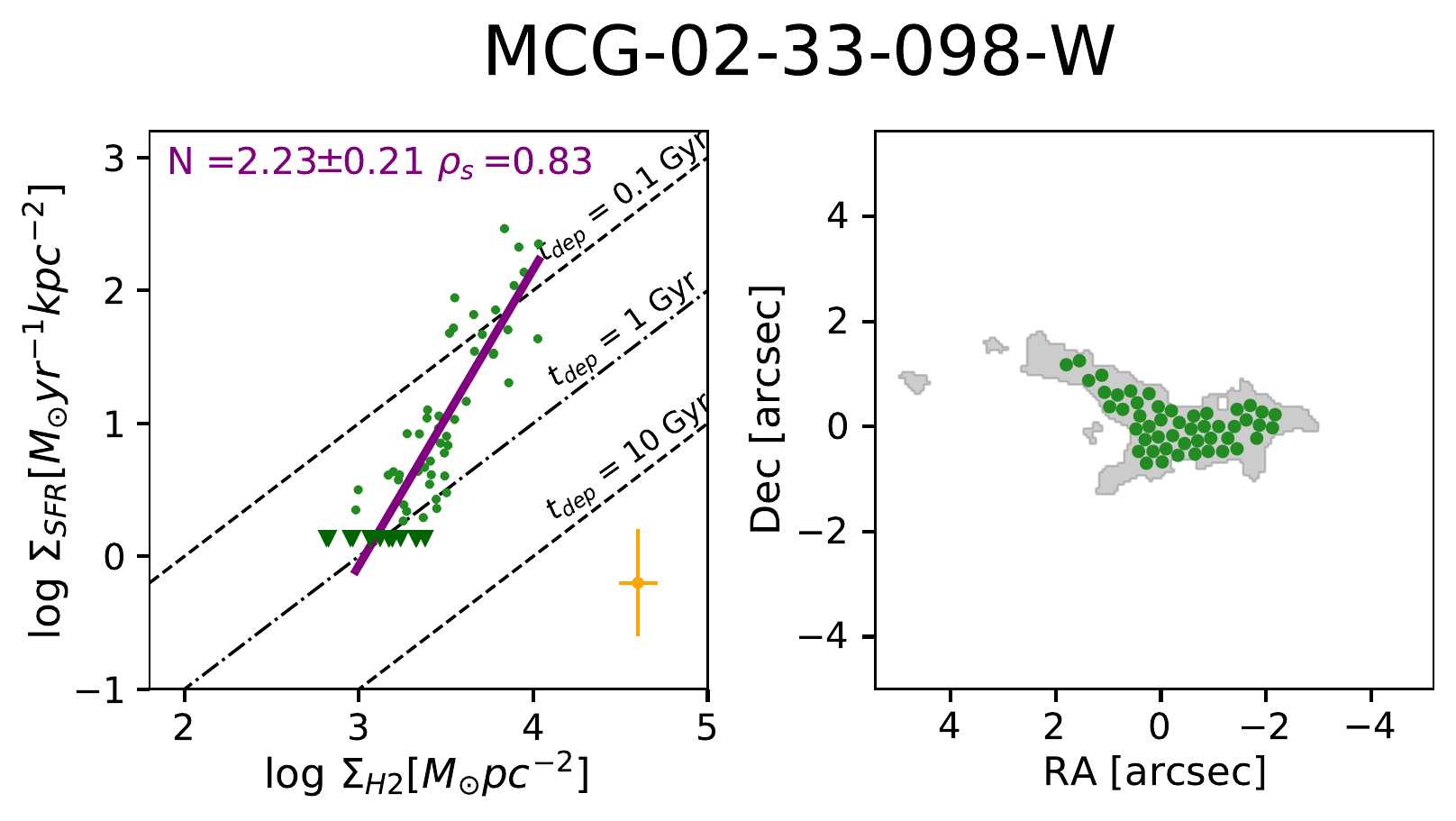}
    \includegraphics[width=.8\linewidth]{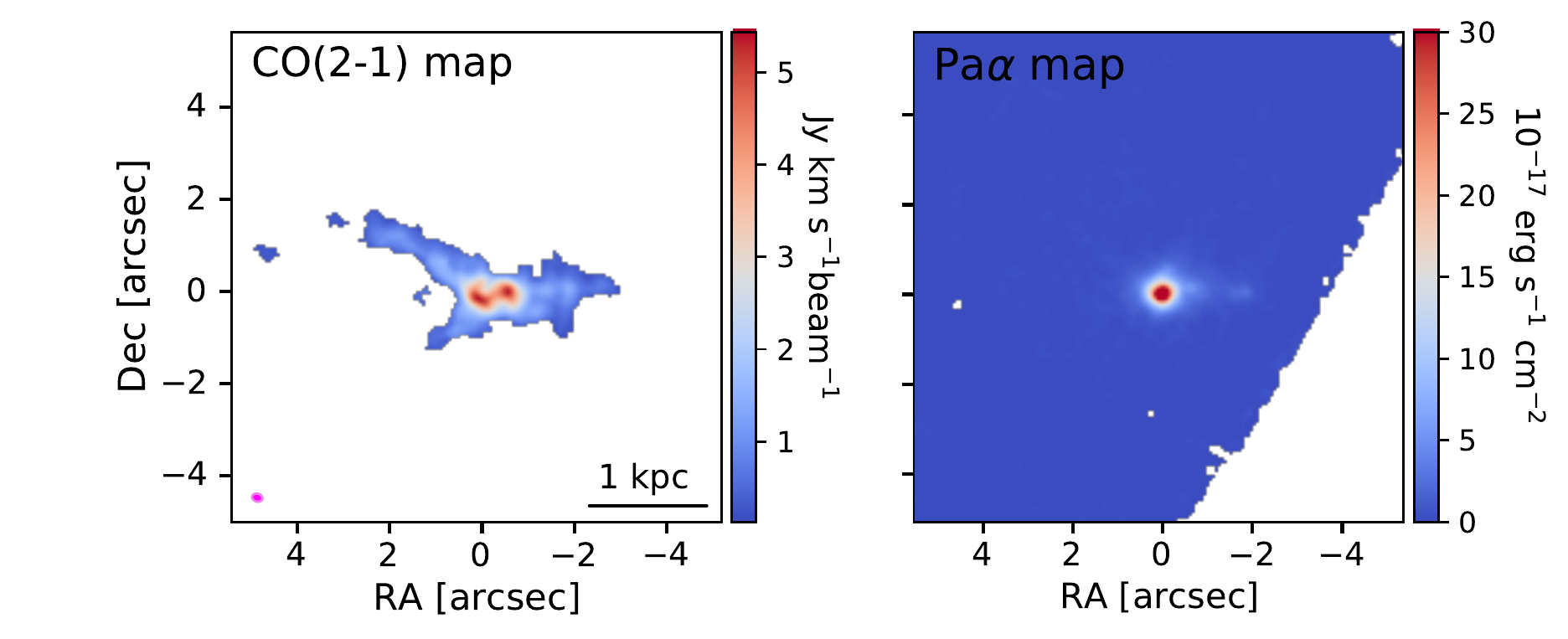}
\caption{Continued.}
  
\end{figure*}

\begin{figure*}[htp]
   \centering
   \addtocounter{figure}{-1}
    \includegraphics[width=.7\linewidth]{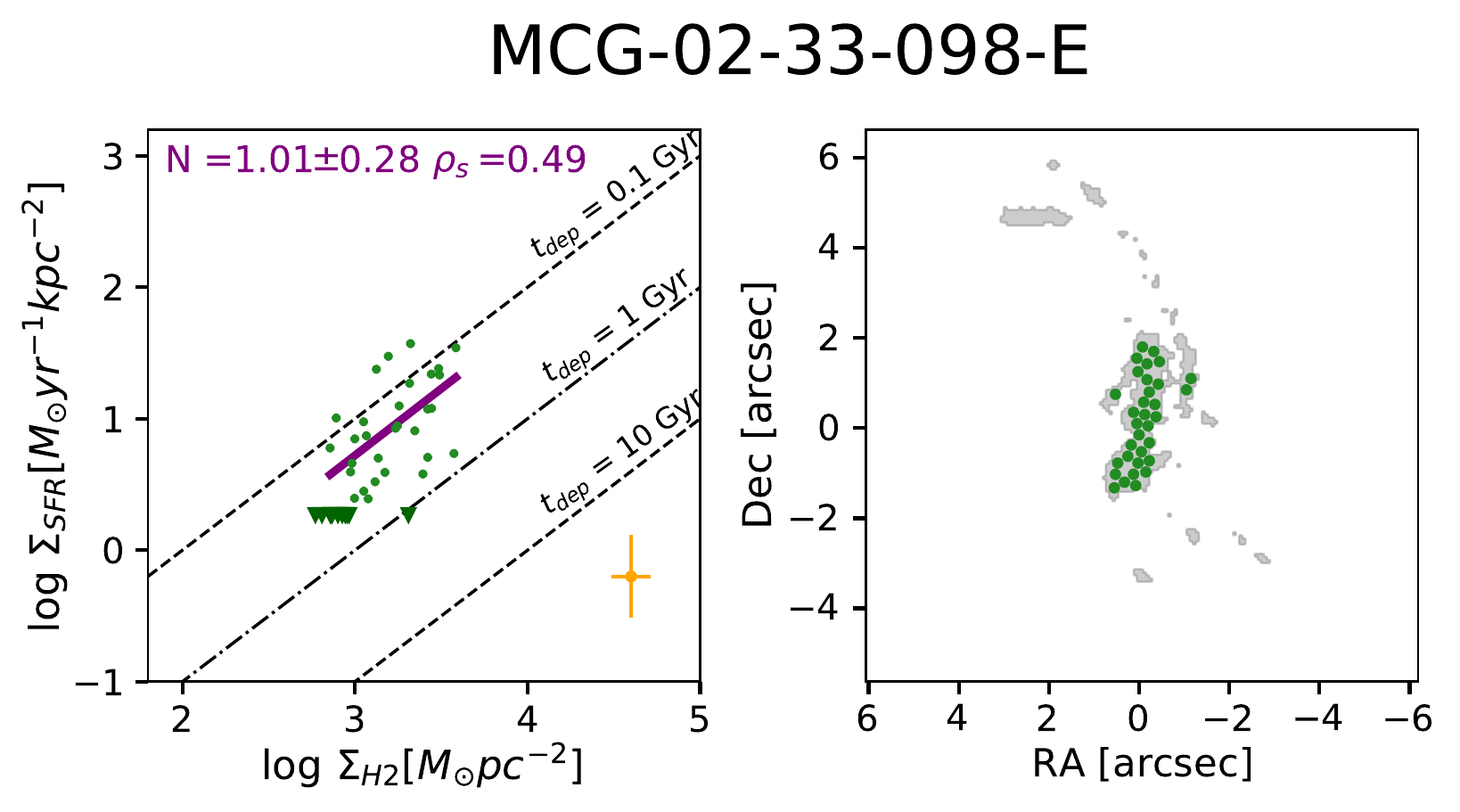}
     \includegraphics[width=.805\linewidth]{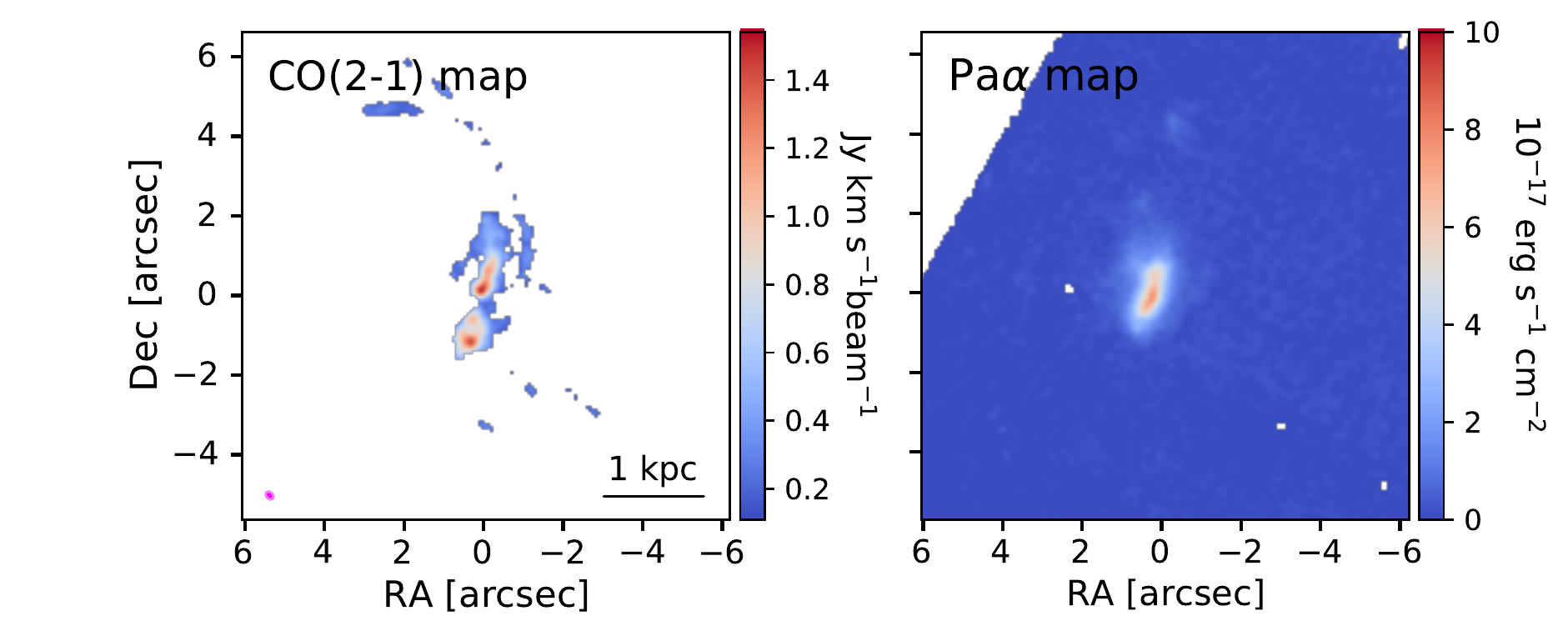}  
 \caption{Continued.}   
\end{figure*}

\begin{figure*}[htp]
   \centering
   \addtocounter{figure}{-1}
    \includegraphics[width=.7\linewidth]{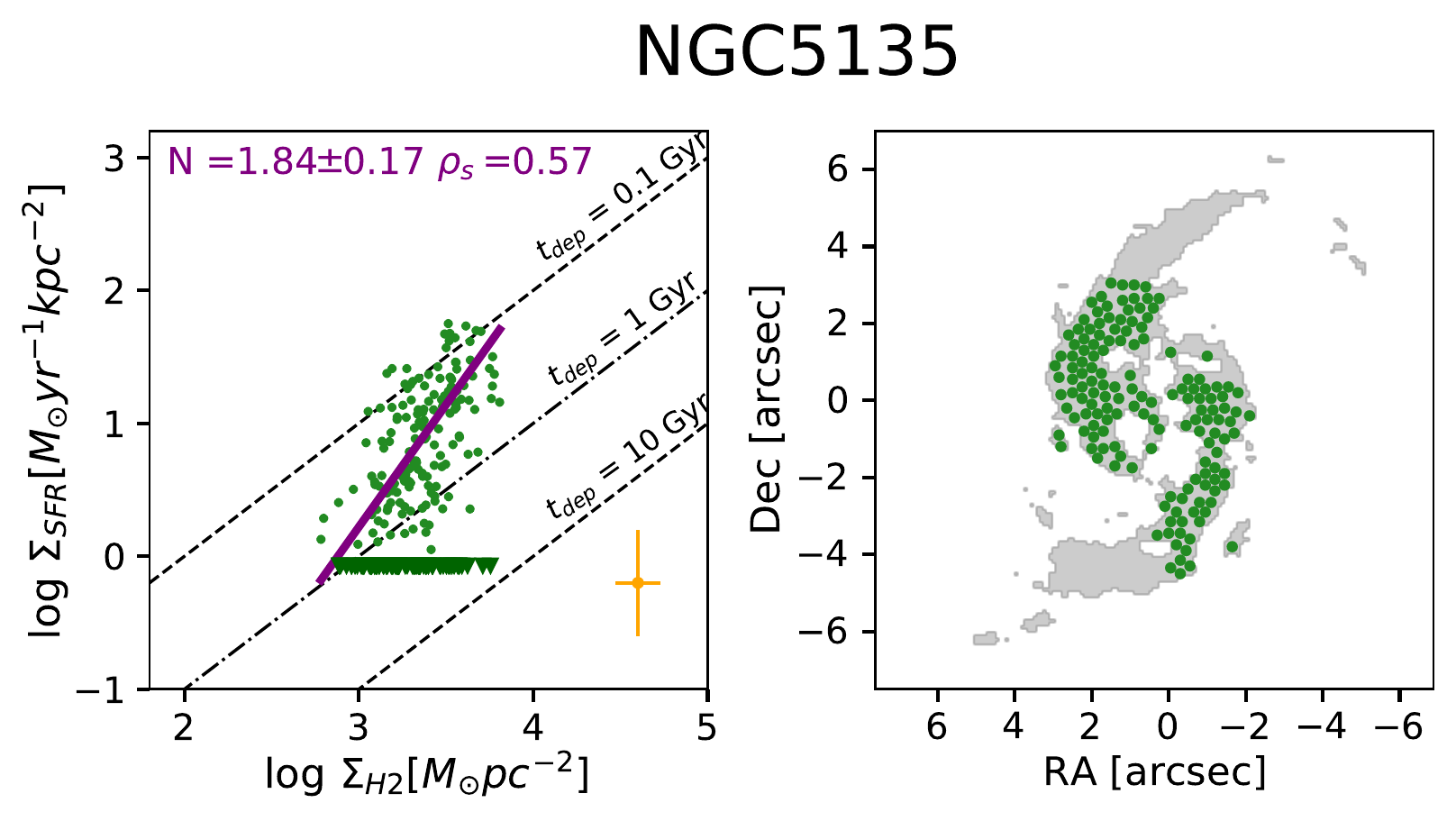}
    \includegraphics[width=.8\linewidth]{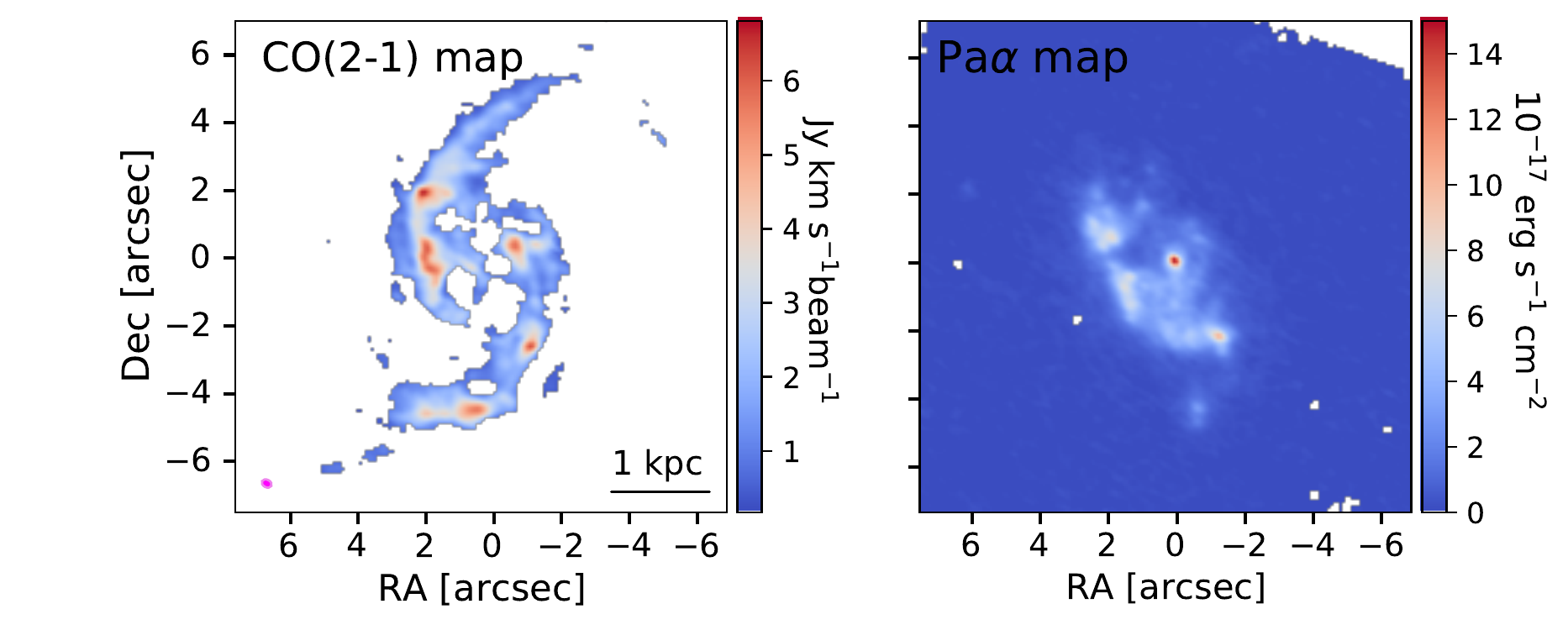}
 \caption{Continued.}   
\end{figure*}

\begin{figure*}[htp]
   \centering
   \addtocounter{figure}{-1}
    \includegraphics[width=.7\linewidth]{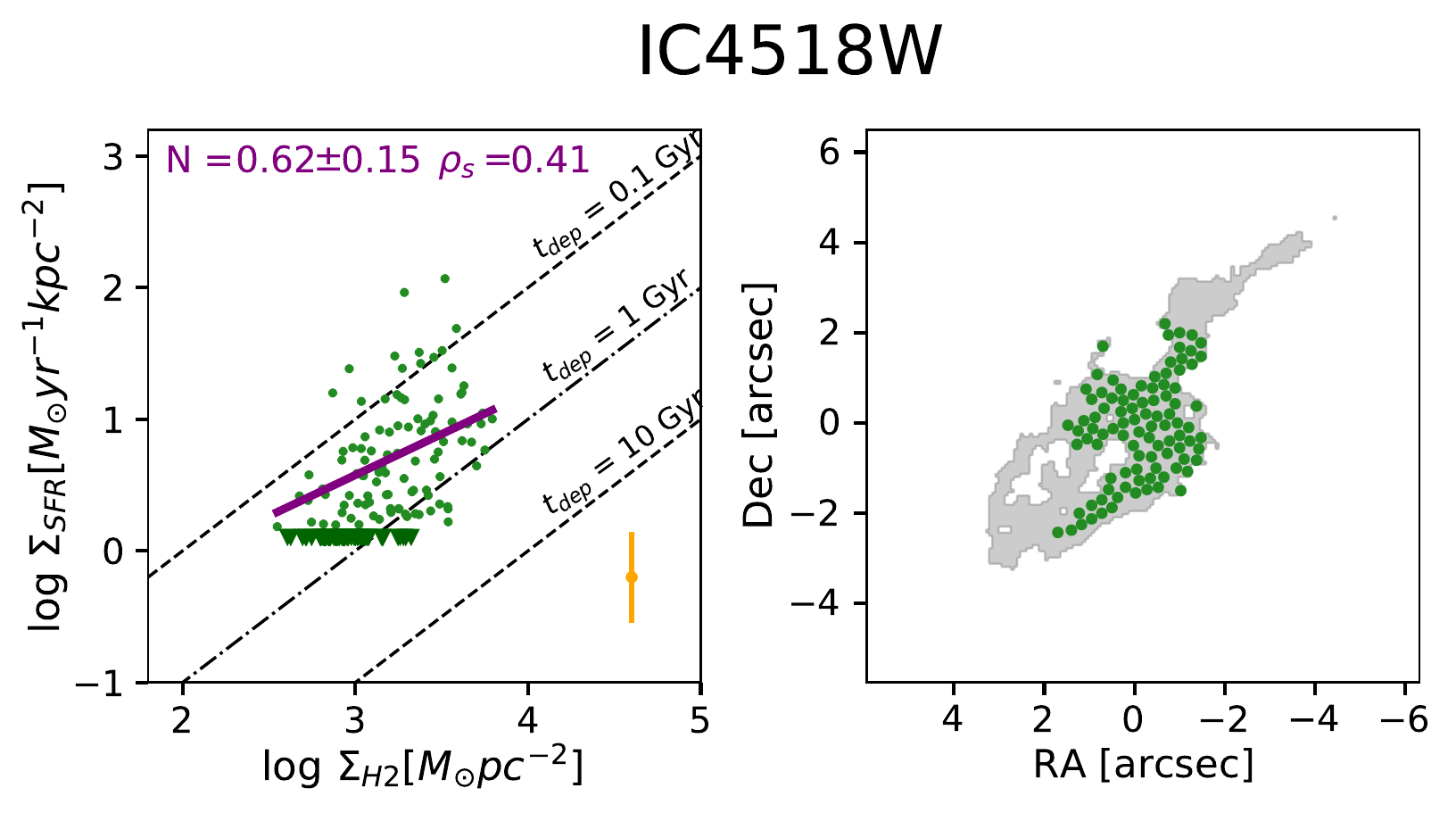}
    \includegraphics[width=.845\linewidth]{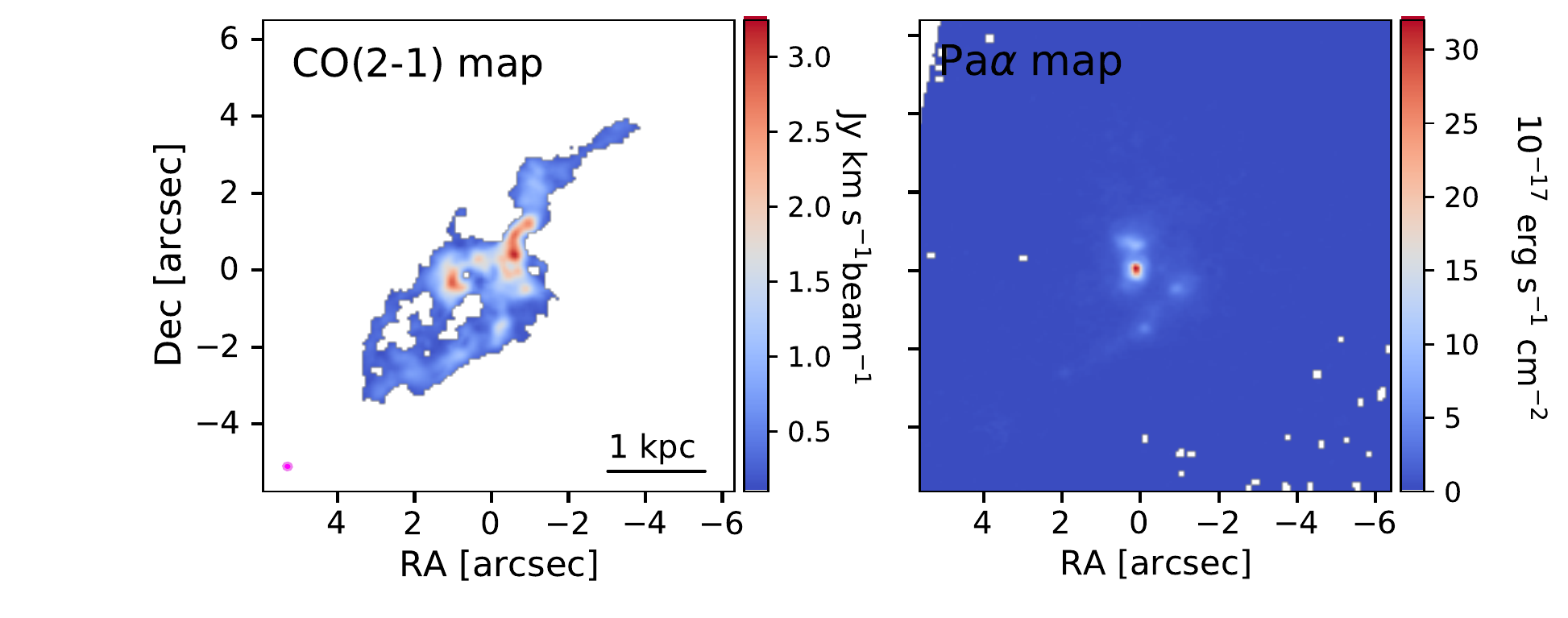}
\caption{Continued.}  
\end{figure*}

\begin{figure*}[htp]
   \centering
   \addtocounter{figure}{-1}
    \includegraphics[width=.7\linewidth]{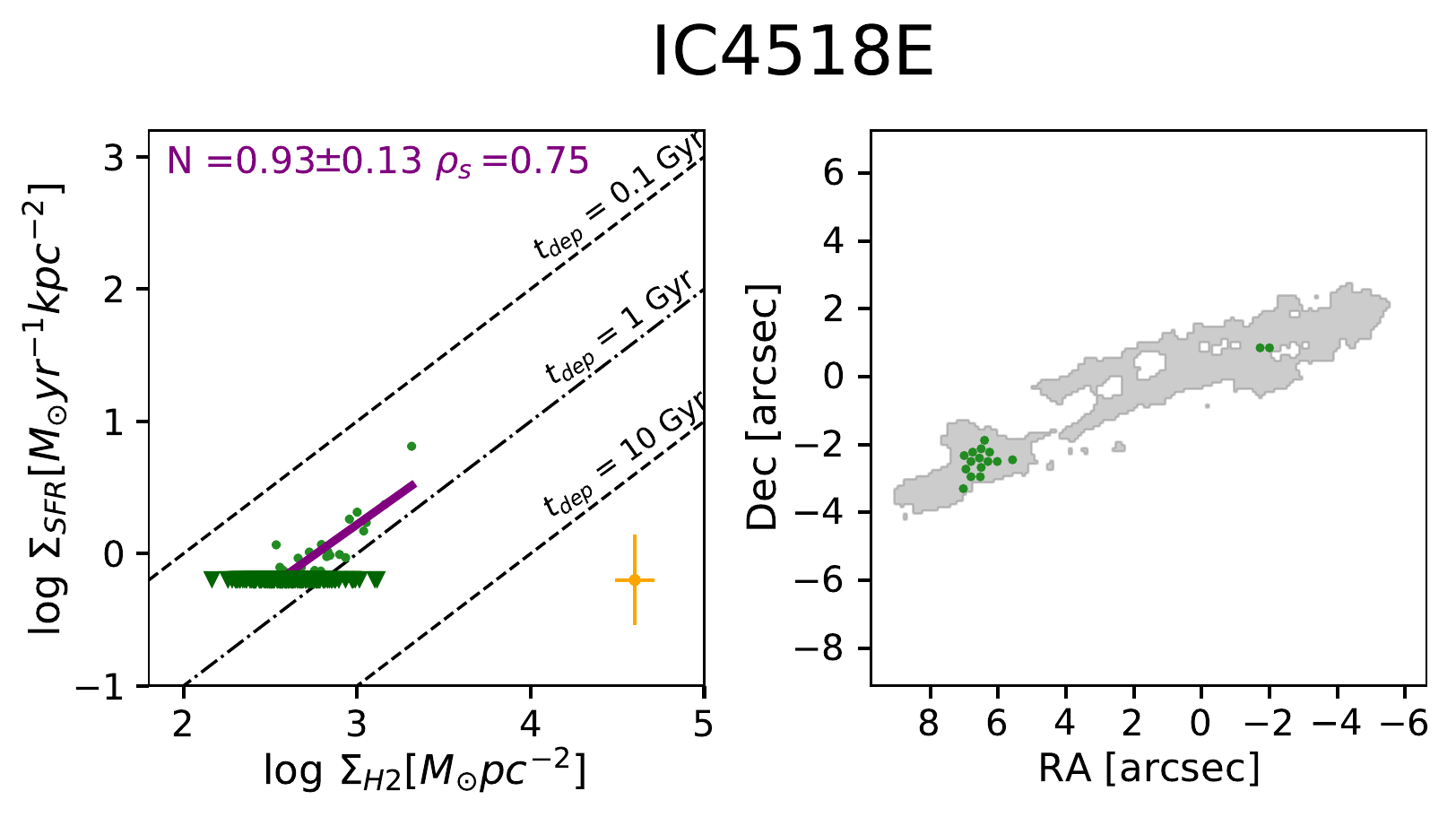}
    \includegraphics[width=.845\linewidth]{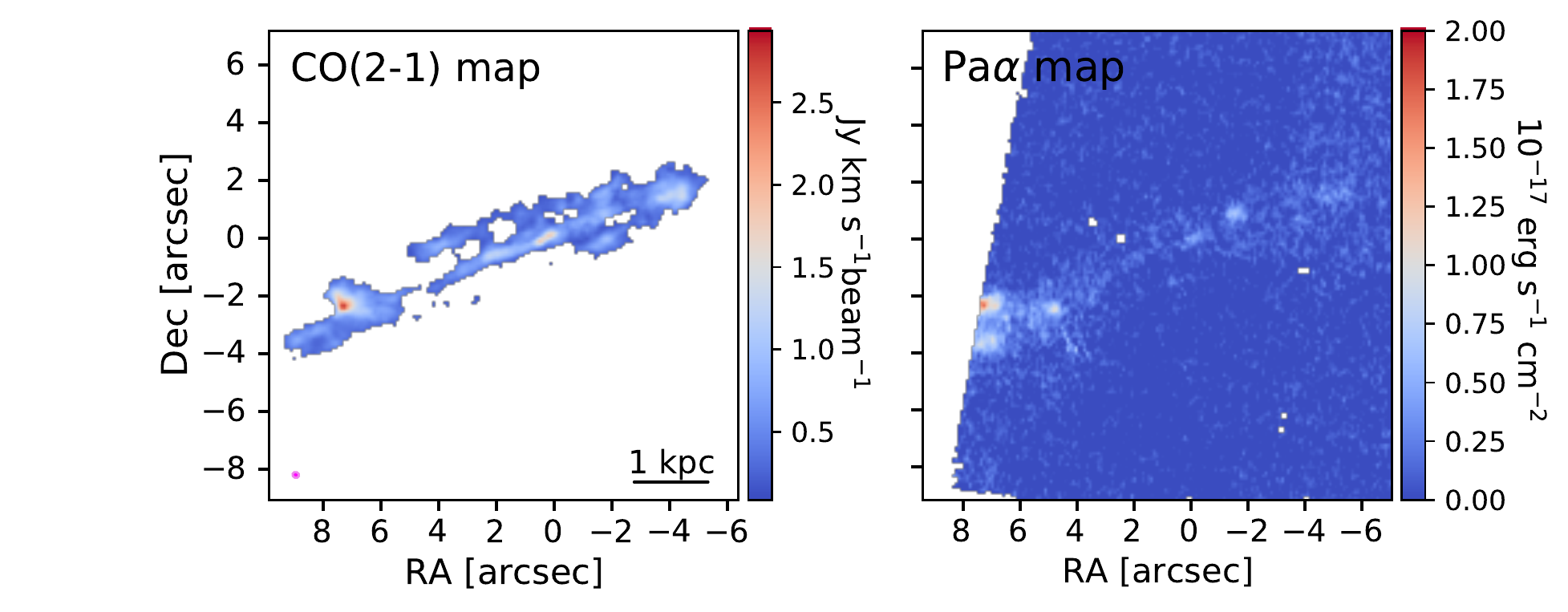}
    \caption{Continued.}
  
\end{figure*}

\begin{figure*}[htp]
   \centering
   \addtocounter{figure}{-1}
    \includegraphics[width=.7\linewidth]{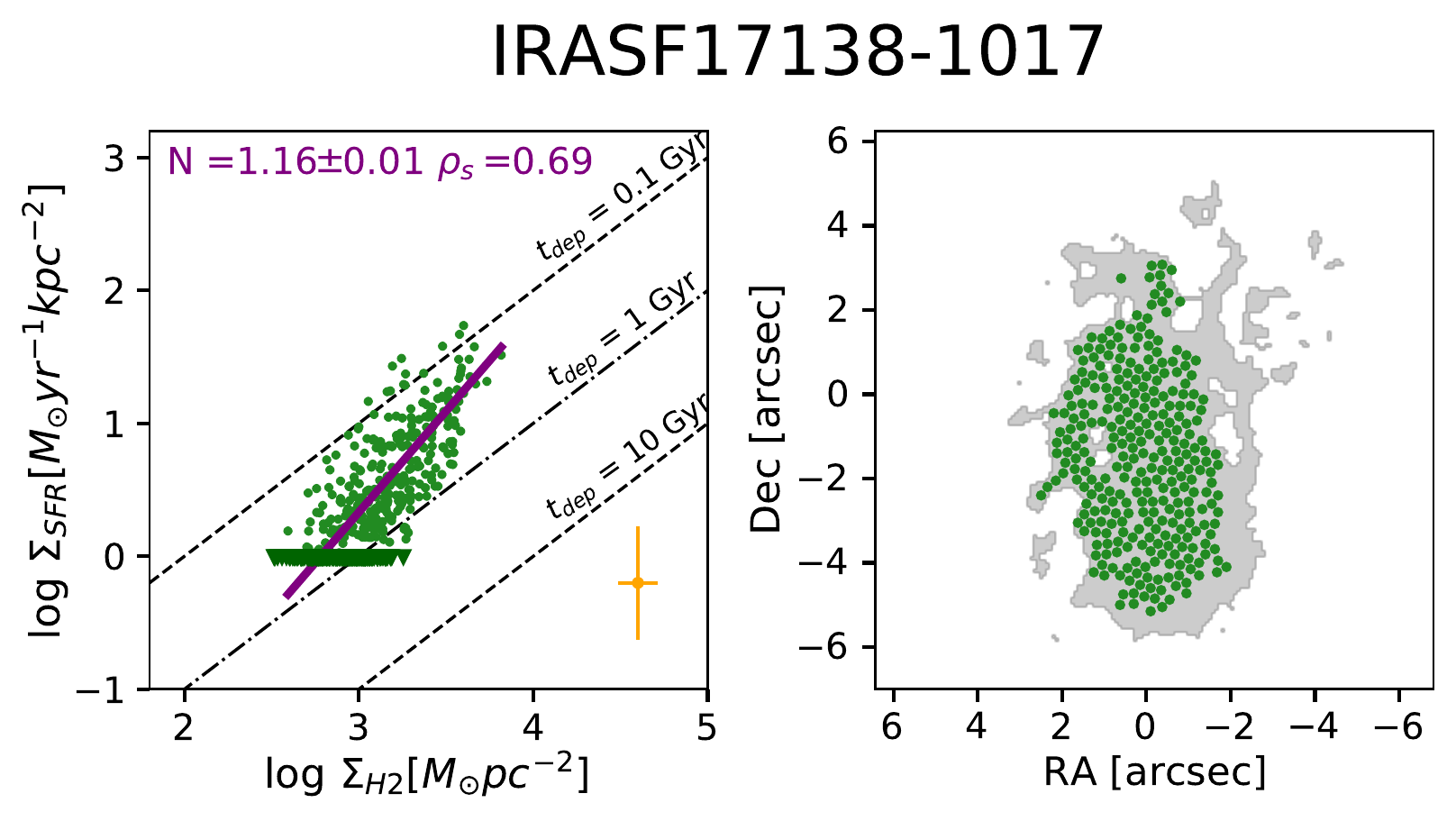}
    \includegraphics[width=.81\linewidth]{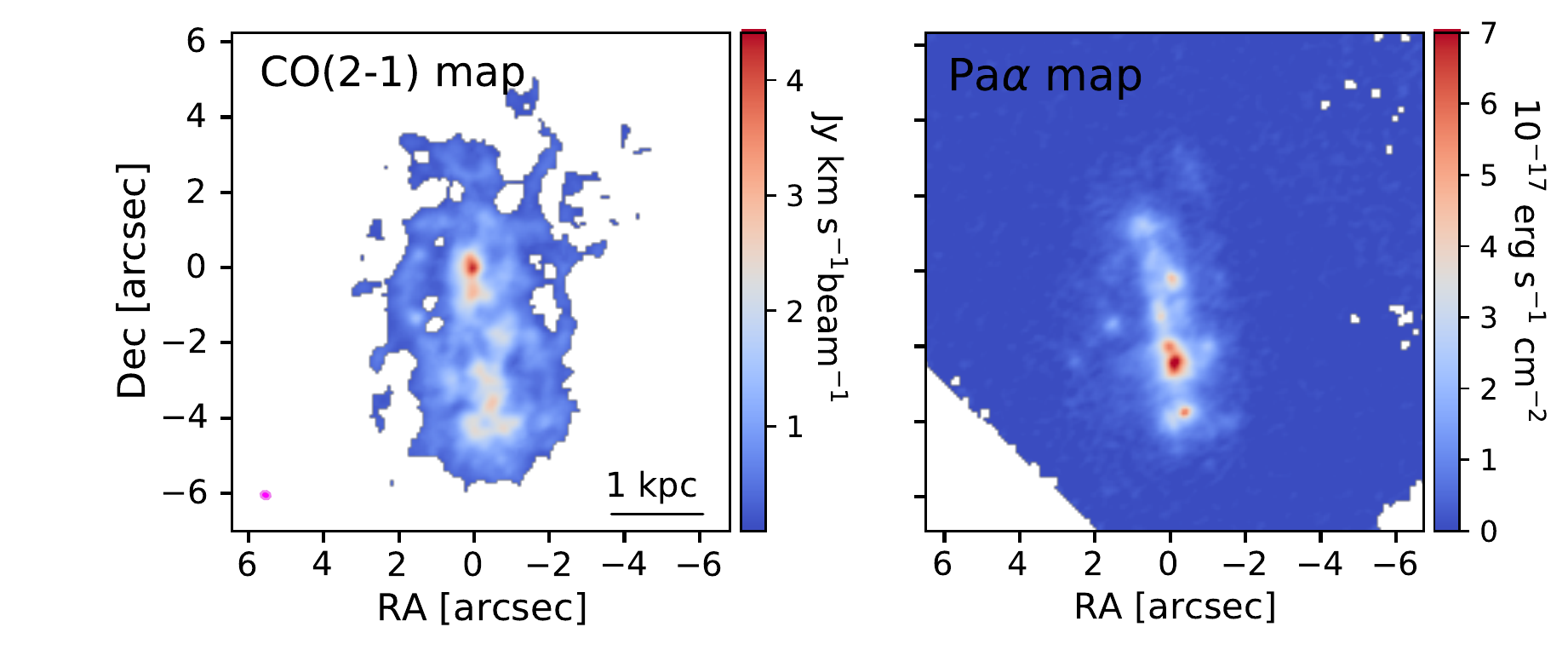}
    \caption{Continued.}  
\end{figure*}

\begin{figure*}[htp]
   \centering
   \addtocounter{figure}{-1}
    \includegraphics[width=.7\linewidth]{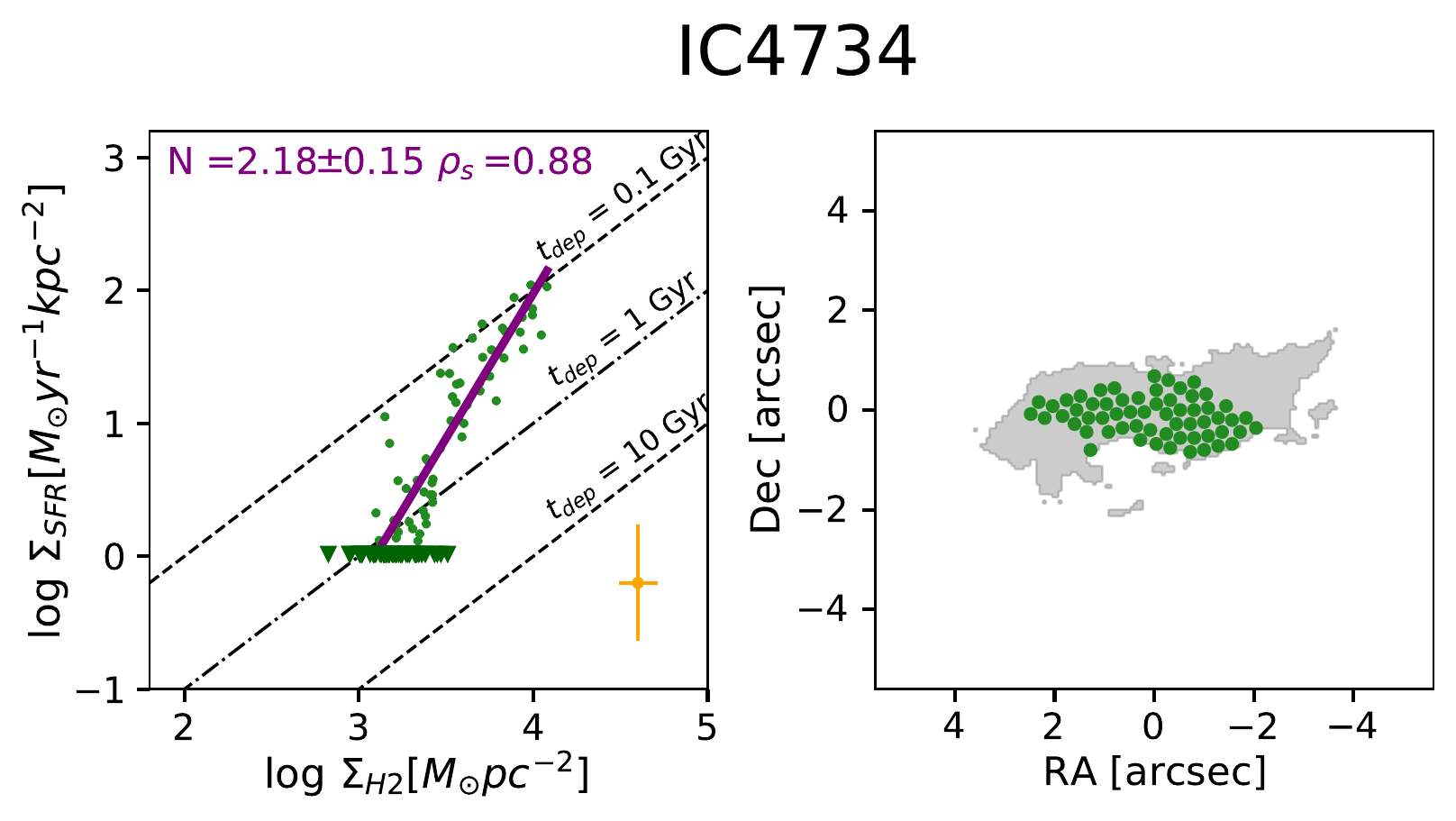}
    \includegraphics[width=.81\linewidth]{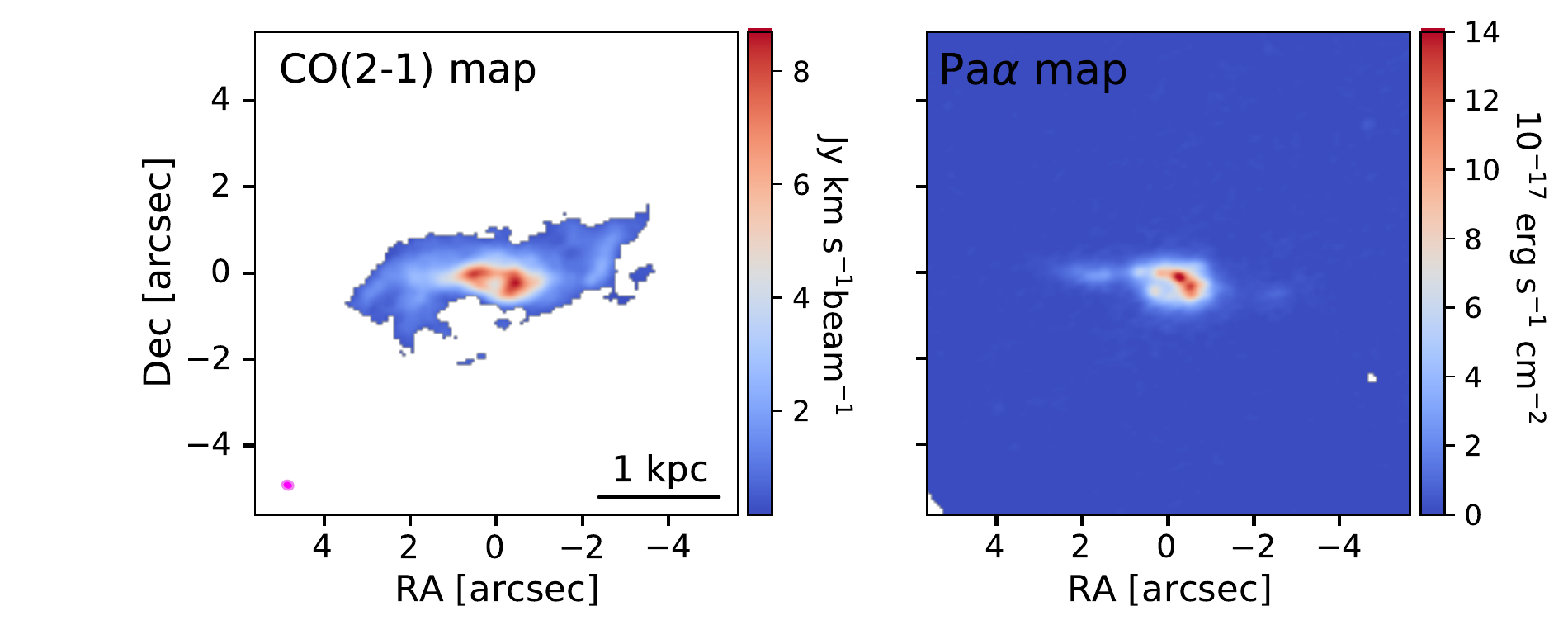}
    \caption{Continued.}  
\end{figure*}

\begin{figure*}[htp]
   \centering
   \addtocounter{figure}{-1}
    \includegraphics[width=.7\linewidth]{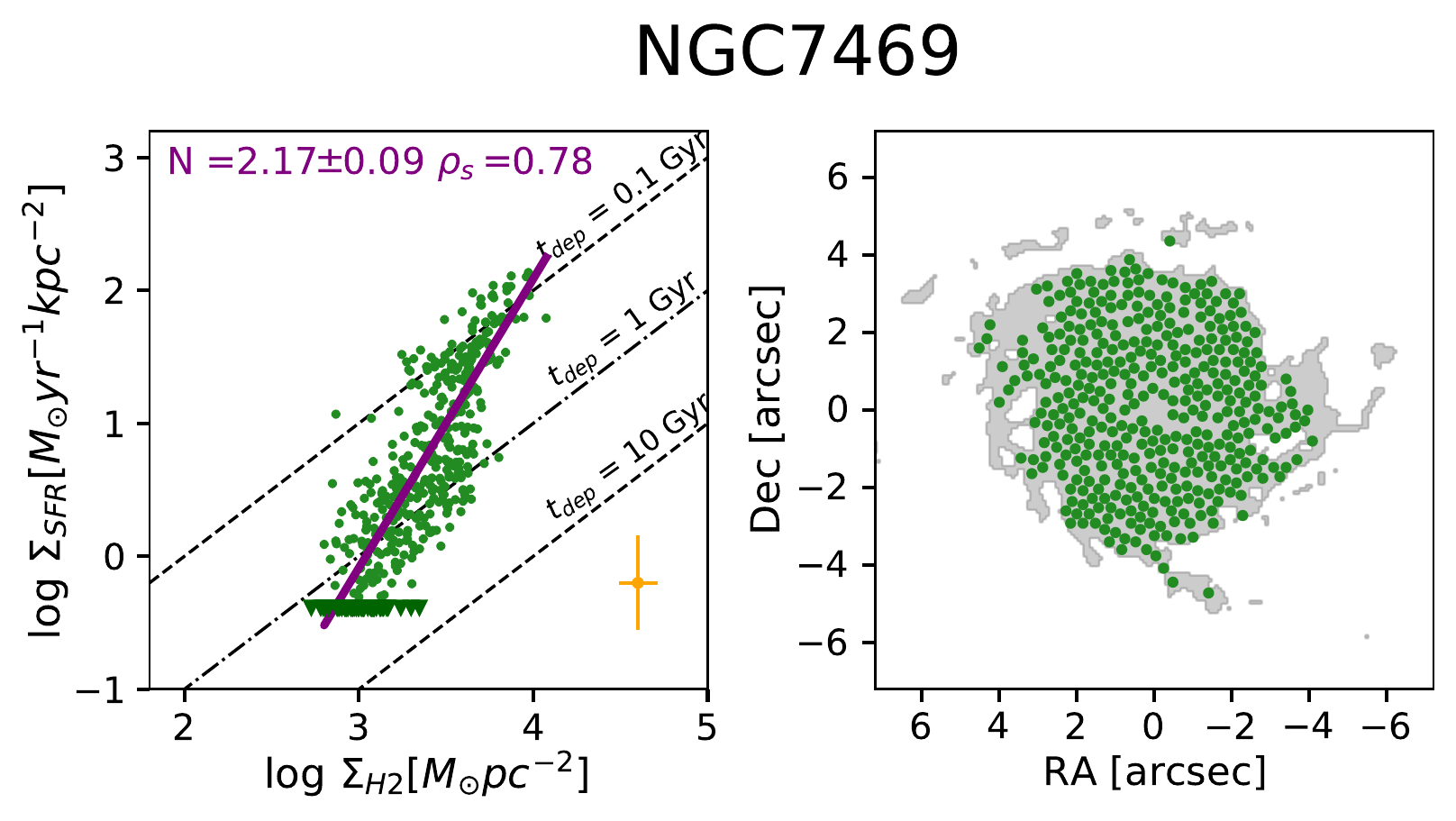}
    \includegraphics[width=.81\linewidth]{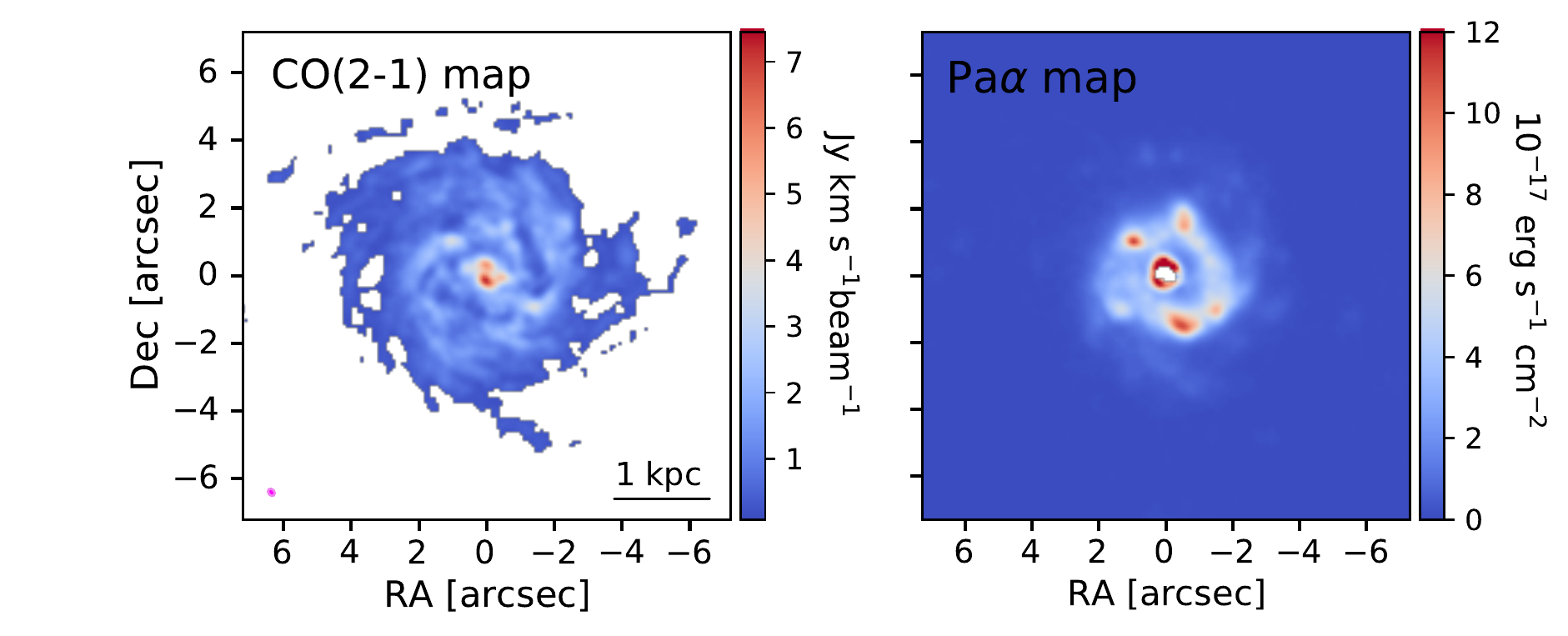}
    \caption{Continued.} 
\end{figure*}

\end{appendix}

\end{document}